\documentclass{aa}  

\usepackage{graphicx}
\usepackage{caption}
\usepackage{xcolor}
\usepackage{enumitem}

\usepackage{txfonts}
\usepackage{hyperref}
\begin{document}

   \title{Coupling between stellar and HI lopsidedness in Milky Way-type galaxies from the Auriga Superstars cosmological simulations}

   \authorrunning{A. Dolfi et al.}
   \titlerunning{Stellar-HI lopsidedness in Auriga Superstars}


   \author{Arianna Dolfi, 
          \inst{1}
          Facundo A. G\'omez,
          \inst{1}
          Rebekka Bieri,
          \inst{2}
          Francesca Fragkoudi,
          \inst{3}
          Robert J. J. Grand,
          \inst{4}
          Antonela Monachesi,
          \inst{1}
          Ruediger Pakmor,
          \inst{6}
          Freeke van de Voort,
          \inst{5}
          }

   \institute{Departamento de Astronom\'ia, Universidad de La Serena, Av. Ra\'ul Bitr\'an 1305, La Serena, Chile\\
              \email{arianna.dolfi@userena.cl} 
   \and Universität Zürich, Institut für Astrophysik, Winterthurerstrasse 190, 8057 Zürich, Switzerland 
   \and Institute for Computational Cosmology, Department of Physics, Durham University, South Road, Durham DH1 3LE, UK 
   \and Astrophysics Research Institute, Liverpool John Moores University, 146 Brownlow Hill, Liverpool, L3 5RF, UK 
   \and Cardiff Hub for Astrophysics Research and Technology, School of Physics and Astronomy, Cardiff University, Queen’s Buildings, Cardiff CF24 3AA, UK 
   \and Max-Planck-Institut für Astrophysik, Karl-Schwarzschild-Str. 1, D-85748, Garching, Germany}

   \date{Received XXX; accepted XXX}

 
   \abstract
   {Lopsidedness is a common feature in disk galaxies, but its origin and evolution remain poorly understood. Previous works have generally focused on either the stellar or gas lopsidedness, separately. However, a combined analysis of stellar and gas lopsidedness can be a powerful tracer of the mechanisms triggering lopsidedness, the recent galactic evolutionary history and the environment in which galaxies reside.}
   {In this work, we study lopsidedness in both the density and kinematics of the stellar and atomic hydrogen (HI) components of nine Milky Way-type galaxies from the Auriga Superstars cosmological zoom-in simulations. The high stellar mass resolution ($m_{\mathrm{star}}=800\, \mathrm{M}_{\odot}$) improves the visibility of features in the stellar disk and halo, reducing the effect of noise on the measurement of lopsidedness. Furthermore, the better sampling of the gravitational potential allows us to investigate how different dynamical processes affect the evolution of galactic disks with unprecedented details in a fully cosmological context.}
   {We quantify morphological and kinematical lopsidedness by measuring the amplitude of the first Fourier mode ($m=1$) of the face-on projected mass distribution and radial velocity map, consistently for stars and gas and over the same spatial region, between $0.5$ and $1\, R_{\mathrm{opt}}$, where $R_{\mathrm{opt}}$ is the stellar optical radius.}
   {At $z=0$, we find a strong correlation between the morphological lopsidedness of the old stellar component ($>0.5\, \mathrm{Gyr\, old}$) and HI, making it a tracer of distortions in the global gravitational potential. In contrast, the young stellar population ($<0.5\, \mathrm{Gyr\, ago}$) traces asymmetric star formation along the strong spiral arms. Morphological and kinematical lopsidedness are strongly correlated in the stellar component, but they are weakly correlated in the HI gas, which is typically characterized by stronger kinematical than morphological asymmetries. We also find an anti-correlation between stellar disk lopsidedness and bar strength, with the strongly barred galaxies typically hosting the most symmetric disks and higher central stellar mass densities.}
   {By tracing the time evolution of lopsidedness over the last $7\, \mathrm{Gyr}$, we find that tidal interactions with massive satellites (mass-ratio $>1$:$50$) produce a coherent lopsided response in both stars and HI gas. In contrast, smooth gas accretion primarily induces lopsidedness in the HI gas and in the young stellar population, while the total stellar component remains largely symmetric. Finally, we find that lopsidedness in individual galaxies is often driven by multiple mechanisms over their evolutionary histories, leading to a complex coupling and decoupling between the distinct tracers. Overall, our results demonstrate that lopsidedness is not merely a morphological feature, but a powerful diagnostic of the coupling between internal disk evolution, gas accretion, and environmental interactions across cosmic time.}

   \keywords{ Galaxies: evolution -- Galaxies: interactions -- Galaxies: kinematics and dynamics -- Galaxies: spiral -- Galaxies: structure }

   \maketitle
%

\section{Introduction}
\label{sec:introduction}
Disk galaxies often display asymmetric mass and light distribution, known as lopsidedness \citep{Baldwin1980}, in which the outer regions are more extended on one side than on the opposite side. Lopsidedness is also observed in the kinematics where the rotation curve rises more steeply on one side of the galaxy than on the opposite one. Observational studies have shown that, both in the local Universe and at high-redshift (i.e. $z\lesssim3$), at least one third of the galaxies have lopsided stellar disks \citep{Zaritsky1997,Bournaud2005,LeBail2025}. 
Lopsidedness is observed not only in the spatial distribution of the old stellar component traced in the near-infrared \citep{Block1994,Rix1995}, but also in both the spatial distribution and kinematics of atomic hydrogen gas, i.e. HI \citep{Richter1994,Schoenmakers1997,Haynes1998,Swaters1999}.
Specifically, \citet{Richter1994} found that at least $50\%$ of galaxies show strong asymmetries in their global HI velocity profiles, whose shape is influenced by both the spatial distribution and the kinematics (i.e. an asymmetry in the global HI profile can result from either a lopsided spatial distribution and/or lopsided kinematics). Since lopsidedness is stronger at larger radii, it is not surprising that a larger fraction of lopsided galaxies is identified in HI, which is more extended than the stellar component.

Despite lopsidedness being a common phenomenon in galactic disks, its origin and evolution are still not well understood. One of the main processes suggested as the origin of lopsidedness are tidal interactions \citep{Beale1969,Weinberg1995}. These are expected to generate a strong but relatively short-lived lopsided perturbation, lasting no more than $\sim2\, \mathrm{Gyr}$. However, the observed lack of correlation between lopsidedness and the presence of nearby satellite galaxies \citep{Bournaud2005,Wilcots2010} suggested that other processes must be at play as well. For example, later studies suggested that lopsidedness can arise as a result of the disk response to a previously tidally perturbed dark matter halo \citep{Weinberg1995,Jog1997,Jog2002} or from torques acting as a result of an off-centered disk with respect to the dark matter halo \citep{Noordermeer2001}. Furthermore, \citet{Bournaud2005} found that late-type disk galaxies are generally more lopsided in the stellar density distribution than early-type disks in low-density environments. Considering that repetitive tidal encounters and mergers can heat the stellar disk and trigger multiple starbursts, ultimately producing a gas-poor lenticular galaxy \citep{Bournaud2004,Bekki2011,Deeley2021}, a tidally-induced scenario appears inconsistent with the observed larger fraction of lopsided late-type disks. For this reason, \citet{Bournaud2005} suggested asymmetric gas accretion as the origin of lopsidedness.
Using cosmological simulations from the IllustrisTNG project, \citet{Lokas2022} showed that lopsided galaxies tend to be characterized by higher star formation rates, bluer colors, lower metallicity and higher gas fraction than symmetric galaxies at $z=0$, consistent with the asymmetric gas accretion scenario. Similar conclusions were also reached by \citet{Dolfi2024}, who showed that lopsided galaxies have typically experienced more significant recent net gas accretion rates onto their galactic disk than symmetric galaxies. This is also consistent with the finding that lopsided galaxies have a distinct star formation history with respect to their symmetric counterparts at $z=0$ \citep{Dolfi2023}. They are characterized by a steadier mass growth until recent times, possibly due to the continuous accretion of gas that kept their star formation rate overall constant up to $z=0$.
However, lopsidedness is not only observed in the young but also in the old stellar component, meaning that recent asymmetric gas accretion cannot be the only process that generates lopsidedness. 

A lopsided ($m=1$) perturbation results in a shift of the disk center of mass with respect to its density cusp, which then acts as an internal force on the original center. This affects not only the density distribution but also the orbits of the particles in the disk and introduces kinematic asymmetries. In the plane of the disk, non-circular motions produce a radial velocity component $V_{\mathrm{R}}$ that appears as a dipolar feature (i.e. $V_{\mathrm{R}}>0$ on one side of the disk and $V_{\mathrm{R}}<0$ on the opposite side), while the tangential velocity $V_{\Phi}$ shows an amplitude difference between the two halves of the galaxy, with one rotation curve rising more steeply than the other \citep{Jog1997,Jog2002}. For this reason, if lopsidedness arises as a result of the disk response to an external perturbation potential, the galaxy is expected to show both morphological and kinematical asymmetries. 

The combined study of stellar and gas lopsidedness in both the density and kinematics can provide important constraints on the mechanisms that trigger a lopsided perturbation. Processes such as gas accretion or ram-pressure stripping that act primarily on the gas are expected to produce a strong lopsidedness in the gas without a corresponding stellar response. Specifically, only a response in the young stellar component is expected as new stars are born from an underlying asymmetric gas distribution. On the contrary, processes such as tidal interactions are expected to produce a similar lopsidedness in both stars and gas, since both respond to the same gravitational perturbation potential.

Previous observational studies have typically characterized lopsidedness either in the stars or gas components. In a recent work, \citet{Feng2025} studied the ionized gas velocity maps of a sample of disk galaxies from the MaNGA survey to quantify the presence of kinematic asymmetries and the correlation with global galaxy properties. They found that galaxies with large kinematic asymmetries tend to lie either above or below the star forming main sequence (SFMS) and below the mass-metallicity relation, regardless of their position on the SFMS. In particular, galaxies with low star formation rates (SFR) and high kinematic asymmetries tend to have low stellar masses (i.e. $\log_{10}\, \mathrm{M}_{*}/\mathrm{M}_{\odot}<10.2$ within the sample), low gas metallicity, and high HI gas content. For this reason, they suggest that the accretion of metal-poor gas is the main driver of the gas velocity map asymmetries in these galaxies, with the reduced star formation rate in the lower-mass systems likely associated with less efficient cooling. In another recent work, \citet{Ghosh2025} studied the origin and evolution of a merger-induced lopsided perturbation in the stellar component of a Milky Way-like galaxy, using numerical simulations. They showed that the interaction with a merging satellite galaxy produces a significant lopsided perturbation in both the stellar density and stellar velocity distributions. Furthermore, they showed that the lopsided amplitude, as well as the correlation between stellar density and stellar kinematic lopsidedness, is stronger for the metal-rich stellar component (i.e. $\mathrm{[Fe/H]}>0$) than for the metal-poor one. They argue that this is because the metal-rich stellar component has lower velocity dispersion (i.e. kinematically colder) than the metal-poor one, thus it is more susceptible to external perturbations. This latter result is in agreement with the finding from the previous simulation works of \citet{VarelaLavin2023} and \citet{Fontirroig2025}. Using cosmological simulations, they showed that there is a strong correlation between lopsidedness and the internal properties of the galaxies with lopsided galaxies being characterized by lower central stellar mass density and being less gravitationally cohesive than symmetric galaxies.

Only the later work of \citet{Bilimogga2025} attempted to study the correlation between morphological and kinematical asymmetries in HI. They also investigated the correlation between morphological asymmetries in the stellar component and in the HI gas in their observed galaxy sample, located in two distinct group environments. They found that there is no straightforward correlation between morphological and kinematical asymmetries. Specifically, \citet{Bilimogga2025} studied the correlation between the asymmetry in the global HI velocity profile and in the HI spatial density, finding that a galaxy with high asymmetry in the global HI profile can have either an asymmetric HI velocity field with a symmetric HI spatial distribution or an asymmetric HI spatial distribution with an asymmetric or symmetric HI velocity field. Furthermore, they noted that an asymmetry in the HI spatial density is not necessarily associated with an asymmetry in the stellar density based on the visual inspection of the optical images, consistent with the lack of correlation between stellar and gas morphological lopsidedness found by \citet{Lokas2022} in the IllustrisTNG simulations. Although a detailed quantification of the stellar asymmetry was not performed in \citet{Bilimogga2025}, their results indicate that different lopsidedness triggering mechanisms might be at play to produce different levels of morphological and kinematical asymmetries in the stellar and gas components.

A study investigating the correlation between the stellar and gas lopsidedness both in the density and velocity distributions is still missing. In this work, we carry out such study using the sample of Milky Way-type galaxies from the novel Auriga Superstars simulations \citep{Grand2017,Pakmor2025_Superstars}. The high stellar mass resolution (i.e. $m_{\mathrm{star}}=800\, \mathrm{M}_{\odot}$) of these simulations allows us to resolve the stellar disk structure with unprecedented detail and to study its evolution in a full cosmological context by investigating the effects of different dynamical processes (e.g. tidal interactions and/or gas accretion) on the properties of galactic disks. 

The paper is structured as follows. In Sec. \ref{sec:data}, we describe the simulations from the Auriga project, the implementation of the Superstars method as described in \citet{Pakmor2025_Superstars}, and the resulting Superstars galaxy sample. In Sec. \ref{sec:presentday_stellar_and_gas_lopsidedness}, we characterize the lopsidedness of the stellar and HI gas disks at $z=0$ and study the correlation between morphological and kinematical asymmetries in both the stellar and HI gas components. In Sec. \ref{sec:drivers_lopsidedness}, we investigate the evolution of lopsidedness and identify the dominant physical mechanisms driving the detected morphological and kinematical asymmetries in the stellar and HI gas components of our galaxies. Finally, in Sec. \ref{sec:conclusions}, we provide a summary of the results and our conclusions. 
In Appendix \ref{sec:convergence_lopsidedness}, we study the effect of changing the stellar mass particle resolution on the measurement of lopsidedness, while, in Appendix \ref{sec:temporal_evolution}, we visualize the stellar and HI density distributions of a sub-sample of our galaxies at consecutive output times.

\section{The data}
\label{sec:data}

\subsection{The Auriga project}
\label{sec:auriga}
The original Auriga project consists of a suite of cosmological gravo-magnetohydrodynamical zoom-in simulations run with the moving-mesh code \texttt{AREPO} \citep{Springel2010}. These halos were selected from a dark-matter-only counterpart of the Eagle simulations at $z=0$ to be relatively isolated and have halo mass between $0.5 \times 10^{10} < \mathrm{M}_{200}/\mathrm{M}_{\odot} < 2 \times 10^{12}$ \citep{Grand2017,Grand2024}. This halo mass range thus includes $26$ dwarf galaxies (i.e. $0.5 \times 10^{10} < \mathrm{M}_{200}/\mathrm{M}_{\odot} < 5 \times 10^{11}$) and $30$ Milky Way-type galaxies (i.e. $0.5 \times 10^{12} < \mathrm{M}_{200}/\mathrm{M}_{\odot} < 2 \times 10^{12}$).

The Auriga halos have been simulated at different resolutions. The original Auriga simulations (L4, hereafter) trace the evolution of all halos from redshift $z=127$ out to $z=0$, with a mass resolution of $m_{\mathrm{DM}}=3\times 10^{5}\, \mathrm{M_{\odot}}$ for dark matter and $m_{\mathrm{star/gas}}=5\times 10^{4}\, \mathrm{M_{\odot}}$ for star particles and gas cells \citep{Grand2017}. The softening length of the stellar and dark matter particles evolves with time up to $z=1$ and corresponds to a comoving softening length of $750\, \mathrm{cpc}$, while the maximum physical softening length at $z<1$ is set to $375\, \mathrm{pc}$. For the gas cells, the physical softening length is scaled by the radius of the gas cell, with a minimum limit set to a value of $375\, \mathrm{pc}$ as for the stellar particles \citep{Grand2017,Grand2024}. 

The Auriga simulations include models for primordial and metal-line cooling, for the multiphase interstellar medium, stochastic star formation and stellar evolution, for stellar feedback, as well as for the seeding and growth of supermassive black holes and feedback from active galactic nuclei, and for magnetic fields. See \citet{Grand2017} for a detailed description.
The simulations assume a $\Lambda$ Cold Dark Matter model with the following cosmological parameters taken from the Planck Collaboration XVI \citep{Planck2014}: Hubble constant $H_{0}=100h\, \mathrm{kms^{-1}Mpc^{-1}}$ where $h=0.6777$, matter density $\Omega_{\mathrm{m}}=0.307$, baryonic mass density $\Omega_{\mathrm{b}}=0.048$, dark energy density $\Omega_{\Lambda}=0.693$ \citep{Grand2017,Grand2024}. 
These simulations are now publicly available\footnote{\url{https://wwwmpa.mpa-garching.mpg.de/auriga/index.html}.}, and a detailed description of the data is provided in \citet{Grand2024}. 

\subsection{Auriga Superstars}
\label{sec:auriga_superstars}
The Auriga Superstars simulations (Superstars, hereafter) consists of nine Milky Way-type halos re-simulated with an improved stellar and dark matter particle mass resolution. The mass resolution of the gas cells is kept at the original L4 resolution (i.e. $m_{\mathrm{gas}}=5\times10^{4}\, \mathrm{M}_{\odot}$), because the galaxy formation model implemented in Auriga was calibrated at this resolution.
Within the Superstars simulations, the dark matter particle mass resolution is $m_{\mathrm{DM}}=5\times10^{4}\, \mathrm{M}_{\odot}$ (i.e. a factor of $8\times$ better than that at the original L4 resolution), while the stellar particle mass resolution is $m_{\mathrm{star}}=8\times10^{2}\, \mathrm{M}_{\odot}$ (i.e. a factor of $64\times$ better than that at the original L4 resolution and than the gas cell resolution).

The increased dark matter particle mass resolution is set at the beginning of the simulation when the initial conditions are created.
On the other hand, the increased stellar particle mass resolution relative to the gas cell resolution is obtained according to the following method. When a gas cell is converted into a stellar particle, $64$ stellar particles are created rather than just one. The total mass of the gas cell is then divided equally among the stellar particles that were created. As a result, the mass resolution of the stellar particles can be much better than that of the gas. 
The newly formed stellar particles inherit the velocity of the parent gas cell plus an additional "kick" velocity term, so they evolve following different trajectories. The requirement is that this additional velocity is large enough such that the stellar particles born from the same gas cell are not bound together and small enough such that the properties of the galaxies do not change. The value of the "kick" velocity is unique to each stellar particle and is randomly drawn from a normal distribution with mean zero. The standard deviation is set to the minimum between the local velocity dispersion, computed from the parent gas cell and its direct neighbors, and the local sound speed. A detailed description and implementation of the Superstars method is given in \citet{Pakmor2025_Superstars}.

The implementation of the Superstars method improves the visibility of substructures and features in both the stellar disk and halo, without significantly changing the global properties of the galaxies. \citet{Pakmor2025_Superstars} studied several properties of the galactic disks, such as the stellar mass surface density, the velocity dispersion, the optical radius and the disk scale height, and found that differences between the Superstars and L4 models are small and within the range of expected variability between the models due to random noise \citep{Pakmor2025_Superstars,Pakmor2025}. Similarly, several global properties of the galaxies, such as the halo and stellar mass and the star formation rate, are also converged between the Superstars and L4 models \citep{Pakmor2025_Superstars}.
In addition to the improved mass resolution, Superstars includes high-cadence outputs of the positions and velocities of the stellar particles every $5\, \mathrm{Myr}$.

The improved stellar mass resolution of Superstars allows to resolve faint features in the stellar disk outskirts and halo, reducing the effect of noise on the measurement of lopsidedness. In fact, in Appendix \ref{sec:convergence_lopsidedness}, we see that the lower resolution lopsided profiles from the original Auriga simulations are generally noisier at larger radii than the Superstars lopsided profiles. Furthermore, the better sampling of the gravitational potential and the high-cadence outputs make Superstars the ideal simulation to study how different dynamical processes can affect the properties and evolution of galactic disks with unprecedented details in a fully cosmological context. In this work, we analyze a pre-final production run of the nine Milky Way-type halos modeled with the Superstars method, where an about twice times larger than intended value of the "kick" velocity was used. However, as shown in the figure A1 of the Appendix of \citet{Pakmor2025_Superstars}, this change in the "kick" velocity does not significantly affect the galaxies' star formation histories, nor the stellar surface density and vertical velocity dispersion within the disk optical radius. For this reason, we do not expect our results to be significantly affected by the different value of the "kick" velocity used.

\begin{figure*}[!htbp]
    \centering
    \includegraphics[width=0.45\textwidth]{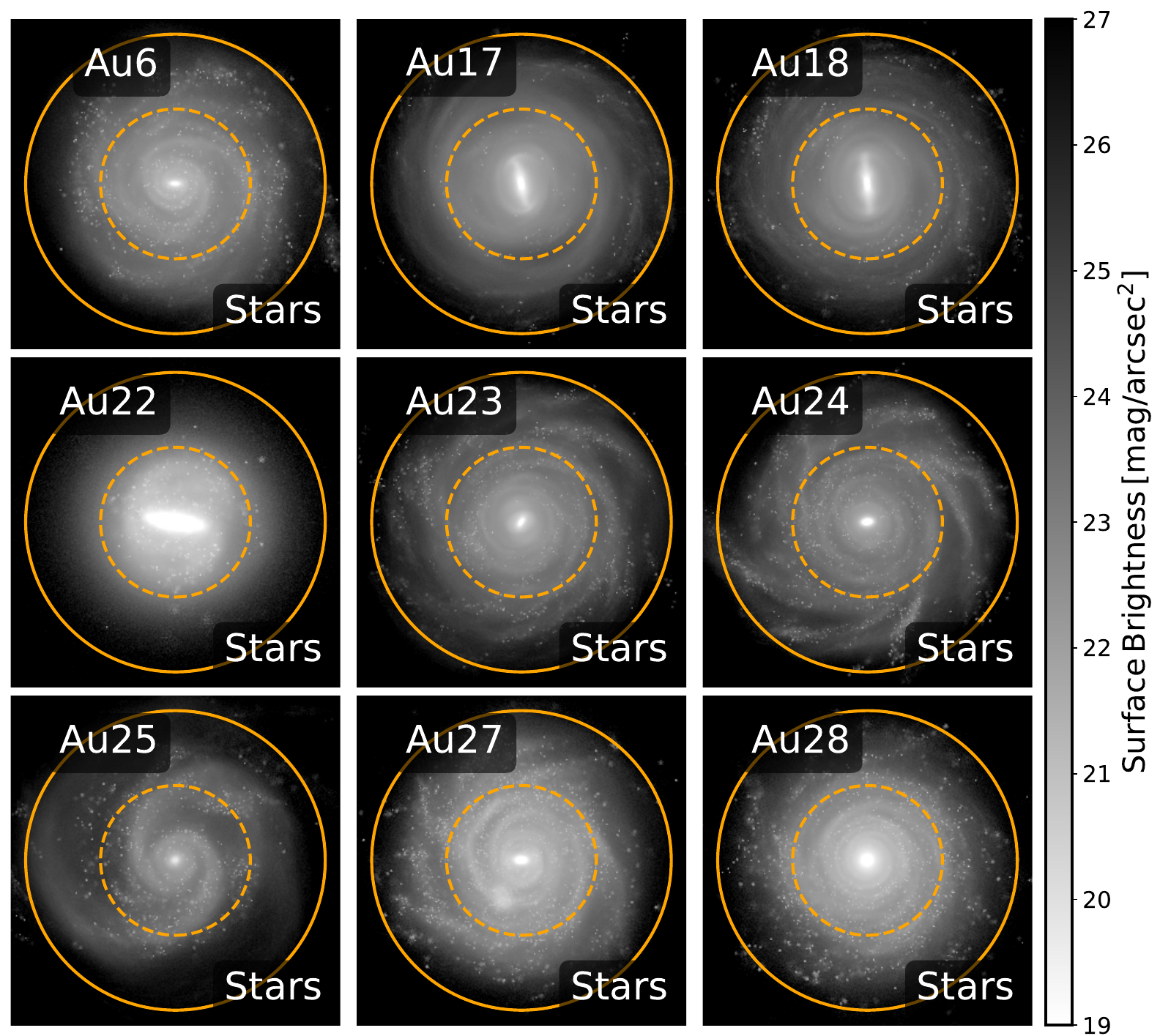}
    \includegraphics[width=0.45\textwidth]{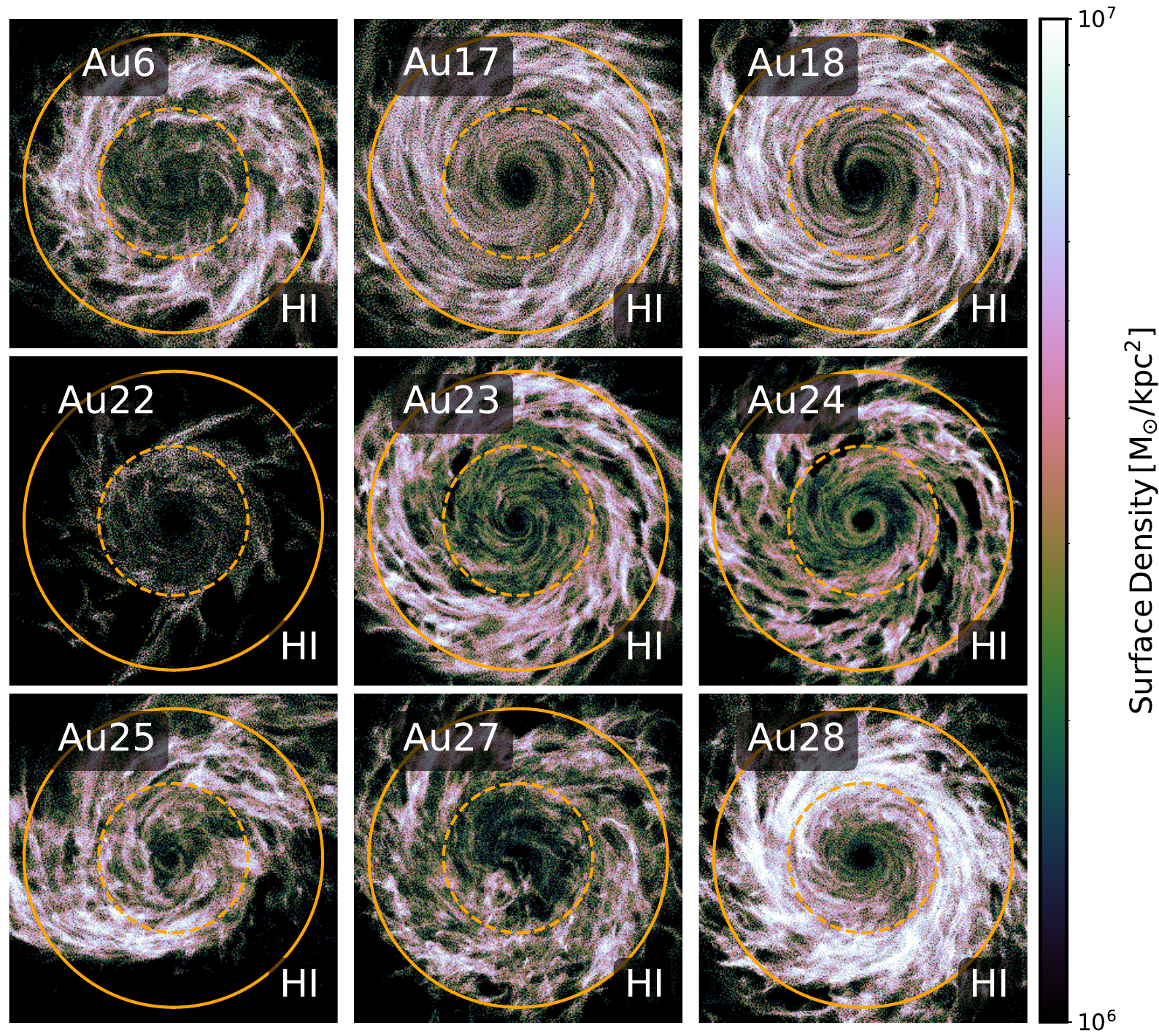}
    \caption{{\it Top panels:} Present-day face-on $V$-band stellar surface brightness maps of the nine galaxies in Superstars. The dashed and solid orange circles define the region at $0.5$ and $1\, R_{\mathrm{opt}}$, respectively, where $R_{\mathrm{opt}}$ is calculated as described in Sec. \ref{sec:stellar_disk}. {\it Bottom panels:} Same as the top panels, but showing the present-day face-on $\mathrm{HI}$ surface density maps. It is worth noting the presence of a satellite galaxy within $0.5\, R_{\mathrm{opt}}$ in Au27, suggesting that this galaxy is undergoing an interaction.}
    \label{fig:sample_visualization_superstars}
\end{figure*}

\subsection{The galaxy sample}
\label{sec:sample_selection}

\subsubsection{Characterization of the stellar disks}
\label{sec:stellar_disk}
In this work, we analyze the nine available models in Superstars. This includes galaxies with disk-to-total mass ratio ($\mathrm{D/T}$), ranging from  $0.35 < \mathrm{D/T} < 0.7$. Here, disk fractions are calculated considering all star particles with circularity $\epsilon > 0.7$ (see \citealt{Grand2017}).

Similarly to \citet{VarelaLavin2023} and \citet{Dolfi2023}, for each galaxy, we define the extension of its stellar disk at $z=0$ as the radius, $R_{\mathrm{opt}}$, where the $V$-band surface brightness profile drops to $26.5\, \mathrm{mag/arcsec^{2}}$. This is found to be a good limit to define the extension of the stellar disk out to its outskirts. We also define the disk height, $h_{90}$, as the distance above and below the galactic plane enclosing $90\%$ of the total stellar mass located within a projected radius of $1\, R_{\mathrm{opt}}$. The measurement of both $R_{\mathrm{opt}}$ and $h_{90}$ is performed after rotating each galaxy to a face-on orientation. To do this, we center the galaxy on the position of its most bound particle and calculate its total stellar angular momentum using only the young stellar particles ($\leq1\, \mathrm{Gyr\, old}$) contained within a sphere of radius $0.2\, R_{200}$, where $R_{200}$ is the virial radius of the galaxy. We then align it with the $z$-axis of a Cartesian coordinate system. In this way, the face-on disk corresponds to the $xy$ projection, while the edge-on disk corresponds to the $xz$ and $yz$ projections.

\subsubsection{Characterization of the HI gas disks}
\label{sec:HI_gas_disk}
For each galaxy, we also calculate the atomic hydrogen (i.e. $\mathrm{HI}$) mass fraction of the individual gas cells. To do this, we follow the approach described in detail in section 3 of \citet{Marinacci2017}. Firstly, we calculate the neutral hydrogen ($\mathrm{HI}+\mathrm{H_{2}}$) gas mass. We use the value of the neutral hydrogen fraction estimated directly from the \texttt{AREPO} gas cooling module for each cell \citep{Vogelsberger2013}. Then, we determine the molecular hydrogen (i.e. $\mathrm{H_{2}}$) gas mass using the empirical relation from \citet{Blitz2006}, where the ratio between the molecular and atomic hydrogen volumetric densities (i.e. $R_{\mathrm{mol}}$) is defined as:

\begin{equation}
    R_{\mathrm{mol},i} = \left(\frac{P}{P_{0}}\right)^{\alpha}, {\rm with}\,  
    P = [(\gamma - 1) \rho]u
\end{equation}

\noindent with $P$ the gas pressure of each gas cell defined by the effective equation of state, $\rho$ the mass density of the gas cell and $u$ its thermal energy per unit mass \citep{Springel2003}. $P_{0}=1.7\times 10^{4}\, \mathrm{Kcm^{-3}}$ and $\alpha=0.8$ are free parameters that we fix to the values determined by \citet{Leroy2008} based on a set of observations of nearby galaxies. The fraction of the molecular hydrogen is then calculated as:

\begin{equation}
    f_{\mathrm{mol},i} = \frac{R_{\mathrm{mol},i}}{(R_{\mathrm{mol},i} + 1)}.
\end{equation}

\noindent Finally, the $\mathrm{HI}$ gas mass is calculated as:

\begin{equation}
    \mathrm{M}_{\mathrm{HI},i} = (1 - f_{\mathrm{mol},i})f_{\mathrm{neutr},i}X_{\mathrm{H},i}\mathrm{M}_{\mathrm{gas},i},
\end{equation}

\noindent where $f_{\mathrm{mol},i}$ and $f_{\mathrm{neutr},i}$ are the molecular and neutral hydrogen gas mass fractions, respectively, $X_{\mathrm{H},i}$ is the hydrogen mass fraction and $\mathrm{M}_{\mathrm{gas},i}$ is the total gas mass of the individual gas cell. 

In Fig. \ref{fig:sample_visualization_superstars}, we show the face-on projection of the present-day $V$-band stellar surface brightness maps (top panels) and of the $\mathrm{HI}$ surface density maps (bottom panels) of the nine Superstars galaxies. The galaxy labels used here are the same as they were defined in the original Auriga simulations by \citet{Grand2017}. The dashed and solid orange circles define the outer disk region between $0.5$ and $1\, R_{\mathrm{opt}}$, respectively, measured from the stellar surface brightness profiles, as described at the beginning of the section. We see that all galaxies show a disk morphology, extending out to $\sim 1\, R_{\mathrm{opt}}$, in both stellar and HI components. Only Au22 shows a compact stellar disk, as well as the lowest HI surface density and the least extended HI spatial distribution, while Au25 shows signs of a perturbed HI disk. Some galaxies (i.e. Au17, Au18, Au22) also show a very strong stellar bar in the inner regions within $0.5\, R_{\mathrm{opt}}$. 

\section{The present-day stellar and HI gas lopsidedness}
\label{sec:presentday_stellar_and_gas_lopsidedness}
As described in detail in the introduction (see Sec. \ref{sec:introduction}), distinct dynamical processes can trigger different responses in the stellar and gas components. In this section, we study the correlation between stellar and HI gas lopsidedness in both density and kinematics for the nine Superstars galaxies, focusing on $z=0$. 
We highlight here that, when studying global galaxy properties, we will be using both Superstars and original L4 Auriga galaxies (see Sec. \ref{sec:present_day_lopsidedness_correlation}, \ref{sec:present_day_lopsidedness_bars}, and \ref{sec:kinematics_morphology_lopsidedness_correlation}) to increase the sample statistics. However, when analyzing density and velocity maps (see Sec. \ref{sec:sample_selection} and \ref{sec:present_day_velocity_maps}) or the lopsidedness time evolution (see Sec. \ref{sec:drivers_lopsidedness}), we will be using only the Superstars galaxies due to their higher mass and temporal resolutions.

\subsection{Measuring morphological lopsidedness}
\label{sec:morphological_lopsidedness}
Following the previous works of \citet{VarelaLavin2023} and \citet{Dolfi2023,Dolfi2024}, we calculate the morphological lopsidedness by measuring the amplitude of the first Fourier mode, $\mathrm{A}_{1}$, from the galaxy stellar or gas mass distribution in concentric radial annuli and normalizing it by the zeroth mode, $\mathrm{A}_{0}$, after rotating the galaxy to a face-on orientation, as described in Sec. \ref{sec:stellar_disk}. We define equally-spaced concentric, radial annuli with a width of $0.5\, \mathrm{kpc}$ and a height of $2|h_{90}|$ from the galactic plane, extending from the center of the galaxy out to $1\, R_{\mathrm{opt}}$. Finally, to quantify the global morphological lopsidedness of the galaxy (i.e. $\langle \mathrm{A_{1}} \rangle$, hereafter), we calculate the average $A_{1}$ in the radial range between $0.5$-$1\, R_{\mathrm{opt}}$. Galaxies with $\langle \mathrm{A_{1}} \rangle > 0.1$ are classified as lopsided, while those with $\langle \mathrm{A_{1}} \rangle < 0.1$ are considered symmetric. The value $\langle \mathrm{A_{1}} \rangle = 0.1$ is typically used in observations as the threshold separating lopsided and symmetric galaxies \citep{Bournaud2005}.

\subsection{Measuring kinematical lopsidedness}
\label{sec:kinematical_lopsidedness}
As described in the introduction (Sec. \ref{sec:introduction}) a tidally-induced lopsided perturbation is expected to show both morphological and kinematical asymmetries. Therefore, by quantifying the kinematical lopsidedness and studying its correlation with the morphological counterpart, we aim to verify whether this is the case and to constrain the mechanisms responsible for triggering mechanisms.

To quantify the kinematical lopsidedness in the stellar and HI gas component of our galaxy sample, we decompose the velocity of each particle into its components using cylindrical coordinates: tangential $V_{\Phi}$, radial $V_{\mathrm{R}}$, vertical $V_{\mathrm{z}}$, after rotating the galaxy to a face-on orientation. In this work, we will focus on the radial component of the velocity: $V_{\mathrm{R}}$. Similar to the morphological lopsidedness, we calculate the kinematical lopsidedness by measuring the amplitude of the first Fourier mode, $A_{1,\mathrm{kin}}$, from the radial velocity map of the galaxy, using the same concentric radial annuli defined in Sec. \ref{sec:morphological_lopsidedness}. Finally, the global kinematical lopsidedness of the galaxy (i.e. $\langle \mathrm{A}_{1,\mathrm{kin}} \rangle$, hereafter) is calculated as the average $A_{1,\mathrm{kin}}$ in the radial range between $0.5$-$1\, R_{\mathrm{opt}}$, as for the morphological lopsidedness. For the kinematical lopsidedness, we adopt the same threshold, $\langle \mathrm{A_{1,\mathrm{kin}}} \rangle = 0.1$, as used for the morphological lopsidedness to distinguish between lopsided and symmetric galaxies, as described in Sec. \ref{sec:morphological_lopsidedness}.

We note here that we measure the morphological (see Sec. \ref{sec:morphological_lopsidedness}) and kinematical lopsidedness of both the stellar and HI gas components over the same spatial region, which is defined by the extension of the stellar disk reaching out to $1\, R_{\mathrm{opt}}$ and with a vertical extent of $2|h_{90}|$ from the galactic plane, calculated as described in Sec. \ref{sec:stellar_disk}. 

\begin{figure}[!htbp]
    \centering
    \includegraphics[width=0.49 \textwidth]{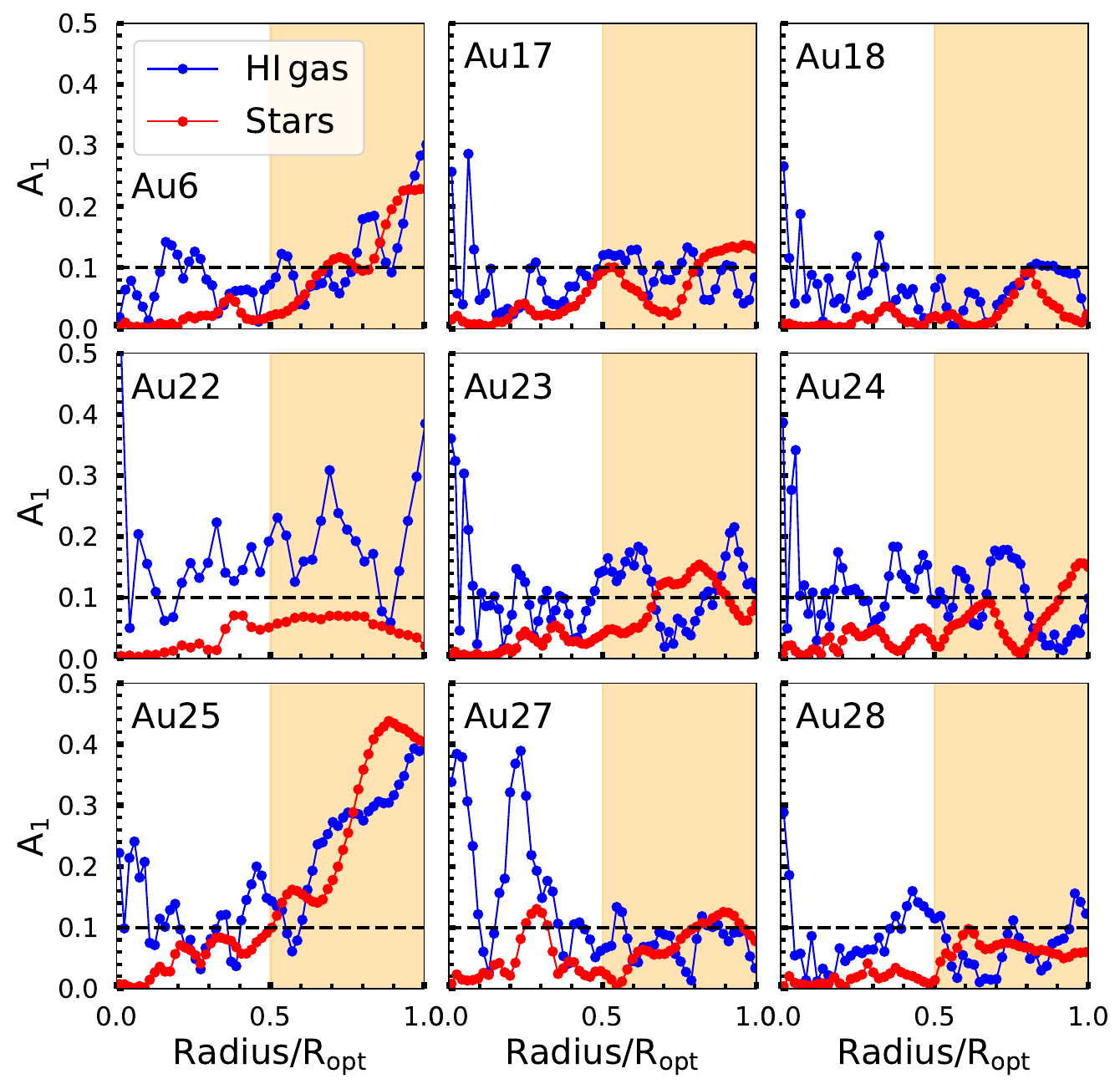}
    \caption{Present-day morphological lopsided profiles of the stellar and $\mathrm{HI}$ mass distributions within the galactic disks of the nine galaxies in Superstars. The shaded orange region mark the radial interval between $0.5$-$1\, R_{\mathrm{opt}}$ used to quantify the global lopsidedness of the galaxy. The horizontal dashed line represents the lopsidedness threshold $\langle \mathrm{A_{1}} \rangle = 0.1$.}
    \label{fig:present_day_lopsided_profiles}
\end{figure}

\subsection{Superstars: The present-day $\mathrm{A}_{1}$ density profiles}
\label{sec:present_day_lopsidedness}
In Fig. \ref{fig:present_day_lopsided_profiles}, we show the present-day $\mathrm{A_{1}}$ profiles of the stellar and HI gas mass distributions within the galactic disk for the Superstars galaxies, calculated as described in Sec. \ref{sec:morphological_lopsidedness}.
We see that, while the HI gas shows greater fluctuations in its profiles than the stars, there is an overall good agreement between the profiles of the two tracers. 
Specifically, we see that the two lopsided galaxies (i.e. Au6 and Au25) that show increasing stellar $\mathrm{A_{1}}$ profiles as we move to the disk outskirts, also show increasing $\mathrm{A}_{1}$ profiles when considering the HI density distribution. On the other hand, galaxies with symmetric stellar distribution also typically show symmetric HI distributions. The only exception is Au22, which shows a discrepancy between the stellar and HI $\mathrm{A_{1}}$ profiles, suggesting a specific driving mechanism for the lopsidedness that affects only the gas distribution.
We also see that, while the stars do not show any significant lopsidedness in the inner galactic regions within $0.5\, R_{\mathrm{opt}}$, the HI gas lopsided amplitude shows peaks in almost all of the galaxies. These strong and localized lopsided features, confined to the inner galactic regions and appearing only in the HI gas, could arise from a combination of effects. First, they may be partly driven by noise associated with the low $\mathrm{HI}$ gas mass surface density within the inner $\sim2$-$3\, \mathrm{kpc}$. This is visible in Fig. \ref{fig:sample_visualization_superstars}, where a central hole appears in the HI gas density maps. In addition, bar-driven gas asymmetries may also contribute in the case of those galaxies where we see a strong stellar bar in the central regions (e.g. Au17, Au18, Au22; see Fig. \ref{fig:sample_visualization_superstars}. The bar strength, quantified by the maximum value of the $m=2$ Fourier coefficient within the inner $10\, \mathrm{kpc}$ \citep{Fragkoudi2025}, typically peaks within $5\, \mathrm{kpc}$ and can be taken as an approximate estimate of the region where the bar dominates. This roughly corresponds to the radial range where we observe peaks in the HI gas lopsided amplitude. Therefore, the bar rotation may be driving these gas asymmetries. Finally, strong AGN activity as in the case of Au22 \citep{Grand2017} may also contribute to these central HI asymmetries. On the contrary, in the case of Au27, the presence of a satellite galaxy within $0.5\, R_{\mathrm{opt}}$ may be responsible for both the central gas and stellar asymmetries seen in Fig. \ref{fig:present_day_lopsided_profiles}.

We note that, in this work, we study the lopsidedness of the stellar and HI gas over the same spatial region, defined by the stellar $R_{\mathrm{opt}}$ calculated as described in Sec. \ref{sec:stellar_disk}. However, observational studies have shown that HI disks can be $2$-$3$ times more extended than the optical stellar disk \citep{Bosma1981} and frequently exhibit lopsided asymmetries in the outer galactic regions \citep{Bournaud2005,Jog2009_review}. We therefore inspected the HI lopsided profiles beyond the stellar $R_{\mathrm{opt}}$, out to $R_{\mathrm{HI}}^{95}$, defined as the radius enclosing $95\%$ of the total HI gas mass. Although these results are not shown in the paper, we find that eight out of nine of the galaxies exhibit lopsided HI distributions in the outer regions out to $R_{\mathrm{HI}}^{95}$.

\begin{figure}[!htbp]
    \centering
    \includegraphics[width=0.5\textwidth]{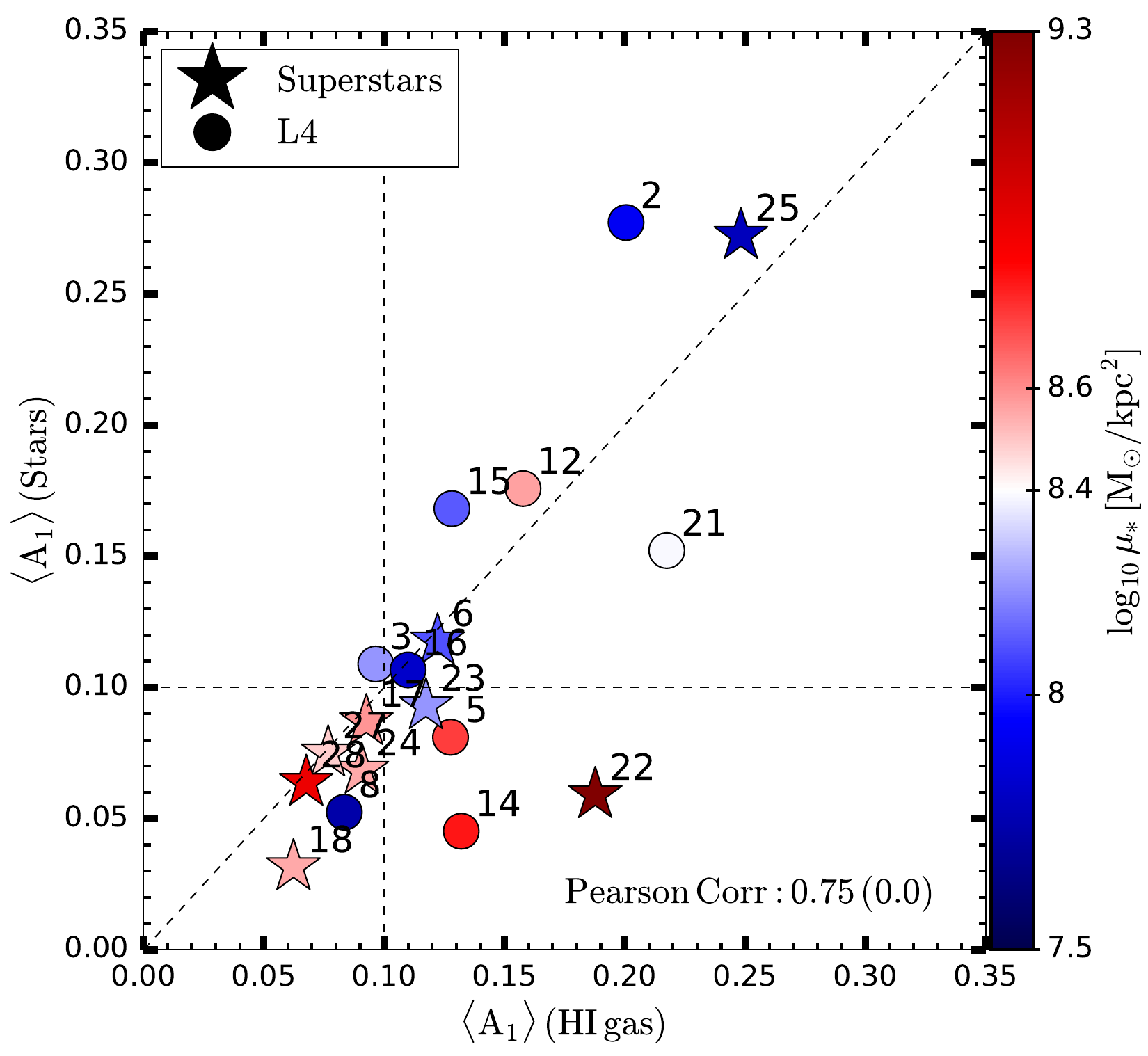}
    \caption{Present-day global lopsidedness of the stellar and $\mathrm{HI}$ disks of the nine galaxies in Superstars. We also include nine additional L4 disk galaxies with $\mathrm{D/T}>0.5$ that are not modeled in Superstars (circle symbols). The value of the Pearson correlation coefficient is reported in the figure, with the p-value indicated within brackets. Markers are color-coded by the central stellar mass density $\mu_{*}$ of each galaxy, calculated as described in Sec. \ref{sec:present_day_lopsidedness_correlation}. The global lopsidedness is measured as described in Sec. \ref{sec:morphological_lopsidedness}. The diagonal dashed line indicates the one-to-one relation, while the vertical and horizontal dashed lines represent the lopsidedness threshold $\langle \mathrm{A}_{1} \rangle = 0.1$.}  
    \label{fig:present_day_lopsidedness}
\end{figure}

\begin{figure}[!htbp]
    \centering
    \includegraphics[width=0.5\textwidth]{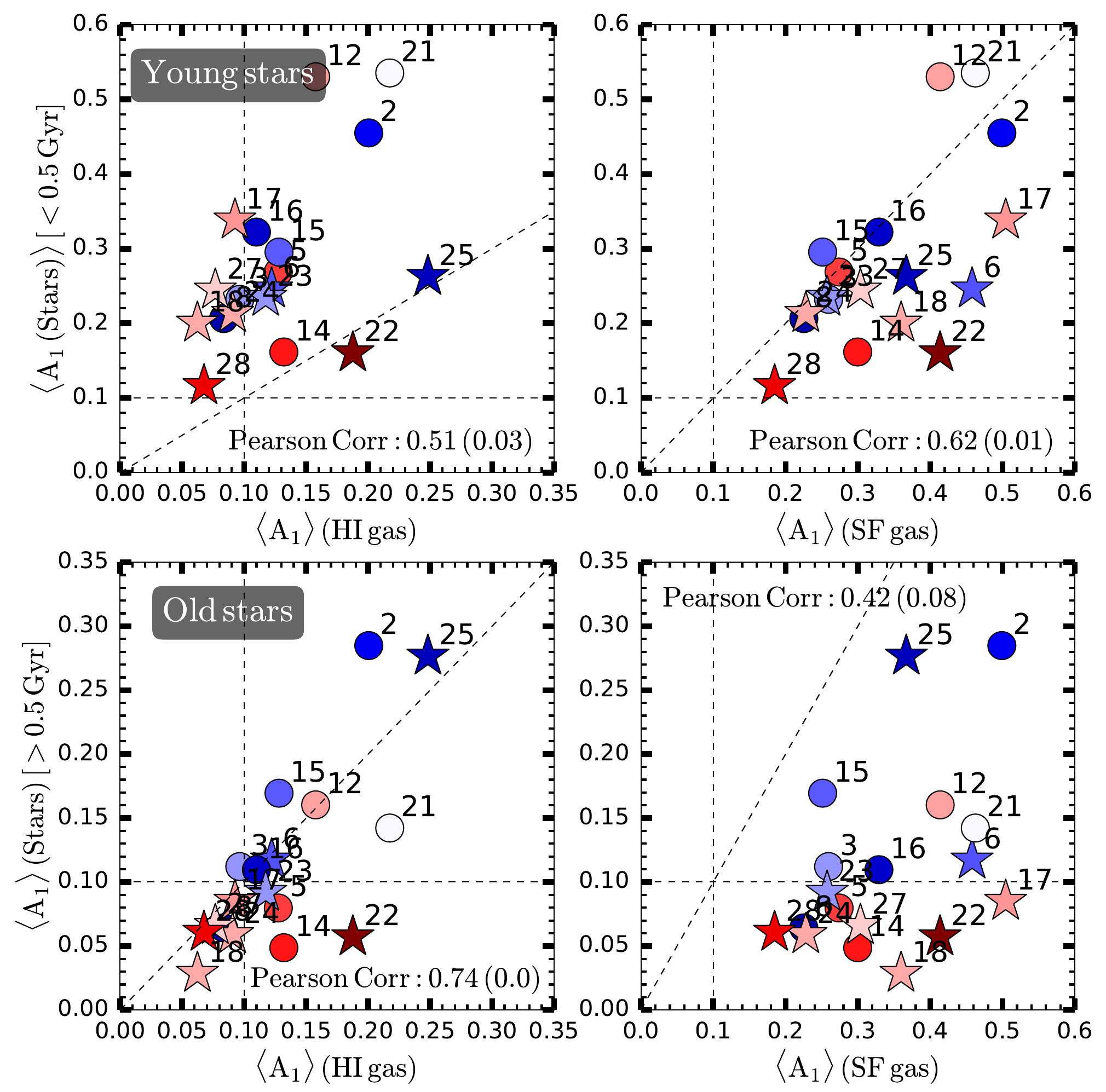}
    \caption{{\it Top panels}: Present-day global lopsidedness of the young stellar component ($<0.5\, \mathrm{Gyr\, old}$) compared with that of the HI ({\it left}) and star-forming ({\it right}) gas components. {\it Bottom panel}: Same as the top panels for the old stellar component ($>0.5\, \mathrm{Gyr\, old}$). The symbols have the same meaning as in Fig. \ref{fig:present_day_lopsidedness}.}
    \label{fig:lopsidedness_present_day_split}
\end{figure}

\subsubsection{The stellar-HI density lopsidedness correlation}
\label{sec:present_day_lopsidedness_correlation}
In Fig. \ref{fig:present_day_lopsidedness}, we show the present-day $\langle \mathrm{A_{1}} \rangle$ of the stellar and HI gas disks for the Superstars galaxies. We also include here the nine L4 disk galaxies with $\mathrm{D/T}>0.5$ that are not modeled with Superstars, as mentioned in Sec. \ref{sec:convergence_lopsidedness}. Galaxies are color-coded by their central stellar mass density, $\mu_{*}$, defined in Sec. \ref{sec:convergence_lopsidedness}. The diagonal dashed black line indicates the one-to-one relation between the stellar and HI gas asymmetry. 

We find a strong correlation between the $\langle \mathrm{A_{1}} \rangle$ in the stars and HI gas disks (i.e. galaxies with large lopsidedness in the stars have also typically large lopsidedness in the HI gas, and vice-versa), suggesting that the two tracers are dynamically coupled. The strong positive linear correlation between stellar and HI gas lopsidedness is also supported by the high value of the Pearson coefficient reported in Fig. \ref{fig:present_day_lopsidedness}. Furthermore, we see that the two most strongly lopsided galaxies (i.e. Au2 and Au25) are among those with the lowest central stellar mass density and that, although there are clear exceptions, the central stellar mass density tends to increase as we move along the diagonal line towards the most symmetric galaxies. This is consistent with the strong anti-correlation found between lopsided amplitude and central stellar mass density in both observations \citep{Reichard2008} and simulations \citep{Fontirroig2025}. Therefore, as two-thirds of the Superstars galaxies have central stellar mass densities ($\mu_{*}$) above the sample average, this explains our moderate bias toward symmetry at the present day. In Fig. \ref{fig:present_day_lopsidedness}, we see that there are three galaxies from the original L4 simulations (i.e. Au8, Au12, and Au21), which represent an exception to the anti-correlation observed between lopsided amplitude and central stellar mass density. Two of these galaxies (i.e. Au12 and Au21) have high $\langle \mathrm{A_{1}} \rangle$ in both stars and HI gas, and relatively high central stellar mass density. These galaxies have experienced a significant recent interaction ($<5\, \mathrm{Gyr\, ago}$) with a satellite of total mass $\gtrsim10^{10}\, \mathrm{M_{\odot}}$, as also shown by \citet{Gomez2017}, which likely triggered their lopsidedness and produced the observed misalignment between the inner disk and the outer shells of the dark matter halo (see figure 8 of \citealt{Gomez2017}). On the other hand, Au8 has one of the lowest central stellar mass density, but also a low $\langle \mathrm{A_{1}} \rangle$ in both stars and HI gas. Visual inspection of the evolutionary history of this galaxy reveals that it is currently experiencing an interaction with a satellite. However, the pericentric passage of this satellite primarily triggered a symmetric two-armed spiral ($m=2$) pattern in both the stellar and HI gas density distributions, rather than inducing a lopsided ($m=1$) asymmetry. Finally, another exception in Fig. \ref{fig:present_day_lopsidedness} is Au22, which is characterized by the highest central stellar mass density and shows high $\langle \mathrm{A_{1}} \rangle$ only in the HI gas. As already mentioned in Sec. \ref{sec:present_day_lopsidedness}, this might suggest a particular driving mechanism for the lopsidedness that primarily affects the gas distribution without a corresponding stellar response, as well as the lack of a significant massive interaction capable of perturbing the compact stellar disk of this galaxy. We will further explore the triggering mechanisms of lopsidedness in the Superstars galaxies in Sec. \ref{sec:drivers_lopsidedness}. 

In the left panels of Fig. \ref{fig:lopsidedness_present_day_split}, we separately show the correlation between $\langle \mathrm{A_{1}} \rangle$ of young ($<0.5\, \mathrm{Gyr\, old}$; top panels) and intermediate$+$old-age ($>0.5\, \mathrm{Gyr\, old}$; bottom panel)\footnote{Hereafter, referred to as old stars for simplicity.} stars and that of the HI gas. We choose an age cut of $<0.5\, \mathrm{Gyr\ old}$ for the young stars in order to trace recent star formation. In the bottom-left panel, a strong positive correlation between the stellar and HI gas lopsidedness persists even after removing the youngest stars, as also supported by the high Pearson coefficient. This indicates that the old stellar component is dynamically coupled to the HI gas and traces asymmetries in the overall mass distribution of galaxies. The correlation between the lopsided amplitude of the old stellar component and the star-forming gas is weaker likely because older stars have had sufficient time to phase-mix (see bottom-right panel of Fig. \ref{fig:lopsidedness_present_day_split}). 
On the other hand, in the top panels, we see that the correlation between the young stellar and HI gas lopsidedness is less strong, with the young stars typically showing higher $\langle \mathrm{A_{1}} \rangle$ amplitude than the HI gas. This is due to the fact that lopsidedness in the young stars is more closely related to asymmetries in the star-forming gas (see top-right panel of Fig. \ref{fig:lopsidedness_present_day_split}). This means that strong lopsidedness in the young stars (and not in the old stellar component) is an indicator of asymmetric star formation.

The results presented in Fig. \ref{fig:present_day_lopsidedness} and \ref{fig:lopsidedness_present_day_split} appear to contrast with those from \citet{Lokas2022}, who did not find any clear correlation between the stellar and gas lopsidedness in their sample of disk galaxies selected from the IllustrisTNG (TNG100) simulations (see figure 10 in \citealt{Lokas2022}). However, several factors may account for these differences. First, \citet{Lokas2022} studied the lopsided distribution of the total gas component, whereas we focus specifically on neutral hydrogen. Second, \citet{Lokas2022} measured the global lopsidedness of both stars and gas within $1$-$2\, R_{h}$, while we quantify lopsidedness in the outskirts of the galactic disk between $0.5$-$1\, R_{\mathrm{opt}}$. The different radial intervals used to measure lopsidedness are likely to affect its strength (see also \citealt{Dolfi2023}), particularly for the gas component, which is typically more extended than the stellar disk. Finally, \citet{Lokas2022} analyzed lopsidedness in the TNG100 simulation, which has a mass resolution of $m_{\mathrm{DM}}=7.5\times10^{6}\, \mathrm{M_{\odot}}$ for dark matter and $m_{\mathrm{gas}}=1.4\times10^{6}\, \mathrm{M_{\odot}}$ for gas \citep{Nelson2019}. This is approximately one and two orders of magnitude lower than the dark matter and gas mass resolution of the original Auriga simulation, respectively (see Sec. \ref{sec:auriga}). The relatively low resolution of the TNG100 simulation is likely to affect the measurement of $\langle \mathrm{A}_{1} \rangle$.

\begin{figure}[!htbp]
    \centering
    \includegraphics[width=0.5\textwidth]{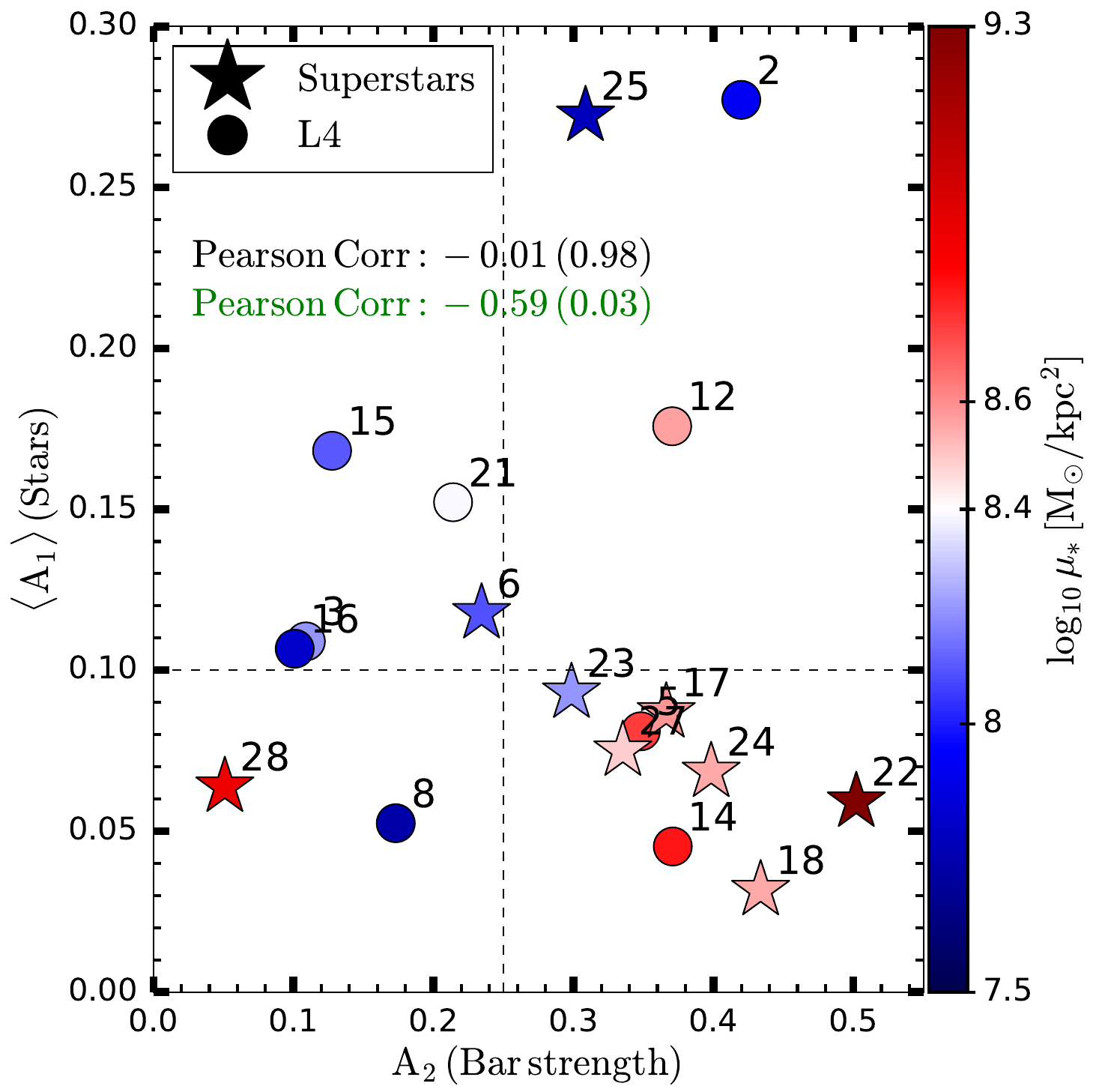}
    \caption{Correlation between the present-day global lopsidedness of the outer stellar disk and the present-day central bar strength of the nine Superstars and nine L4 galaxies. Two values of the Pearson coefficient are reported: the top one is obtained considering all the galaxies, while the bottom one excludes four outliers (i.e. Au2, Au8, Au25, Au28). The symbols have the same meaning as in Fig. \ref{fig:present_day_lopsidedness}. Markers are color-coded by the central stellar mass density $\mu_{*}$ of each galaxy, calculated as described in Sec. \ref{sec:present_day_lopsidedness_correlation}. The bar strength is measured as described in Sec. \ref{sec:present_day_lopsidedness_bars}, following \citet{Fragkoudi2025}. The horizontal dashed line indicates the lopsidedness threshold $\langle \mathrm{A}_{1} \rangle = 0.1$, while the vertical dashed line at $A_{2}\geq0.25$ represents the threshold used in \citet{Fragkoudi2025} to classify barred and unbarred galaxies.}
    \label{fig:present_day_lopsidedness_bars}
\end{figure}

\subsubsection{The stellar density lopsidedness-bar connection}
\label{sec:present_day_lopsidedness_bars}
Previous works that studied the presence of morphological features in disk galaxies, such as stellar bars and lopsidedness, have shown that strong lopsidedness is more easily induced in galaxies that have low central stellar mass density and are less gravitationally cohesive \citep{Reichard2008,VarelaLavin2023}. Furthermore, lopsided galaxies are also found to have assembled their stellar mass, on average, at later times than symmetric galaxies \citep{Dolfi2023}. On the contrary, barred galaxies are typically found to be more baryon-dominated and to have assembled their stellar mass at early times \citep{Fragkoudi2025}. This means that galaxies that are prone to develop a stellar bar should be less likely to develop a lopsided perturbation based on their internal properties.

In Fig. \ref{fig:present_day_lopsidedness_bars}, we show the correlation between the stellar $\langle \mathrm{A_{1}} \rangle$ and the bar strength at $z=0$ for the Superstars galaxies, as well as for the nine L4 disk galaxies with $\mathrm{D/T}>0.5$ that are not modeled with Superstars. The bar strength (i.e. $\langle \mathrm{A}_{2} \rangle$, hereafter) is quantified by measuring the amplitude of the $m=2$ Fourier mode (i.e. $A_{2}$) from the galaxy stellar mass distribution in concentric radial annuli, and by taking the maximum value of $A_{2}$ within the inner $10\, \mathrm{kpc}$ (see \citealt{Fragkoudi2025} for a detailed description of how the bar strength is calculated). We find an anti-correlation between the stellar density lopsidedness in the outer galactic disk and the bar strength in the inner galactic regions (i.e. galaxies with high $\langle \mathrm{A_{1}} \rangle$ tend to have low $\langle \mathrm{A_{1}} \rangle$, and vice versa), although clear exceptions exist. When considering the full galaxy sample, the Pearson coefficient indicates no significant correlation (top value in Fig. \ref{fig:present_day_lopsidedness_bars}). However, after removing the clear outliers (i.e. Au2, Au8, Au12, Au25, Au28), we find a moderate anti-correlation (bottom value in Fig. \ref{fig:present_day_lopsidedness_bars}). 
Additionally, we see that the most strongly barred galaxies with the most strongly symmetric disks are typically characterized by high central stellar mass density, whereas the lopsided galaxies are typically characterized by low central stellar mass density, consistent with previous results. 

Regarding the outliers in the anti-correlation between stellar density lopsidedness and bar strength mentioned above, \citet{Fragkoudi2025}, who studied the bar formation in the original Auriga simulations, showed that Au2 hosts a strong bar that formed at early times $\sim7\, \mathrm{Gyr\, ago}$, while Au12 and Au25 host a bar that formed more recently at $\sim2.5$ and $<1\, \mathrm{Gyr\, ago}$, respectively. All these three galaxies have also experienced a strong and recent interaction with a massive satellite (i.e. $\mathrm{M_{tot}} > 10^{10.5}\, \mathrm{M_{\odot}}$; see \citealt{Gomez2017}) within the last $5\, \mathrm{Gyr}$. This interaction could have triggered the subsequent formation of both morphological features (i.e. stellar bar$+$stellar density lopsidedness) in Au12 and Au25. However, Au2 hosts an ancient bar and one of the most extended disks in the L4 simulation \citep{Grand2017}. As shown in \citet{Grand2017}, the extended disk of Au2 results from a merger occurring in the plane of the disk between $\sim8$ and $\sim7\, \mathrm{Gyr\, ago}$. During this period, the bar also formed. This co-planar satellite accretion may therefore have triggered both the bar formation and the disk growth. Once the disk became significantly extended ($\sim6.5\, \mathrm{Gyr\, ago}$) subsequent satellite interactions occurring over the last $5\, \mathrm{Gyr}$ could more easily perturb the outer galactic disk due to its lower self-gravity (very extended disk and low central stellar mass density), without significantly affecting the inner regions.

In a previous work, \citet{Lokas2022} also found that lopsided stellar disks tend to have weaker bars (i.e. low value of <$\mathrm{A_{2}}$>) than symmetric ones from the analysis of a sample of disk galaxies from the IllustrisTNG simulations, consistent with our results. However, these results appear in disagreement with those from \citet{Bournaud2005}, who showed that stellar density lopsidedness is correlated with the presence of $m=2$ asymmetries, such as bars and spiral arms, in their observational galaxy sample. We note, however, that \citet{Bournaud2005} measured the amplitude of <$\mathrm{A_{2}}$> within $0.5$-$1.5$ disk scalelengths ("inner" regions) and $1.5$-$2.5$ disk scalelengths ("outer" regions), which means that they are measuring $m=2$ asymmetries at larger radii (beyond $0.5$ disk scalelengths) than in \citet{Lokas2022} and in this work. Therefore, \citet{Bournaud2005} is also likely measuring spiral arm features rather than just bars.

Overall, the results shown in Fig. \ref{fig:present_day_lopsidedness_bars} suggest that the observed anti-correlation between the presence of stellar density lopsidedness and the presence of a bar arises as a result of the internal properties of the galaxies, such as their central stellar mass distribution, which is found to be a key parameter in determining whether a galaxy is susceptible to lopsidedness/bar formation. In a follow-up work, we will investigate in more details the origin of the anti-correlation between stellar density lopsidedness and bar strength, as well as the formation and survival of both morphological features in galaxies. 

\begin{figure*}[!htbp]
    \centering
    \includegraphics[width=0.45\textwidth]{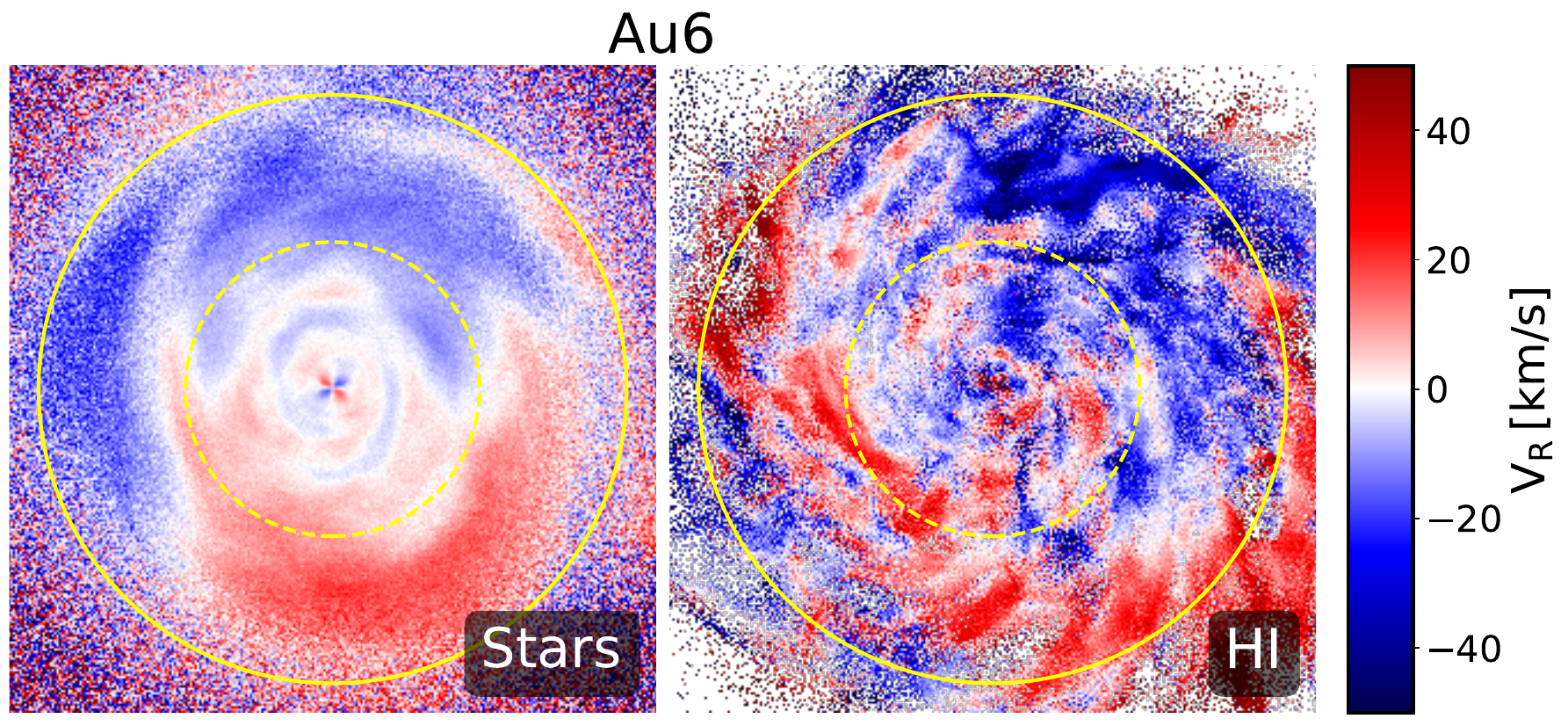}
    \includegraphics[width=0.45\textwidth]{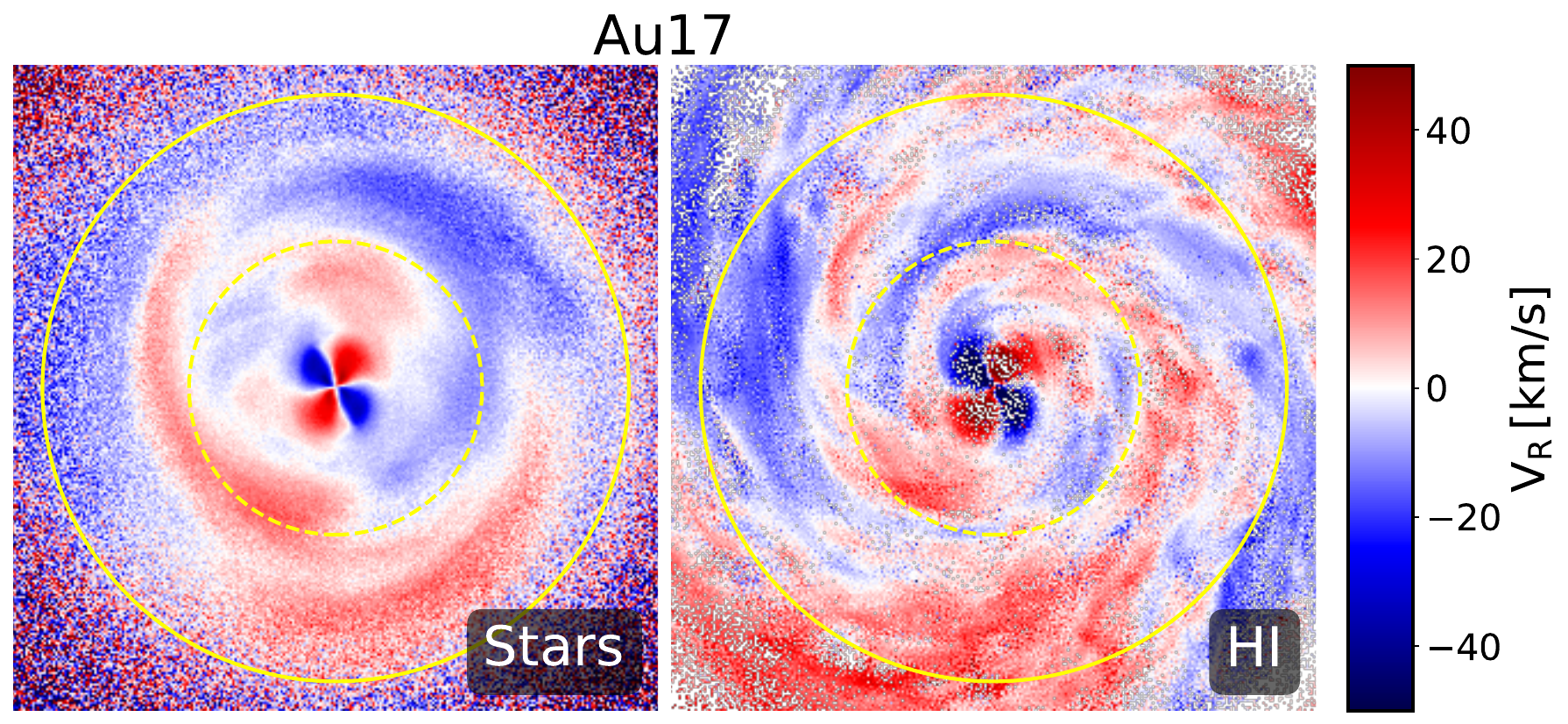}
    \includegraphics[width=0.45\textwidth]{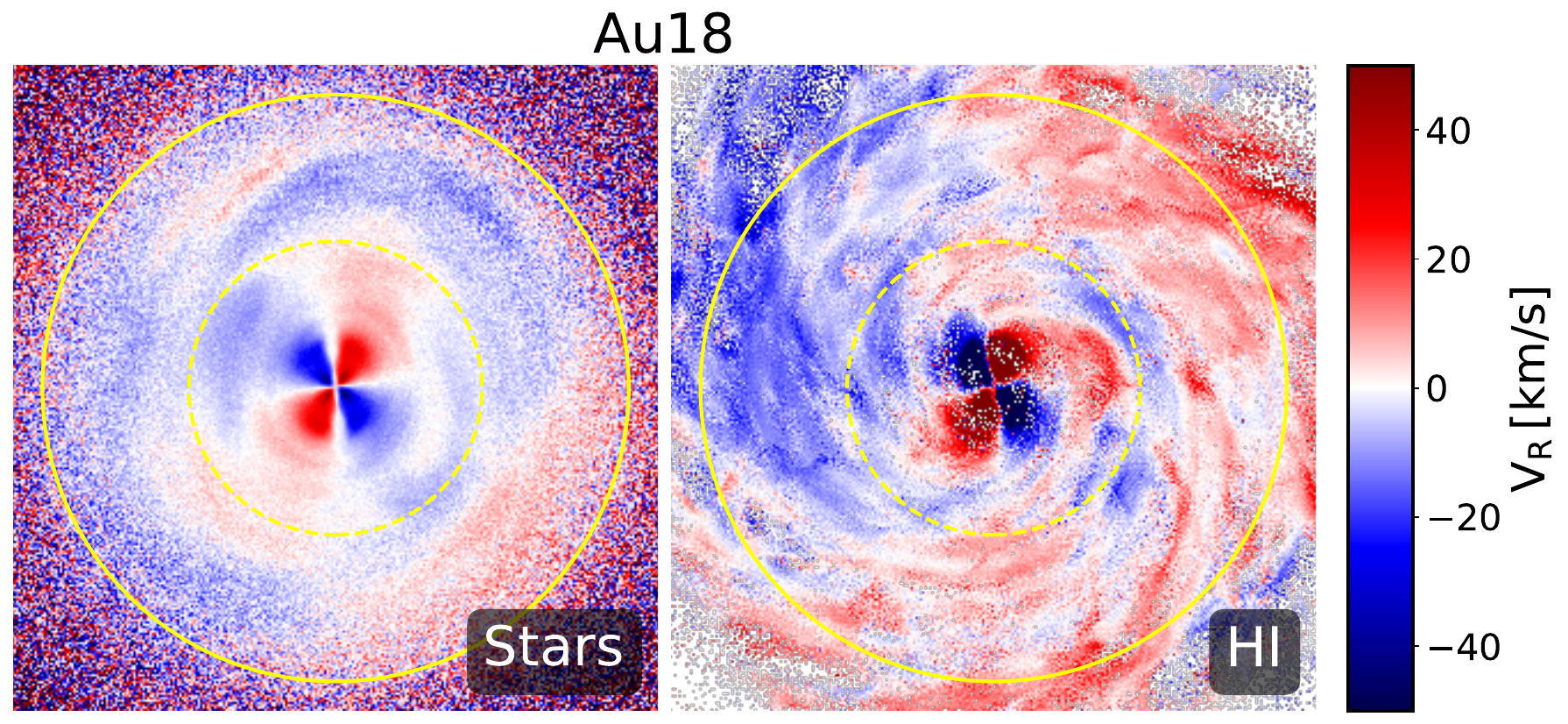}
    \includegraphics[width=0.45\textwidth]{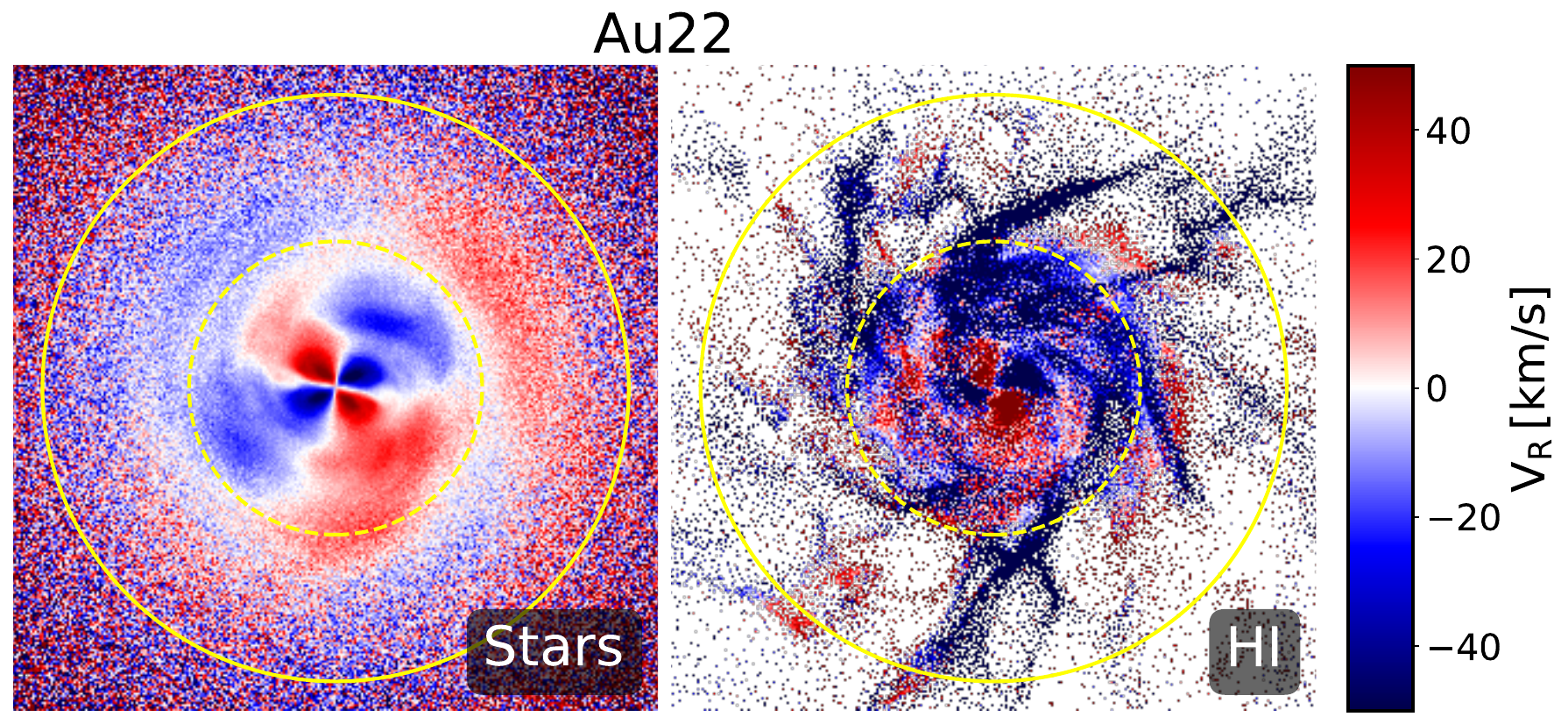}
    \includegraphics[width=0.45\textwidth]{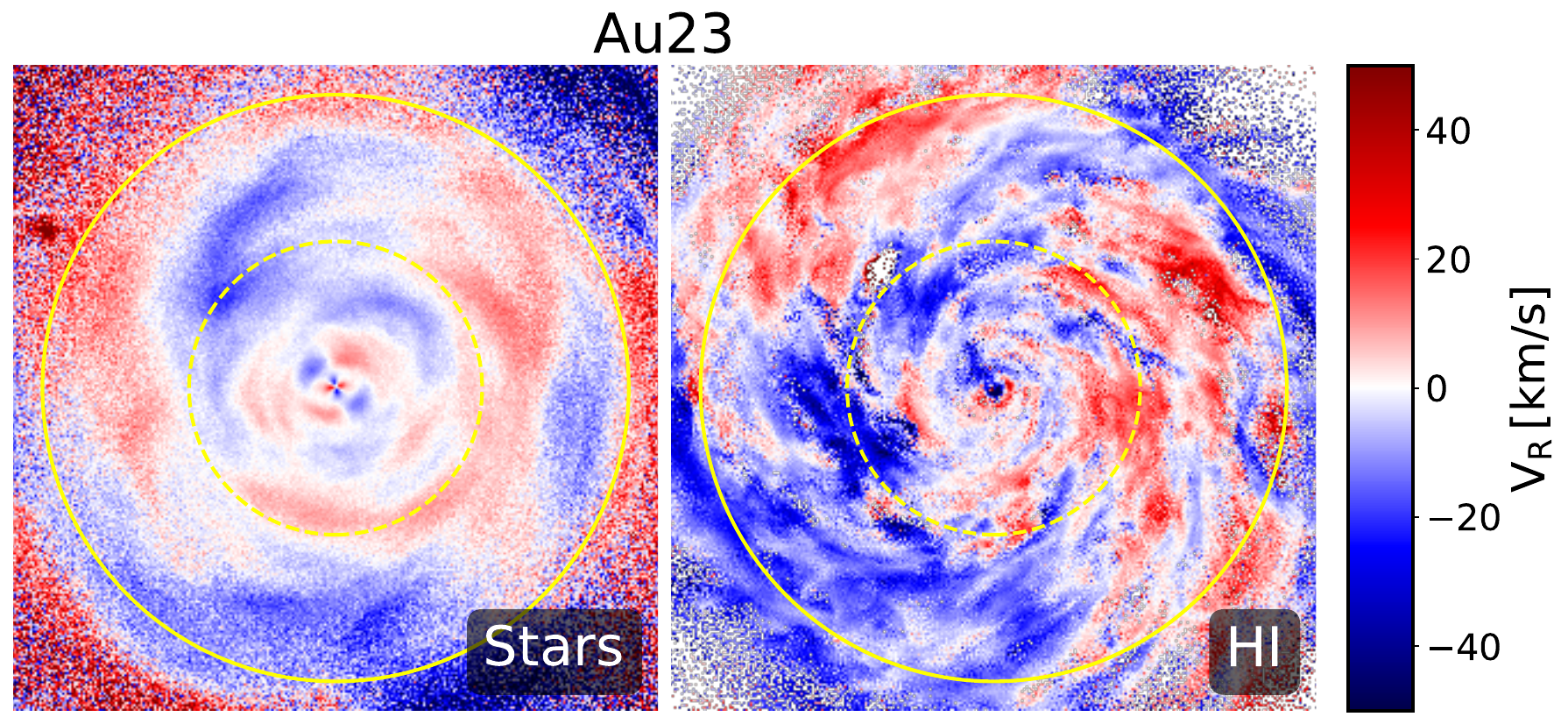}
    \includegraphics[width=0.45\textwidth]{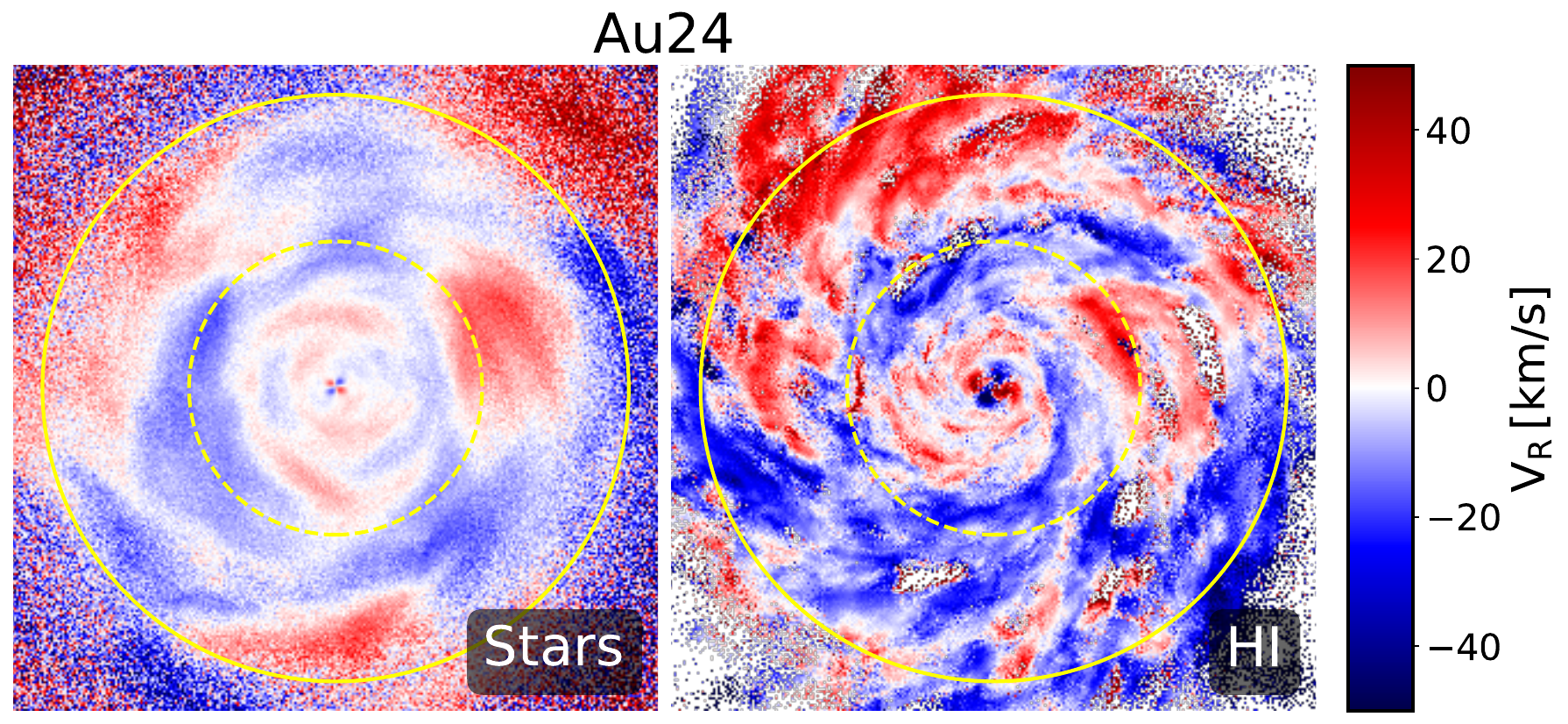}
    \includegraphics[width=0.45\textwidth]{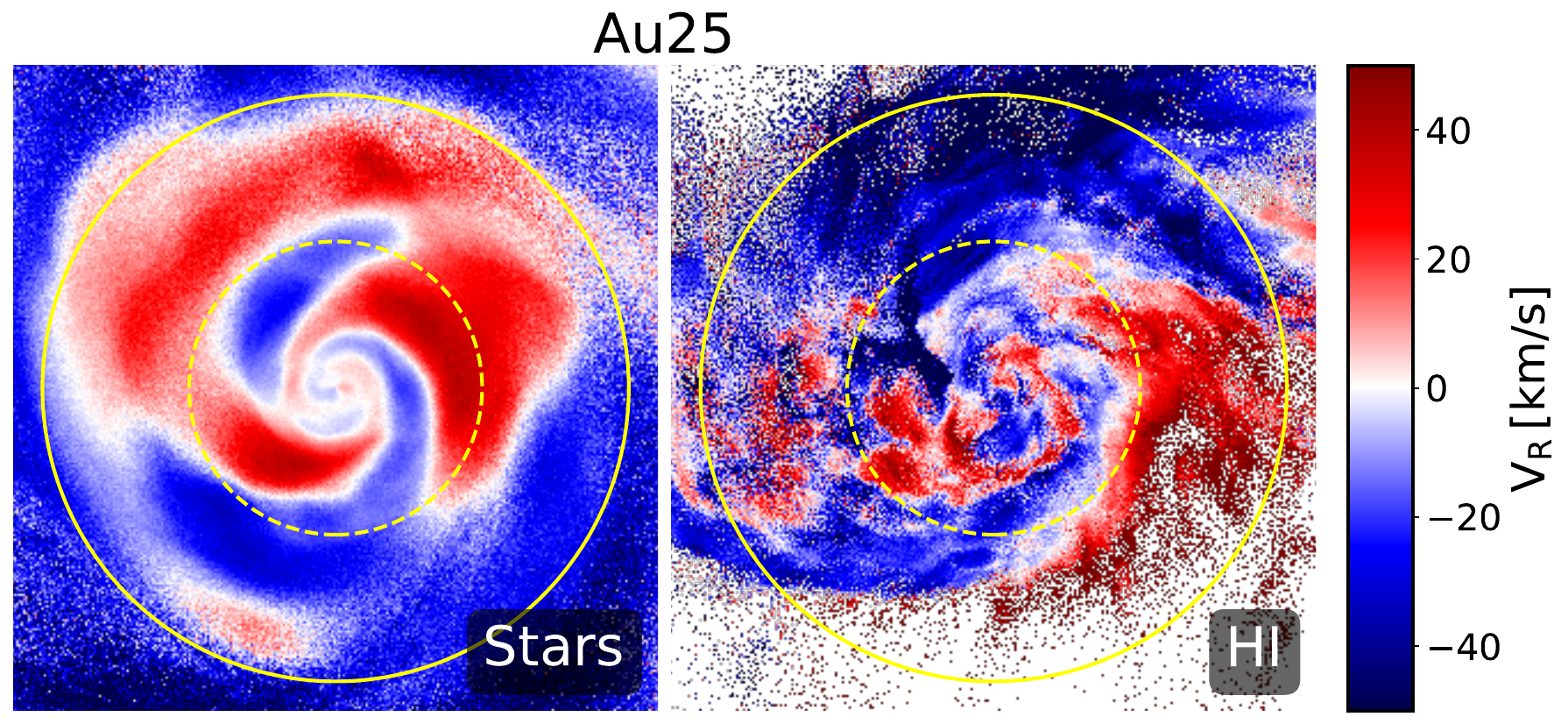}
    \includegraphics[width=0.45\textwidth]{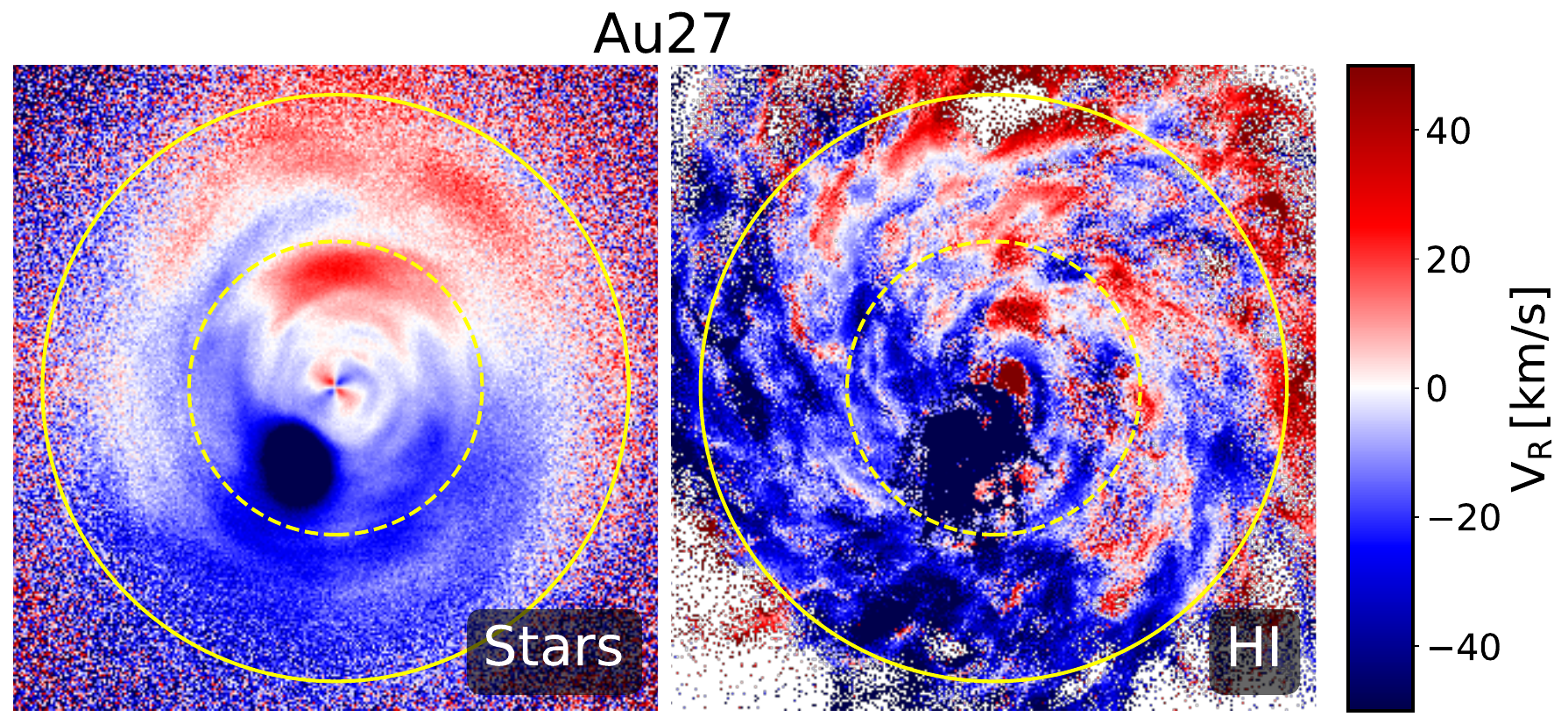}
    \includegraphics[width=0.45\textwidth]{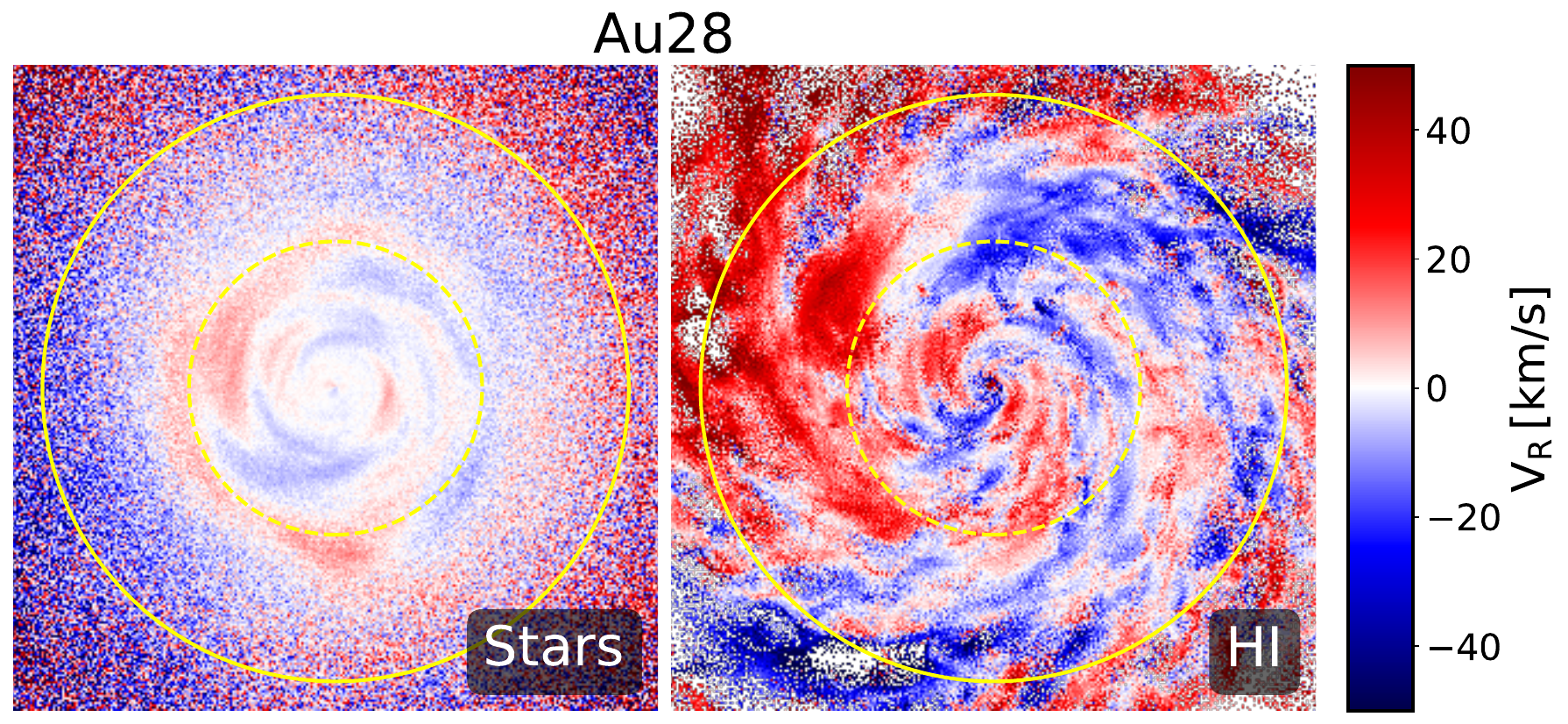}
    \caption{Face-on projections of the present-day radial velocity maps of the stars and HI disks of the nine galaxies in Superstars. The radial velocity map of each galaxy is obtained as described in Sec. \ref{sec:kinematical_lopsidedness}. The dashed and solid yellow circles define the region at $0.5$ and $1\, R_{\mathrm{opt}}$, respectively, where $R_{\mathrm{opt}}$ is calculated as described in Sec. \ref{sec:stellar_disk}.}
    \label{fig:present_day_velocity_maps}
\end{figure*}

\begin{figure*}[!htbp]
    \centering
    \includegraphics[width=0.45\textwidth]{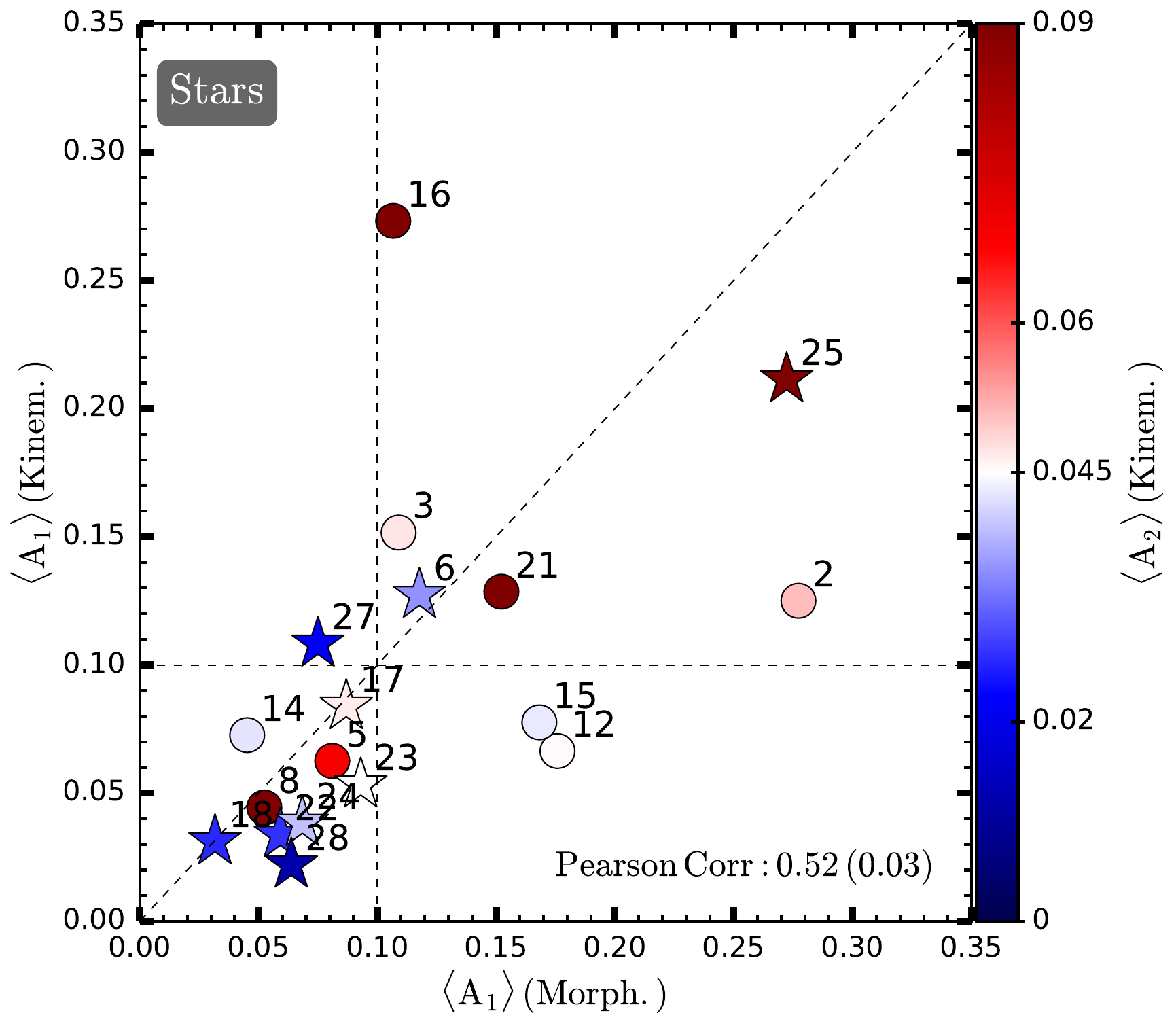}
    \includegraphics[width=0.445\textwidth]{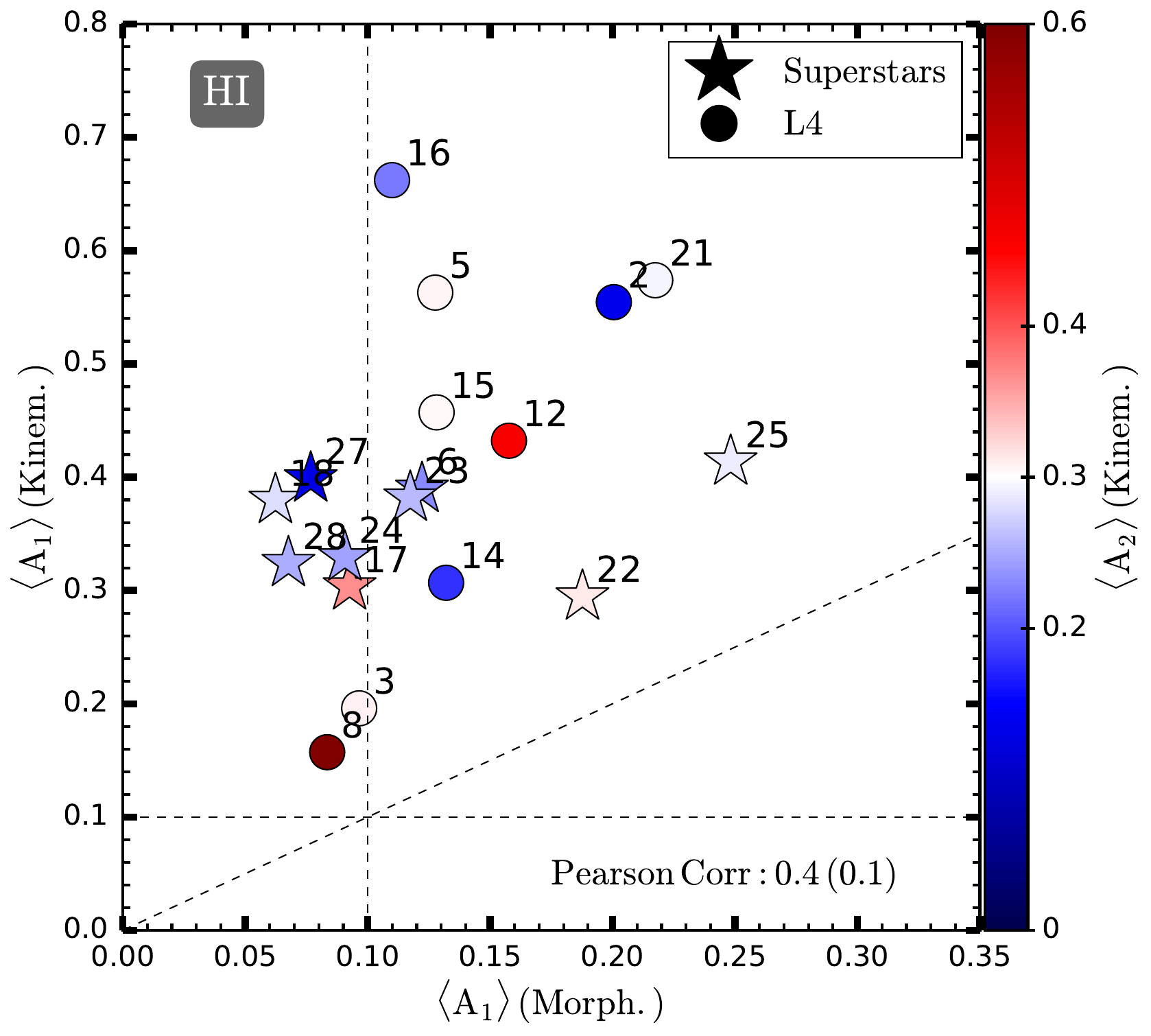}
    \caption{Present-day global kinematical lopsidedness compared with the present-day global morphological lopsidedness of the stellar ({\it left}) and HI ({\it right}) disks of the nine Superstars and nine L4 galaxies. The value of the Pearson coefficient quantifying the strength of the correlation is reported in each figure, together with the p-value indicated within brackets. The symbols have the same meaning as in Fig. \ref{fig:present_day_lopsidedness}. Markers are color-coded by the corresponding amplitude of the $m=2$ Fourier mode calculated from the face-on projected present-day radial velocity maps of the stellar and HI disks between $0.5$ and $1\, R_{\mathrm{opt}}$. The global morphological and kinematical lopsidedness are measured for both stars and HI as described in Sec. \ref{sec:morphological_lopsidedness} and Sec. \ref{sec:kinematical_lopsidedness}, respectively, over the same spatial extent.  The diagonal dashed line indicates the one-to-one relation, while the vertical and horizontal dashed lines represent the lopsidedness threshold $\langle \mathrm{A}_{\mathrm{1,kin}} \rangle = \langle \mathrm{A}_{\mathrm{1}} \rangle = 0.1$.}
    \label{fig:kinematics_morphology_lopsidedness_correlation}
\end{figure*}

\subsection{Superstars: The present-day $V_{\mathrm{R}}$ maps}
\label{sec:present_day_velocity_maps}
In Fig. \ref{fig:present_day_velocity_maps}, we show the present-day face-on mass-weighted mean radial velocity maps of the stellar and HI gas disks for the Superstars galaxies. We see that almost all galaxies in Fig. \ref{fig:present_day_velocity_maps} show a dipolar feature in the outer regions of the HI radial velocity maps (i.e. between $0.5$-$1\, R_{\mathrm{opt}}$, orange circles), which is a signature of kinematical lopsidedness.
On the contrary, only three galaxies show a dipolar feature in the outer regions of the stellar radial velocity maps (i.e. Au6, Au25, and Au27). Among these, two galaxies (i.e. Au6 and Au25) are also currently lopsided in both the stellar and HI gas density distributions (see Sec. \ref{sec:present_day_lopsidedness}), as expected if lopsidedness arises as the result of an external perturbation potential (see also \citealt{Jog2009_review} for a review). 
For the remaining galaxies, which do not show dipolar features in the outer regions of the stellar radial velocity maps, the different levels of kinematic asymmetries in the stellar and HI gas disks might indicate other lopsidedness triggering mechanisms. In what follows (Sec. \ref{sec:kinematics_morphology_lopsidedness_correlation}), we quantify the kinematical lopsidedness of both the stellar and HI gas disks and examine its correlation with the corresponding morphological lopsidedness, in order to investigate the different mechanisms that may trigger lopsidedness.

From the radial velocity maps in Fig. \ref{fig:present_day_velocity_maps}, we also see that many of them shows a quadrupole feature in the inner galactic regions within $0.5\, R_{\mathrm{opt}}$ in both the stellar and HI gas velocity maps. In particular, three galaxies (i.e. Au17, Au18, Au22) show the strongest quadrupole signal, which is indicative of the presence of a stellar bar. 

\begin{figure*}[!htbp]
    \centering
    \includegraphics[width=1\textwidth]
    {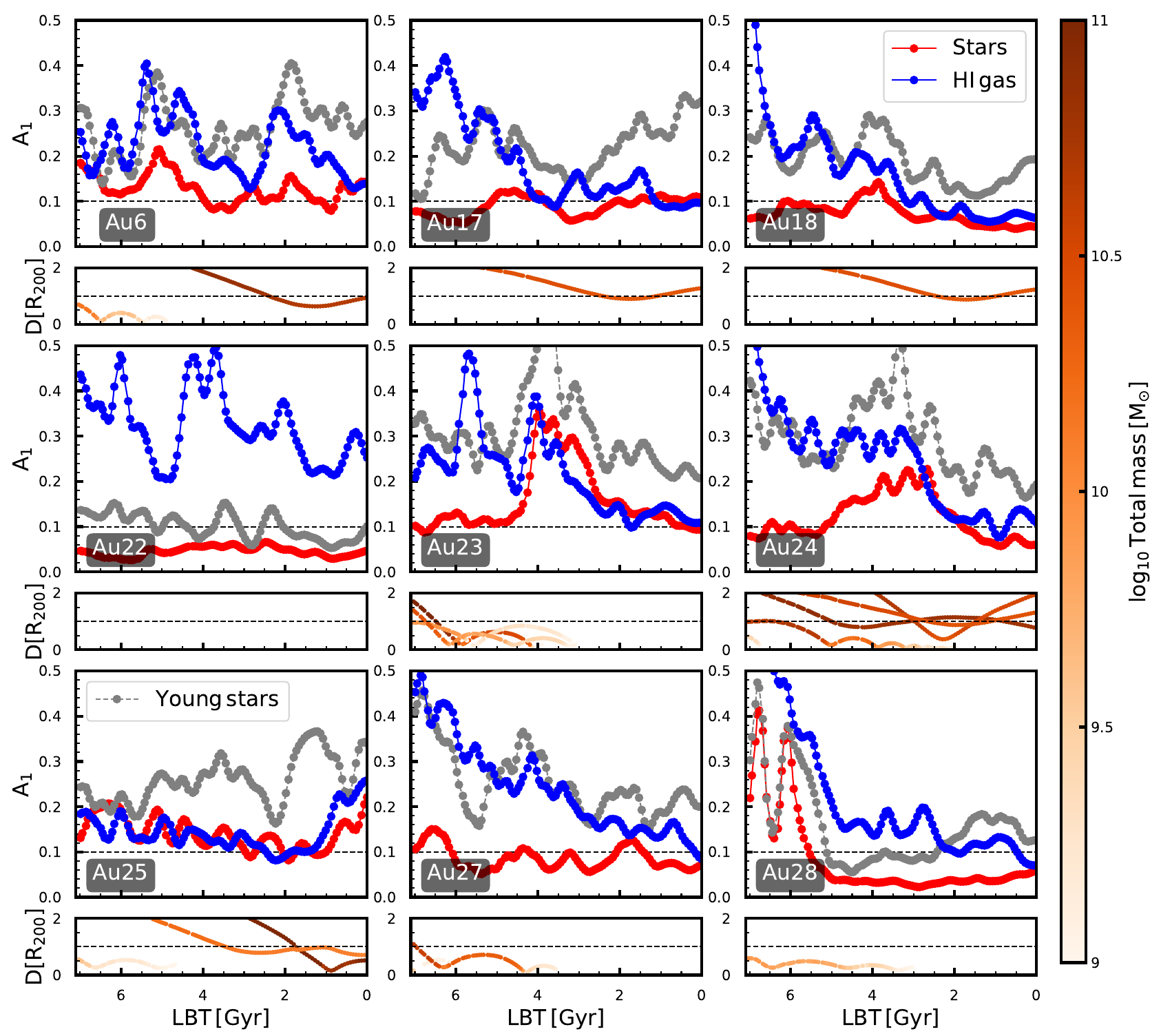}
    \caption{Global morphological lopsidedness of the stellar ({\it red curve}) and HI ({\it blue curve}) disks as a function of lookback time of the nine Superstars galaxies. We consider a time window between the present day and $7\, \mathrm{Gyr\, ago}$. We also show the time evolution of the global morphological lopsidedness of the young stellar component ($<0.5\, \mathrm{Gyr}$; {\it gray curve}). A small Gaussian smoothing ($\mathrm{sigma}=2$) is applied to the lopsidedness curves of each component. The global morphological lopsidedness at each time is measured. The horizontal dashed line represents the lopsidedness threshold $\langle \mathrm{A}_{\mathrm{1}} \rangle = 0.1$. 
    The small rectangular panel below each galaxy shows the relative distance of the most significant interacting and merging satellites over the last $7\, \mathrm{Gyr}$ from the center of the main host galaxy at each time. Distances are normalized by the virial radius $R_{200}$ of the main host at each time and are shown in linear scale. The satellite orbits are color-coded by the total mass of the satellite galaxy at each time. We consider only satellite galaxies with a total mass-ratio $>1$:$50$ at the time they first cross the virial radius $R_{200}$ of the main host. Furthermore, we show the satellite orbits only until their total mass drops below $\mathrm{M}_{\mathrm{tot}}=10^{9}\, \mathrm{M}_{\odot}$.}
    \label{fig:drivers_lopsidedness}
\end{figure*}

\subsubsection{Kinematic-morphological lopsidedness correlation}
\label{sec:kinematics_morphology_lopsidedness_correlation}
In Fig. \ref{fig:kinematics_morphology_lopsidedness_correlation}, we show the correlation between the present-day kinematic $\langle \mathrm{A_{1,kin}} \rangle$ and morphological $\langle \mathrm{A_{1}} \rangle$ lopsidedness of the stellar and HI gas disks for the Superstars galaxies, as well as for the L4 disk galaxies with $\mathrm{D/T}>0.5$, measured as described in Sec. \ref{sec:morphological_lopsidedness} and \ref{sec:kinematical_lopsidedness}. Galaxies are color-coded by the average amplitude of the kinematic Fourier $\mathrm{m}=2$ coefficient calculated from the face-on projected radial velocity maps of both stars and HI gas in the outer regions of the galactic disk between $0.5$-$1\, R_{\mathrm{opt}}$. A non-zero kinematic $\langle A_{2,\mathrm{kin}} \rangle$ in the outer galaxy regions is typically associated with spiral arms.    

For the stellar component, we find that kinematical and morphological lopsidedness are well correlated, indicating that stellar lopsidedness likely arises from a perturbed gravitational potential that coherently affects both stellar orbits and their spatial distribution. In contrast, for the HI gas, we find a weaker correlation between kinematical and morphological lopsidedness, with the lopsided signature being generally stronger in the kinematics than in the morphology. Previous works have associated strong gas kinematic asymmetries with gas accretion \citep{Feng2025}. However, this does not exclude the possibility that lopsidedness results from external interactions, given the strong correlation that we find between the stellar and HI gas morphological lopsidedness in Sec. \ref{sec:present_day_lopsidedness_correlation}. Therefore, the results shown in Fig. \ref{fig:kinematics_morphology_lopsidedness_correlation} suggest that, independent of whether lopsidedness is generated through tidal interactions or gas accretion, its imprint is stronger in the HI gas kinematics than in its morphology.

In a previous work, \citet{Bilimogga2025} studied the correlation between the asymmetry in the global HI velocity profiles and the HI morphological asymmetry in their observed galaxy sample. Specifically, they found that a galaxy with a strongly asymmetric HI velocity field can either have a symmetric or asymmetric HI density distribution.
In this work, we find similar results. In fact, as shown in the right-hand panel of Fig. \ref{fig:kinematics_morphology_lopsidedness_correlation}, we find galaxies such as Au27 and Au28 characterized by a symmetric HI density distribution, $\mathrm{A_{1}}(\mathrm{HI\, Morph})\lesssim0.1$, but still asymmetric HI kinematics, $\mathrm{A_{1}}(\mathrm{HI\, Kinem})>0.1$, as well as galaxies such as Au6 and Au25 characterized by both an asymmetric HI density distribution, $\mathrm{A_{1}}(\mathrm{HI\, Morph})>0.1$, and asymmetric HI kinematics, $\mathrm{A_{1}}(\mathrm{HI\, Kinem})>0.1$. However, while \citet{Bilimogga2025} reported a lack of correlation between stellar and HI density lopsidedness, we find that the two are strongly correlated, as previously shown in Fig. \ref{fig:present_day_lopsidedness}. It is important to note, however, that \citet{Bilimogga2025} classified optical asymmetries based on visual inspection. Therefore, it is not specified whether stellar and HI asymmetries were measured over the same spatial extent. As discussed in Sec. \ref{sec:present_day_lopsidedness_correlation}, the different spatial regions used to quantify stellar and HI asymmetries might affect the resulting correlation, given that lopsidedness is stronger at larger radii and that HI is generally more extended than the stellar component. Alternatively, differences in the environments where the galaxies reside might also affect the correlation between stellar and HI asymmetries. In fact, while \citet{Bilimogga2025} studied HI asymmetries in galaxies located in group environments, this work focuses on galaxies in the field. Consequently, the lack of correlation between HI and stellar morphological asymmetries reported by \citet{Bilimogga2025} may also suggest that lopsidedness is triggered by mechanisms, such as ram-pressure stripping, that directly affect the gas but not the stellar component. 

Our results, so far, favor a scenario in which lopsidedness is triggered by tidal interactions and/or gas accretion. In particular, we find: 

\begin{enumerate}[label=\roman*)]

    \item a strong correlation between stellar and HI gas morphological asymmetries (see Sec. \ref{sec:present_day_lopsidedness_correlation}), 
    
    \item a strong correlation between kinematical and morphological asymmetries in the stellar component (see left-hand panel of Fig. \ref{fig:kinematics_morphology_lopsidedness_correlation}), 
    
    \item a weak correlation between kinematical and morphological asymmetries in the HI gas, which is typically characterized by stronger kinematical than morphological asymmetries (see right-hand panel of Fig. \ref{fig:kinematics_morphology_lopsidedness_correlation}).
    
\end{enumerate}

Galaxies with asymmetric $\mathrm{A_{1}}(\mathrm{HI\, Morph})>0.1$ and asymmetric $\mathrm{A_{1}}(\mathrm{HI\, Kinem})>0.1$, which typically have also asymmetric stellar morphology and kinematics, suggest a tidally-induced lopsidedness. On the other hand, galaxies with symmetric $\mathrm{A_{1}}(\mathrm{HI\, Morph})\lesssim0.1$ and asymmetric $\mathrm{A_{1}}(\mathrm{HI\, Kinem})>0.1$, which are typically characterized by symmetric stellar morphology and kinematics, suggest either a gas accretion-induced lopsidedness whose signature appears stronger in the gas kinematics or that gas kinematics retains non-axisymmetric signatures longer.

In the following section, we study the evolution of lopsidedness in the Superstars galaxies and its connection to past interactions, with the aim of identifying the dominant mechanisms driving the observed morphological and kinematical asymmetries.

\section{Drivers of lopsidedness in Superstars}
\label{sec:drivers_lopsidedness}
In order to understand the origin of lopsidedness, we study its evolution as a function of time. We focus here on the morphological lopsidedness in a time window covering the last $7\, \mathrm{Gyr}$ of galactic history up to the present-day. We do not go back at earlier times where there is not a well formed disk yet. In Fig. \ref{fig:drivers_lopsidedness}, we show the global $\langle \mathrm{A}_{1} \rangle$ of the stellar and HI disks as a function of lookback time for the nine Superstars galaxies. We also show the time evolution of the global $\langle \mathrm{A}_{1} \rangle$ of the young stellar component ($<0.5\, \mathrm{Gyr\, ago}$). 
We measure $\langle \mathrm{A}_{1} \rangle$ using the method described in Sec. \ref{sec:morphological_lopsidedness}, after determining the optical radius $R_{\mathrm{opt}}$ and the disk height $h_{90}$ of the stellar disk at each time (see Sec. \ref{sec:stellar_disk}). This ensures that the global $\langle \mathrm{A}_{1} \rangle$ of both the stellar and HI gas components is measured over the same spatial region, defined by the stellar disk extent at each time (i.e. between $0.5$-$1\, R_{\mathrm{opt}}$ and within a vertical extent of $2|h_{90}|$ from the galactic plane).
For each galaxy, in Fig. \ref{fig:drivers_lopsidedness}, we also show the relative distance of the most significant interacting and merging satellites over the last $7\, \mathrm{Gyr}$ from the center of the main host galaxy at each time. Distances are normalized by the virial radius $R_{200}$ of the main host at each time. The satellite orbits are color-coded by the total mass of the satellite galaxy at each time. We consider only satellite galaxies with a total mass-ratio $>1$:$50$ at the time they first cross the virial radius $R_{200}$ of the main host. Furthermore, we show the satellite orbits only until their total mass drops below $\mathrm{M}_{\mathrm{tot}} = 10^{9}\, \mathrm{M}_{\odot}$.

\subsection{Present-day lopsided galaxies}
\label{sec:sample1}
In Fig. \ref{fig:drivers_lopsidedness}, we see that the two present-day lopsided galaxies with consistent stellar and HI gas density asymmetries at $z=0$ (i.e. Au6 and Au25; see Sec. \ref{sec:present_day_lopsidedness}) also show very good agreement between the global $\langle \mathrm{A}_{1} \rangle$ of the stellar and HI gas components at all earlier times over the last $7\, \mathrm{Gyr}$. 

In particular, Au25 shows a mildly decreasing $\langle \mathrm{A}_{1} \rangle$ amplitude over most of its recent history, followed by a steep increase during the last $1\, \mathrm{Gyr}$ up to the present day, consistently in both the stellar and HI gas components. This sharp rise in $\langle \mathrm{A}_{1} \rangle$ occurs just before the deep pericentric passage\footnote{deep pericenter passage here refers to when satellite-host separation reaches its minimum, approaching zero.}, $\sim1\, \mathrm{Gyr\, ago}$, of a massive satellite galaxy with a total mass-ratio in the range $1$:$4$-$1$:$3$ at the time it first crossed the $R_{200}$ of the main host, meaning that the interaction already starts affecting the galaxy morphology before the pericentric passage. In Fig. \ref{fig:Au25} of the Appendix, we show a snapshot sequence to visualize the time of this significant interaction within the last $\sim1\, \mathrm{Gyr\, ago}$.
We therefore conclude that the present-day morphological lopsidedness of Au25 is the result of a tidal interaction with a massive satellite that still survives at $z=0$ and perturbed both the stellar and HI gas density distributions. In previous works, \citet{Gomez2017} showed that this strong interaction also caused a misalignment between the outer dark matter halo and the inner disk in this galaxy using the original Auriga simulation, while \citet{Grand2026} showed that strong two-armed spirals arise from this interaction using the Superstars simulation.

In contrast, Au6 shows two prominent $\langle \mathrm{A}_{1} \rangle$ peaks at $\sim5$ and $\sim2\, \mathrm{Gyr\, ago}$ in both the stellar and HI gas components, indicating that lopsidedness is triggered multiple times in this galaxy. As in the case of Au25, the earlier $\langle \mathrm{A}_{1} \rangle$ peak of Au6 at $\sim5\, \mathrm{Gyr\, ago}$ is associated with an interacting satellite with total mass-ratio in the range $1$:$50$-$1$:$20$ at the time it first crossed the $R_{200}$ of the main host. This satellite undergoes multiple deep pericentric passages at around $6\, \mathrm{Gyr\, ago}$, during which its total mass remains above $10^{9}\, \mathrm{M}_{\odot}$, perturbing both the stellar and HI gas density distributions. Subsequently, the satellite continues orbiting the main host but has lost most of its mass ($\mathrm{M}_{\mathrm{tot}}<10^{9}\, \mathrm{M}_{\odot}$), which is why it does not appear in the bottom panel of Fig. \ref{fig:drivers_lopsidedness}. It eventually merges with the host $\sim0.5\, \mathrm{Gyr\, ago}$, without inducing significant additional perturbations. 
The more recent $\langle \mathrm{A}_{1} \rangle$ peak $\sim2\, \mathrm{Gyr\, ago}$ of Au6 also coincides with the close passage of a fly-by satellite with a total mass-ratio slightly greater than $1$:$20$ at the time of first $R_{200}$ crossing. However, in this case, the satellite does not penetrate deeper than $0.5\, R_{200}$. In Fig. \ref{fig:au6} of the Appendix, we show the snapshot sequence of the fly-by passage of this massive satellite. As a result, the lopsided distribution observed in both the stellar and HI gas density fields may have been induced by dark matter density wakes generated in the halo of the main host by the orbiting satellite \citep{Gomez2016,Garavito2019,Grand2023}. In this scenario, the lopsidedness arises from the response of the disk to the perturbed dark matter halo potential. We will verify this scenario in a follow-up work. Observational studies have reported that the Milky Way has a lopsided HI gas distribution at large galactocentric radii (i.e. $\gtrsim15\, \mathrm{kpc}$) in the direction of the Magellanic Clouds, suggesting that this asymmetry is the result of a dark matter wake induced by this satellite system \citep{Kalberla2008}.

We note that both Au6 and Au25 are characterized by a lower-than-average central stellar mass density, $\mu_{*}$ (see Fig. \ref{fig:present_day_lopsidedness}). For this reason, even the fly-by passage of a lower-mass satellite is likely to induce small perturbations in both the stellar and HI gas density fields.

One interesting difference between Au6 and Au25 concerns the behavior of the young stellar component (i.e. $<0.5\, \mathrm{Gyr\, old}$). In Au6, we find a close correspondence between the lopsidedness amplitudes of the HI gas, young stars, and the total stellar component, suggesting that the gravitational perturbations trigger a coherent response across all components. In contrast, in Au25, we find that the young stellar component is already more strongly lopsided at earlier times ($\sim6$-$2\, \mathrm{Gyr\, ago}$), while both the HI gas and the total stellar component are more weakly lopsided. Inspection of the surface density maps suggests that this behavior reflects the fact that young stars trace asymmetric star formation along the strong spiral arms, meaning that there is more star formation in one of the spirals compared to the other for a two-armed spiral. Therefore, the young stellar component provides a more sensitive tracer of localized, non-axisymmetric overdensities associated with the spiral pattern. This is consistent with the recent work of \citet{Grand2026} who, using the Auriga Superstars simulation, showed that spiral structure can arise as a result of different dynamical processes and evolve on sub-Gyr timescales, leading to significant variations in the morphology and strengths of spiral arms.

\subsection{Present-day symmetric galaxies}
\label{sec:sample2}

\subsubsection{Galaxies with early stellar and HI gas lopsidedness}
Among the present-day symmetric galaxies, Au23, Au24, and Au28 show strong lopsidedness in both stellar and HI gas components at earlier times, i.e. between $\sim5$-$2\, \mathrm{Gyr\, ago}$ for Au23 and Au24, and at $>5\, \mathrm{Gyr\, ago}$ for Au28.

In the case of Au23, only the HI gas density distribution is lopsided at epochs earlier than $5\, \mathrm{Gyr\, ago}$ as a result of smooth gas accretion from two gas-rich fly-by satellites at $\sim6\, \mathrm{Gyr\, ago}$ with total mass-ratios between $1$:$50$ and $1$:$10$ at the time of their first $R_{200}$ crossing.
This gas accretion likely triggers asymmetric star formation, which is traced by the young stellar population. The most massive of the previous two satellites (i.e. total mass-ratio $\sim1$:$10$ at the time of first $R_{200}$ crossing) later interacts more strongly with the main host during a second pericentric passage at $\sim4\, \mathrm{Gyr\, ago}$, producing a coherent lopsided perturbation in the density distributions of the HI gas, young stars, and the total stellar component. In Fig. \ref{fig:au23} of the Appendix, the sequence of snapshots at $\sim6\, \mathrm{Gyr\, ago}$ ({\it top panels}) indicates the epoch of the early gas accretion-induced lopsidedness, while the sequence of snapshots at $\sim4\, \mathrm{Gyr\, ago}$ ({\it bottom panels}) shows the subsequent significant interaction from a massive satellite. Subsequently, the galaxy does not experience any further significant interactions, leading to a gradual damping of its lopsidedness across all components. Using a larger statistical sample from the IllustrisTNG (TNG50) simulation, \citet{Fontirroig2025} showed that $\sim20\%$ of galaxies underwent an early interaction ($\sim6.5\, \mathrm{Gyr\, ago}$) with a massive satellite (mass-ratio $>1$:$20$) that triggered a prominent lopsided perturbation. However, at later times, this lopsidedness gradually decreased and the galaxies eventually became symmetric in the absence of further external interactions (see figure 9 of \citealt{Fontirroig2025}), as we observe for Au23 in this work.

Similarly, Au24 experiences smooth gas accretion from fly-by satellites with total mass-ratios of $\sim1$:$20$ at the time of their first $R_{200}$ crossing at early times ($\gtrsim5\, \mathrm{Gyr\, ago}$), which primarily perturbs the HI gas density distribution and likely triggers asymmetric star formation traced by the young stellar population. At later times, these satellites produce a coherent lopsided perturbation in the density distributions of the HI gas, young stars, and the total stellar components during their subsequent pericentric passages between $\sim5$ and $2\, \mathrm{Gyr\, ago}$, while continuing to accrete gas onto the main host.
At more recent times (i.e. $\lesssim2\, \mathrm{Gyr\, ago}$), Au24 only experienced minor interactions with satellites of total mass-ratios $\sim1$:$50$ at the time of their first $R_{200}$ crossing. However, these satellites' interactions do not induce further significant perturbations due to the fact that Au24 has an above-average central stellar mass density (see Fig. \ref{fig:present_day_lopsidedness}). As a result, there is a gradual decrease in lopsidedness across all components.

In contrast, Au28 experiences a significant interaction with a massive satellite (total mass-ratio $\sim1$:$20$ at the time of its first $R_{200}$ crossing) at early times ($\sim6\, \mathrm{Gyr\, ago}$), which coherently perturbs the density distributions of the HI gas, young stars, and the total stellar component. Following this event, the galaxy undergoes a relatively quiescent evolutionary phase, without further significant interactions and only minor smooth accretion from very low-mass satellites (total mass-ratios $<1$:$50$ at the time of their first $R_{200}$ crossing). The lack of significant interactions, combined with the high central stellar mass density of this galaxy (see Fig. \ref{fig:present_day_lopsidedness}), leads to a rapid decrease in lopsidedness in the stellar component and a more gradual damping of lopsidedness in the HI gas component towards the present-day.
    
\subsubsection{Galaxies with only HI gas lopsidedness}
Finally, the remaining four present-day symmetric galaxies (i.e. Au17, Au18, Au22, and Au27) maintain an overall symmetric stellar disk throughout their evolution, as shown in Fig. \ref{fig:drivers_lopsidedness}. 
In particular, three galaxies (i.e. Au17, Au18, and Au27) exhibit an early perturbation in the HI gas density distribution ($\gtrsim4\, \mathrm{Gyr\, ago}$), which is gradually damped toward the present-day, whereas Au22 maintains a perturbed HI gas density distribution with an approximately constant $\langle \mathrm{A}_{1} \rangle$ throughout its evolution history.

In the case of Au17 and Au18, no significant interactions with nearby satellites occur at early times between $\sim7$-$4\, \mathrm{Gyr\, ago}$. These galaxies are also characterized by above-average central stellar mass densities (see Fig. \ref{fig:present_day_lopsidedness}), which likely stabilizes their stellar disk against external perturbations. For this reason, they maintain an overall symmetric stellar disk, while the lopsidedness in their HI gas density distributions is likely driven by the asymmetric smooth gas accretion from the cosmic web or low-mass satellites. This process may also trigger asymmetric star formation, traced by the young stellar population. At later times, both galaxies experience the pericentric passage ($\sim2\, \mathrm{Gyr\, ago}$) of a satellite with a total mass-ratio of $\sim1$:$50$ at the time of its first $R_{200}$ crossing. However, because of the low satellite mass, its shallow penetration into the main host $R_{200}$, and the relatively high central stellar mass density of these galaxies, this fly-by does not induce significant perturbations in either the stellar or HI disks. Only the young stellar component becomes mildly lopsided at these times, tracing localized, non-axisymmetric overdensities associated with the spiral pattern, similarly to Au25.

A similar scenario also applies to Au22, which has not experienced any significant interactions or mergers within the last $7\, \mathrm{Gyr}$. The lack of strong interactions, combined with the very high central stellar mass density of this galaxy, likely explains why it exhibits the most symmetric stellar disk throughout its history. In this case, continuous smooth gas accretion, either from the cosmic web or low-mass satellite galaxies, likely produces the lopsided HI mass distribution, while both the total and young stellar components trace the underlying smooth mass distribution of the compact stellar disk with no prominent spiral arms. Alternatively, the strong AGN activity presents in this galaxy could be an additional source of perturbation of its HI gas distribution \citep{Grand2017}. In a follow-up study, we will test the gas accretion scenario by using tracer particles to track the time evolution of gas cells in these galaxies.

In contrast, Au27 undergoes several pericentric passages from a gas-rich satellite with a total mass-ratio between $1$:$50$ and $1$:$20$ at the time of its first $R_{200}$ crossing between $\sim7$-$4\, \mathrm{Gyr\, ago}$. These interactions primarily perturb the density distribution of the HI gas and likely trigger asymmetric star formation, traced by the young stellar population that closely follows the HI gas distribution. The total stellar disk remains largely unaffected, likely due to its relatively high central stellar mass density, which stabilizes it against such perturbations. At later times, Au27 mainly experiences smooth accretion from low-mass satellites (total mass-ratios $\lesssim1$:$50$ at the time of their first $R_{200}$ crossing), leading to a gradual decrease in lopsidedness in the density distributions of both the HI gas and the young stellar population.
    
\section{Summary and conclusions}
\label{sec:conclusions}
In this work, we characterize lopsidedness in both the density and kinematics of the stellar and HI gas components of a sample of nine Milky Way-type galaxies from the Auriga Superstars cosmological zoom-in simulations, as well as of a sample from the original Auriga simulations. We quantify both morphological and kinematical lopsidedness, as described in Sec. \ref{sec:morphological_lopsidedness} and \ref{sec:kinematical_lopsidedness}. From the analysis of the correlation between global morphological lopsidedness in the stellar and HI gas components, as well as the correlation between global morphological and kinematical lopsidedness in both the stellar and HI gas components at $z=0$, we find the following results:

\begin{itemize} \itemsep0.25cm
    \item Within the stellar optical radius $R_{\mathrm{opt}}$, we find that an asymmetry in the stellar density distribution is generally accompanied by a corresponding asymmetry in the HI gas density distribution, but the opposite is not necessarily true. In particular, we find that the old stellar component ($>0.5\, \mathrm{Gyr\, old}$) is strongly coupled with the HI gas, such that high $\langle \mathrm{A}_{1} \rangle$ in the stars corresponds to high $\langle \mathrm{A}_{1} \rangle$ in the HI gas, making it a tracer of asymmetries in the overall mass distribution of galaxies. In contrast, the young stellar component ($<0.5\, \mathrm{Gyr\, old}$) is sensitive to asymmetric star formation along the strong spiral arms. These results are discussed in Sec. \ref{sec:present_day_lopsidedness_correlation}.

    \item Morphological and kinematical lopsidedness in the stellar component are also correlated, as expected if lopsidedness arises from a perturbed gravitational potential that coherently affects both stellar orbits and their spatial distribution, consistent with the results from previous numerical simulations \citep{Ghosh2025}. In contrast, we find only a weak correlation between morphological and kinematical lopsidedness in the HI gas, which is typically characterized by stronger kinematical than morphological asymmetries. Taken together, these results suggest that, independent of whether lopsidedness is tidally- and/or gas accretion-induced, its imprint is stronger in the HI gas kinematics than in its morphology (as described in Sec. \ref{sec:kinematics_morphology_lopsidedness_correlation}).

    \item An anti-correlation between lopsidedness in the stellar density distribution of the outer galactic disk and the bar strength in the inner regions is found. In particular, the most strongly barred galaxies typically host the most symmetric disks and are characterized by high central stellar mass density, whereas lopsided galaxies tend to have low central stellar mass density. These results, discussed in Sec. \ref{sec:present_day_lopsidedness_bars}, suggest that the anti-correlation between morphological lopsidedness in the stellar component and the presence of a central bar is linked to the internal properties of the galaxies. In particular, the central stellar mass density likely plays a major role in determining whether the galaxy is more susceptible to developing lopsidedness and/or a bar, consistent with previous results \citep{Fragkoudi2025,Fontirroig2025}.

    \item The overall stellar mass distribution of the galaxy, as well as its central stellar mass density, are not strongly affected by the significantly improved stellar mass resolution of the Superstars method, which is a factor of $64\times$ better (i.e. $m_{\mathrm{star}}=800\, \mathrm{M}_{\odot}$) relative to the gas cell mass resolution (i.e. $m_{\mathrm{gas}}=5\times10^{4}\, \mathrm{M}_{\odot}$). This is demonstrated in Sec. \ref{sec:convergence_lopsidedness}, indicating that lopsidedness is well converged with stellar resolution.
\end{itemize}

Additionally, from the analysis of the time evolution of lopsidedness in the stellar and HI gas density distributions of the Superstars galaxies (see Sec. \ref{sec:drivers_lopsidedness}), we find the following results:

\begin{itemize}
    \item A close correspondence between peaks in the global $\langle \mathrm{A}_{1} \rangle$ of the stellar and HI gas components is typically associated with the pericentric passage of a massive satellite (total mass-ratio $>1$:$50$ at the time of first $R_{200}$ crossing), which triggers a coherent response across all components. However, whether lopsidedness arises through the direct physical encounter or through the indirect response of the disk to the perturbed dark matter halo will be explored in a follow up work. We find that in $5$ out of $9$ galaxies lopsidedness is triggered by a tidal encounter (i.e. Au6, Au23, Au24, Au25, and Au28). 

    \item A lopsided density distribution only in the HI gas is generally consistent with smooth gas accretion, either from gas-rich fly-by satellites or from the cosmic web. We find that in $6$ out of $9$ galaxies lopsidedness is triggered by smooth gas accretion (i.e. Au17, Au18, Au22, Au23, Au24, and Au27). When lopsidedness is driven by gas accretion, the HI gas and young stars show similar lopsided amplitudes, while the total stellar distribution remains largely symmetric. 

    \item When only the young stellar population is lopsided, this is indicative of asymmetric star formation along strong spiral arms (e.g. Au17 and Au25). Therefore, the young stellar component can act as a sensitive tracer of localized, non-axisymmetric overdensities associated with the spiral pattern. On the contrary, when only the HI gas is lopsided at all times (e.g. Au22), it likely indicates smooth gas accretion or strong AGN activity.
\end{itemize}

One interesting finding is that lopsidedness over a galaxy evolutionary history can be driven by a combination of both processes can occur. Gas accretion from a fly-by satellite may first produce lopsidedness only in the HI gas component, while a subsequent tidal interaction with the same or another satellite can later produce lopsidedness in all components (i.e. stars and HI gas). Specific examples include Au23 and Au24. 

In a follow-up work, we will investigate in more detail the role of gas accretion as a driving mechanism of lopsidedness in disk galaxies. By tracking gas cells in the Auriga Superstars simulations, we aim to distinguish accretion from the cosmic web, from satellite galaxies, and from recycled gas associated with galactic fountain flow \citep{Grand2019}, and to measure the direction of gas inflows. This will allow us to determine the origin of gas accretion, determine whether this inflow occurs asymmetrically, and assess how such accretion is connected to the observed coupling or decoupling of lopsidedness in the HI gas, young stellar population, and total stellar components.

Overall, these results show that, depending on whether we observe strong lopsidedness in the HI gas and/or in the stellar component, can provide important clues about the mechanism that triggered the lopsided perturbation and the recent galactic evolutionary history.

\begin{acknowledgements}
AD acknowledges support from the FONDECYT Postdoctorado No. 3250558. RB is supported by the SNSF through the Ambizione Grant PZ00P2$\_$223532. RJJG is supported by an STFC Ernest Rutherford Fellowship (ST/W003643/1). FAG and AM acknowledge support from the ANID BASAL project FB210003, and funding from the HORIZON-MSCA-2021-SE-01 Research and Innovation Programme under the Marie Sklodowska-Curie grant agreement number 101086388. FAG acknowledges support from the ANID FONDECYT Regular grant 1251493. AM acknowledges support from the ANID FONDECYT Regular grant 1251882. FvdV is supported by a Royal Society University Research Fellowship (URF$\textbackslash$R1$\textbackslash$191703 and URF$\textbackslash$R$\textbackslash$241005). 
\end{acknowledgements}

%
\bibliographystyle{aa} 
\bibliography{aa} 

\begin{thebibliography}{50}
\expandafter\ifx\csname natexlab\endcsname\relax\def\natexlab#1{#1}\fi

\bibitem[{{Baldwin} {et~al.}(1980){Baldwin}, {Lynden-Bell}, \&
  {Sancisi}}]{Baldwin1980}
{Baldwin}, J.~E., {Lynden-Bell}, D., \& {Sancisi}, R. 1980, \mnras, 193, 313

\bibitem[{{Beale} \& {Davies}(1969)}]{Beale1969}
{Beale}, J.~S. \& {Davies}, R.~D. 1969, \nat, 221, 531

\bibitem[{{Bekki} \& {Couch}(2011)}]{Bekki2011}
{Bekki}, K. \& {Couch}, W.~J. 2011, \mnras, 415, 1783

\bibitem[{{Bilimogga} {et~al.}(2025){Bilimogga}, {Busekool}, {Verheijen}, \&
  {van der Hulst}}]{Bilimogga2025}
{Bilimogga}, P.~V., {Busekool}, E., {Verheijen}, M.~A.~W., \& {van der Hulst},
  J.~M. 2025, arXiv e-prints, arXiv:2508.01425

\bibitem[{{Blitz} \& {Rosolowsky}(2006)}]{Blitz2006}
{Blitz}, L. \& {Rosolowsky}, E. 2006, \apj, 650, 933

\bibitem[{{Block} {et~al.}(1994){Block}, {Bertin}, {Stockton}, {Grosbol},
  {Moorwood}, \& {Peletier}}]{Block1994}
{Block}, D.~L., {Bertin}, G., {Stockton}, A., {et~al.} 1994, \aap, 288, 365

\bibitem[{{Bosma}(1981)}]{Bosma1981}
{Bosma}, A. 1981, \aj, 86, 1791

\bibitem[{{Bournaud} {et~al.}(2004){Bournaud}, {Combes}, \&
  {Jog}}]{Bournaud2004}
{Bournaud}, F., {Combes}, F., \& {Jog}, C.~J. 2004, \aap, 418, L27

\bibitem[{{Bournaud} {et~al.}(2005){Bournaud}, {Combes}, {Jog}, \&
  {Puerari}}]{Bournaud2005}
{Bournaud}, F., {Combes}, F., {Jog}, C.~J., \& {Puerari}, I. 2005, \aap, 438,
  507

\bibitem[{{Deeley} {et~al.}(2021){Deeley}, {Drinkwater}, {Sweet}, {Bekki},
  {Couch}, {Forbes}, \& {Dolfi}}]{Deeley2021}
{Deeley}, S., {Drinkwater}, M.~J., {Sweet}, S.~M., {et~al.} 2021, \mnras, 508,
  895

\bibitem[{{Dolfi} {et~al.}(2024){Dolfi}, {Gomez}, {Monachesi}, {Tissera},
  {Sifon}, \& {Galaz}}]{Dolfi2024}
{Dolfi}, A., {Gomez}, F.~A., {Monachesi}, A., {et~al.} 2024, arXiv e-prints,
  arXiv:2411.19426

\bibitem[{{Dolfi} {et~al.}(2023){Dolfi}, {G{\'o}mez}, {Monachesi},
  {Varela-Lavin}, {Tissera}, {Sif{\'o}n}, \& {Galaz}}]{Dolfi2023}
{Dolfi}, A., {G{\'o}mez}, F.~A., {Monachesi}, A., {et~al.} 2023, \mnras, 526,
  567

\bibitem[{{Feng} {et~al.}(2025){Feng}, {Shen}, {Chen}, {Dai}, {Yin}, {Cui},
  {Ju}, \& {Li}}]{Feng2025}
{Feng}, S., {Shen}, S., {Chen}, Y., {et~al.} 2025, \apj, 995, 18

\bibitem[{{Fontirroig} {et~al.}(2025){Fontirroig}, {G{\'o}mez}, {Jaque
  Arancibia}, {Dolfi}, \& {Monsalves}}]{Fontirroig2025}
{Fontirroig}, V., {G{\'o}mez}, F.~A., {Jaque Arancibia}, M., {Dolfi}, A., \&
  {Monsalves}, N. 2025, \aap, 699, A118

\bibitem[{{Fragkoudi} {et~al.}(2025){Fragkoudi}, {Grand}, {Pakmor},
  {G{\'o}mez}, {Marinacci}, \& {Springel}}]{Fragkoudi2025}
{Fragkoudi}, F., {Grand}, R. J.~J., {Pakmor}, R., {et~al.} 2025, \mnras, 538,
  1587

\bibitem[{{Garavito-Camargo} {et~al.}(2019){Garavito-Camargo}, {Besla},
  {Laporte}, {Johnston}, {G{\'o}mez}, \& {Watkins}}]{Garavito2019}
{Garavito-Camargo}, N., {Besla}, G., {Laporte}, C. F.~P., {et~al.} 2019, \apj,
  884, 51

\bibitem[{{Ghosh} {et~al.}(2025){Ghosh}, {Di Matteo}, {Jog}, \&
  {Frankel}}]{Ghosh2025}
{Ghosh}, S., {Di Matteo}, P., {Jog}, C.~J., \& {Frankel}, N. 2025, arXiv
  e-prints, arXiv:2511.07549

\bibitem[{{G{\'o}mez} {et~al.}(2017){G{\'o}mez}, {White}, {Grand}, {Marinacci},
  {Springel}, \& {Pakmor}}]{Gomez2017}
{G{\'o}mez}, F.~A., {White}, S. D.~M., {Grand}, R. J.~J., {et~al.} 2017,
  \mnras, 465, 3446

\bibitem[{{G{\'o}mez} {et~al.}(2016){G{\'o}mez}, {White}, {Marinacci},
  {Slater}, {Grand}, {Springel}, \& {Pakmor}}]{Gomez2016}
{G{\'o}mez}, F.~A., {White}, S. D.~M., {Marinacci}, F., {et~al.} 2016, \mnras,
  456, 2779

\bibitem[{{Grand} {et~al.}(2024){Grand}, {Fragkoudi}, {G{\'o}mez}, {Jenkins},
  {Marinacci}, {Pakmor}, \& {Springel}}]{Grand2024}
{Grand}, R. J.~J., {Fragkoudi}, F., {G{\'o}mez}, F.~A., {et~al.} 2024, \mnras,
  532, 1814

\bibitem[{{Grand} {et~al.}(2026){Grand}, {Fragkoudi}, {Pakmor}, {G{\'o}mez},
  {van de Voort}, {Bieri}, \& {Townson}}]{Grand2026}
{Grand}, R. J.~J., {Fragkoudi}, F., {Pakmor}, R., {et~al.} 2026, arXiv
  e-prints, arXiv:2602.15108

\bibitem[{{Grand} {et~al.}(2017){Grand}, {G{\'o}mez}, {Marinacci}, {Pakmor},
  {Springel}, {Campbell}, {Frenk}, {Jenkins}, \& {White}}]{Grand2017}
{Grand}, R. J.~J., {G{\'o}mez}, F.~A., {Marinacci}, F., {et~al.} 2017, \mnras,
  467, 179

\bibitem[{{Grand} {et~al.}(2023){Grand}, {Pakmor}, {Fragkoudi}, {G{\'o}mez},
  {Trick}, {Simpson}, {van de Voort}, \& {Bieri}}]{Grand2023}
{Grand}, R. J.~J., {Pakmor}, R., {Fragkoudi}, F., {et~al.} 2023, \mnras, 524,
  801

\bibitem[{{Grand} {et~al.}(2019){Grand}, {van de Voort}, {Zjupa}, {Fragkoudi},
  {G{\'o}mez}, {Kauffmann}, {Marinacci}, {Pakmor}, {Springel}, \&
  {White}}]{Grand2019}
{Grand}, R. J.~J., {van de Voort}, F., {Zjupa}, J., {et~al.} 2019, \mnras, 490,
  4786

\bibitem[{{Haynes} {et~al.}(1998){Haynes}, {Hogg}, {Maddalena}, {Roberts}, \&
  {van Zee}}]{Haynes1998}
{Haynes}, M.~P., {Hogg}, D.~E., {Maddalena}, R.~J., {Roberts}, M.~S., \& {van
  Zee}, L. 1998, \aj, 115, 62

\bibitem[{{Jog}(1997)}]{Jog1997}
{Jog}, C.~J. 1997, \apj, 488, 642

\bibitem[{{Jog}(2002)}]{Jog2002}
{Jog}, C.~J. 2002, \aap, 391, 471

\bibitem[{Jog \& Combes(2009)}]{Jog2009_review}
Jog, C.~J. \& Combes, F. 2009, Physics Reports, 471, 75

\bibitem[{{Kalberla} \& {Dedes}(2008)}]{Kalberla2008}
{Kalberla}, P.~M.~W. \& {Dedes}, L. 2008, \aap, 487, 951

\bibitem[{{Le Bail} {et~al.}(2024){Le Bail}, {Daddi}, {Elbaz}, {Dickinson},
  {Giavalisco}, {Magnelli}, {G{\'o}mez-Guijarro}, {Kalita}, {Koekemoer},
  {Holwerda}, {Bournaud}, {de la Vega}, {Calabr{\`o}}, {Dekel}, {Cheng},
  {Bisigello}, {Franco}, {Costantin}, {Lucas}, {P{\'e}rez-Gonz{\'a}lez}, {Lu},
  {Wilkins}, {Arrabal Haro}, {Bagley}, {Finkelstein}, {Kartaltepe}, {Papovich},
  {Pirzkal}, \& {Yung}}]{LeBail2025}
{Le Bail}, A., {Daddi}, E., {Elbaz}, D., {et~al.} 2024, \aap, 688, A53

\bibitem[{{Leroy} {et~al.}(2008){Leroy}, {Walter}, {Brinks}, {Bigiel}, {de
  Blok}, {Madore}, \& {Thornley}}]{Leroy2008}
{Leroy}, A.~K., {Walter}, F., {Brinks}, E., {et~al.} 2008, \aj, 136, 2782

\bibitem[{{{\L}okas}(2022)}]{Lokas2022}
{{\L}okas}, E.~L. 2022, \aap, 662, A53

\bibitem[{{Marinacci} {et~al.}(2017){Marinacci}, {Grand}, {Pakmor}, {Springel},
  {G{\'o}mez}, {Frenk}, \& {White}}]{Marinacci2017}
{Marinacci}, F., {Grand}, R. J.~J., {Pakmor}, R., {et~al.} 2017, \mnras, 466,
  3859

\bibitem[{{Nelson} {et~al.}(2019){Nelson}, {Springel}, {Pillepich},
  {Rodriguez-Gomez}, {Torrey}, {Genel}, {Vogelsberger}, {Pakmor}, {Marinacci},
  {Weinberger}, {Kelley}, {Lovell}, {Diemer}, \& {Hernquist}}]{Nelson2019}
{Nelson}, D., {Springel}, V., {Pillepich}, A., {et~al.} 2019, Computational
  Astrophysics and Cosmology, 6, 2

\bibitem[{{Noordermeer} {et~al.}(2001){Noordermeer}, {Sparke}, \&
  {Levine}}]{Noordermeer2001}
{Noordermeer}, E., {Sparke}, L.~S., \& {Levine}, S.~E. 2001, \mnras, 328, 1064

\bibitem[{{Pakmor} {et~al.}(2025{\natexlab{a}}){Pakmor}, {Bieri}, {Fragkoudi},
  {G{\'o}mez}, {Grand}, {Simpson}, {Talbot}, {van de Voort}, \&
  {Werhahn}}]{Pakmor2025}
{Pakmor}, R., {Bieri}, R., {Fragkoudi}, F., {et~al.} 2025{\natexlab{a}},
  \mnras, 543, 1761

\bibitem[{{Pakmor} {et~al.}(2025{\natexlab{b}}){Pakmor}, {Fragkoudi}, {Grand},
  {Simpson}, {G{\'o}mez}, {van de Voort}, {Bieri}, {Trick}, {Werhahn}, \&
  {Talbot}}]{Pakmor2025_Superstars}
{Pakmor}, R., {Fragkoudi}, F., {Grand}, R. J.~J., {et~al.} 2025{\natexlab{b}},
  \mnras, 543, 4355

\bibitem[{{Planck Collaboration} {et~al.}(2014){Planck Collaboration}, {Ade},
  {Aghanim}, {Armitage-Caplan}, {Arnaud}, {Ashdown}, {Atrio-Barandela},
  {Aumont}, {Baccigalupi}, {Banday}, {Barreiro}, {Bartlett}, {Battaner},
  {Benabed}, {Beno{\^\i}t}, {Benoit-L{\'e}vy}, {Bernard}, {Bersanelli},
  {Bielewicz}, {Bobin}, {Bock}, {Bonaldi}, {Bond}, {Borrill}, {Bouchet},
  {Bridges}, {Bucher}, {Burigana}, {Butler}, {Calabrese}, {Cappellini},
  {Cardoso}, {Catalano}, {Challinor}, {Chamballu}, {Chary}, {Chen}, {Chiang},
  {Chiang}, {Christensen}, {Church}, {Clements}, {Colombi}, {Colombo},
  {Couchot}, {Coulais}, {Crill}, {Curto}, {Cuttaia}, {Danese}, {Davies},
  {Davis}, {de Bernardis}, {de Rosa}, {de Zotti}, {Delabrouille}, {Delouis},
  {D{\'e}sert}, {Dickinson}, {Diego}, {Dolag}, {Dole}, {Donzelli}, {Dor{\'e}},
  {Douspis}, {Dunkley}, {Dupac}, {Efstathiou}, {Elsner}, {En{\ss}lin},
  {Eriksen}, {Finelli}, {Forni}, {Frailis}, {Fraisse}, {Franceschi}, {Gaier},
  {Galeotta}, {Galli}, {Ganga}, {Giard}, {Giardino}, {Giraud-H{\'e}raud},
  {Gjerl{\o}w}, {Gonz{\'a}lez-Nuevo}, {G{\'o}rski}, {Gratton}, {Gregorio},
  {Gruppuso}, {Gudmundsson}, {Haissinski}, {Hamann}, {Hansen}, {Hanson},
  {Harrison}, {Henrot-Versill{\'e}}, {Hern{\'a}ndez-Monteagudo}, {Herranz},
  {Hildebrandt}, {Hivon}, {Hobson}, {Holmes}, {Hornstrup}, {Hou}, {Hovest},
  {Huffenberger}, {Jaffe}, {Jaffe}, {Jewell}, {Jones}, {Juvela},
  {Keih{\"a}nen}, {Keskitalo}, {Kisner}, {Kneissl}, {Knoche}, {Knox}, {Kunz},
  {Kurki-Suonio}, {Lagache}, {L{\"a}hteenm{\"a}ki}, {Lamarre}, {Lasenby},
  {Lattanzi}, {Laureijs}, {Lawrence}, {Leach}, {Leahy}, {Leonardi},
  {Le{\'o}n-Tavares}, {Lesgourgues}, {Lewis}, {Liguori}, {Lilje},
  {Linden-V{\o}rnle}, {L{\'o}pez-Caniego}, {Lubin}, {Mac{\'\i}as-P{\'e}rez},
  {Maffei}, {Maino}, {Mandolesi}, {Maris}, {Marshall}, {Martin},
  {Mart{\'\i}nez-Gonz{\'a}lez}, {Masi}, {Massardi}, {Matarrese}, {Matthai},
  {Mazzotta}, {Meinhold}, {Melchiorri}, {Melin}, {Mendes}, {Menegoni},
  {Mennella}, {Migliaccio}, {Millea}, {Mitra}, {Miville-Desch{\^e}nes},
  {Moneti}, {Montier}, {Morgante}, {Mortlock}, {Moss}, {Munshi}, {Murphy},
  {Naselsky}, {Nati}, {Natoli}, {Netterfield}, {N{\o}rgaard-Nielsen},
  {Noviello}, {Novikov}, {Novikov}, {O'Dwyer}, {Osborne}, {Oxborrow}, {Paci},
  {Pagano}, {Pajot}, {Paladini}, {Paoletti}, {Partridge}, {Pasian},
  {Patanchon}, {Pearson}, {Pearson}, {Peiris}, {Perdereau}, {Perotto},
  {Perrotta}, {Pettorino}, {Piacentini}, {Piat}, {Pierpaoli}, {Pietrobon},
  {Plaszczynski}, {Platania}, \& {Pointecouteau}}]{Planck2014}
{Planck Collaboration}, {Ade}, P.~A.~R., {Aghanim}, N., {et~al.} 2014, \aap,
  571, A16

\bibitem[{{Reichard} {et~al.}(2008){Reichard}, {Heckman}, {Rudnick},
  {Brinchmann}, \& {Kauffmann}}]{Reichard2008}
{Reichard}, T.~A., {Heckman}, T.~M., {Rudnick}, G., {Brinchmann}, J., \&
  {Kauffmann}, G. 2008, \apj, 677, 186

\bibitem[{{Richter} \& {Sancisi}(1994)}]{Richter1994}
{Richter}, O.-G. \& {Sancisi}, R. 1994, \aap, 290, L9

\bibitem[{{Rix} \& {Zaritsky}(1995)}]{Rix1995}
{Rix}, H.-W. \& {Zaritsky}, D. 1995, \apj, 447, 82

\bibitem[{{Schoenmakers} {et~al.}(1997){Schoenmakers}, {Franx}, \& {de
  Zeeuw}}]{Schoenmakers1997}
{Schoenmakers}, R.~H.~M., {Franx}, M., \& {de Zeeuw}, P.~T. 1997, \mnras, 292,
  349

\bibitem[{{Springel}(2010)}]{Springel2010}
{Springel}, V. 2010, \mnras, 401, 791

\bibitem[{{Springel} \& {Hernquist}(2003)}]{Springel2003}
{Springel}, V. \& {Hernquist}, L. 2003, \mnras, 339, 289

\bibitem[{{Swaters} {et~al.}(1999){Swaters}, {Schoenmakers}, {Sancisi}, \& {van
  Albada}}]{Swaters1999}
{Swaters}, R.~A., {Schoenmakers}, R.~H.~M., {Sancisi}, R., \& {van Albada},
  T.~S. 1999, \mnras, 304, 330

\bibitem[{{Varela-Lavin} {et~al.}(2023){Varela-Lavin}, {G{\'o}mez}, {Tissera},
  {Besla}, {Garavito-Camargo}, {Marinacci}, \& {Laporte}}]{VarelaLavin2023}
{Varela-Lavin}, S., {G{\'o}mez}, F.~A., {Tissera}, P.~B., {et~al.} 2023,
  \mnras, 523, 5853

\bibitem[{{Vogelsberger} {et~al.}(2013){Vogelsberger}, {Genel}, {Sijacki},
  {Torrey}, {Springel}, \& {Hernquist}}]{Vogelsberger2013}
{Vogelsberger}, M., {Genel}, S., {Sijacki}, D., {et~al.} 2013, \mnras, 436,
  3031

\bibitem[{{Weinberg}(1995)}]{Weinberg1995}
{Weinberg}, M.~D. 1995, \apjl, 455, L31

\bibitem[{{Wilcots}(2010)}]{Wilcots2010}
{Wilcots}, E.~M. 2010, in Astronomical Society of the Pacific Conference
  Series, Vol. 421, Galaxies in Isolation: Exploring Nature Versus Nurture, ed.
  L.~{Verdes-Montenegro}, A.~{Del Olmo}, \& J.~{Sulentic}, 149

\bibitem[{{Zaritsky} \& {Rix}(1997)}]{Zaritsky1997}
{Zaritsky}, D. \& {Rix}, H.-W. 1997, \apj, 477, 118

\end{thebibliography}
%

\appendix 

\section{Stellar mass distribution across models and resolution}
\label{sec:convergence_lopsidedness}

\begin{figure*}[!htbp]
    \centering
    \includegraphics[width=0.33\textwidth]{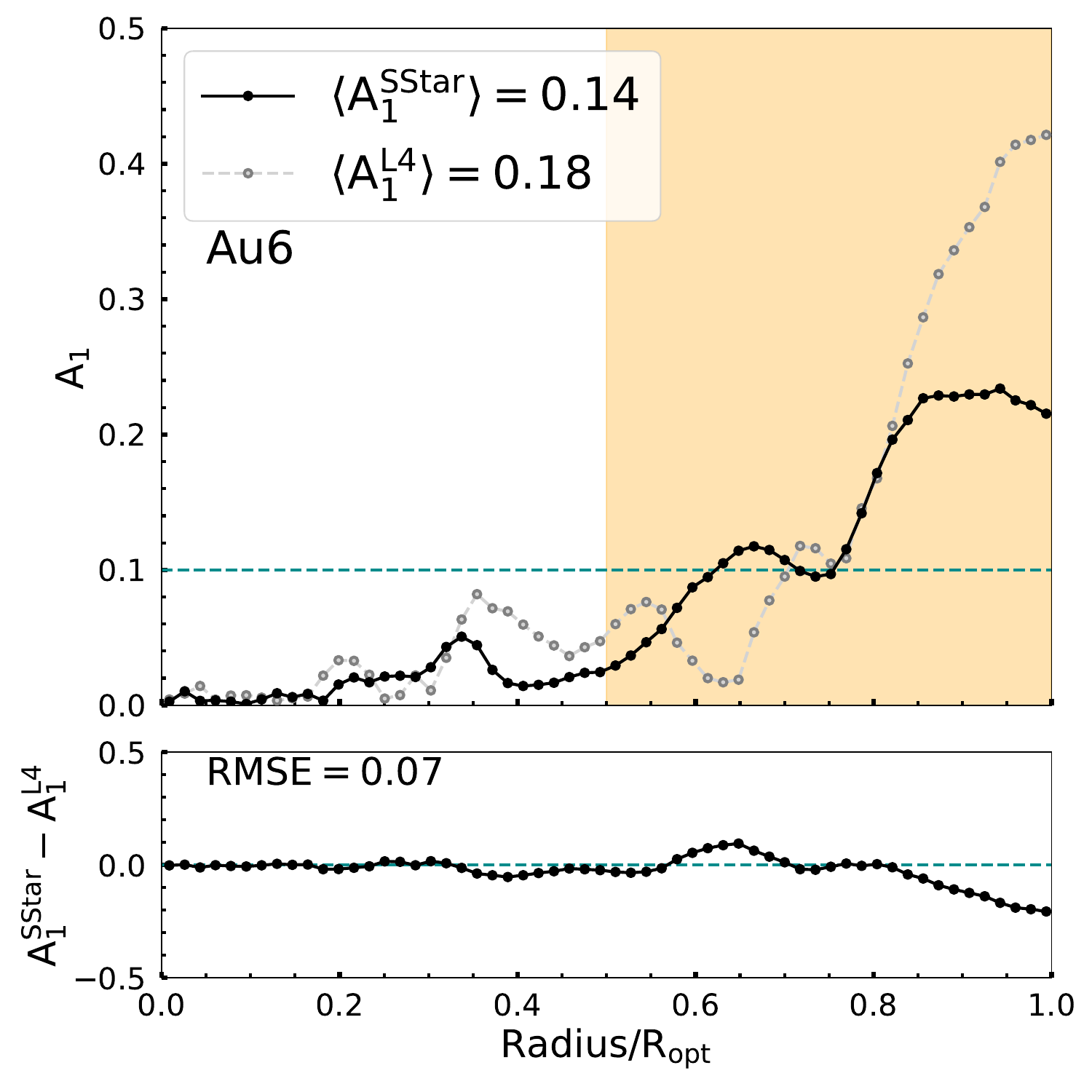}
    \includegraphics[width=0.33\textwidth]{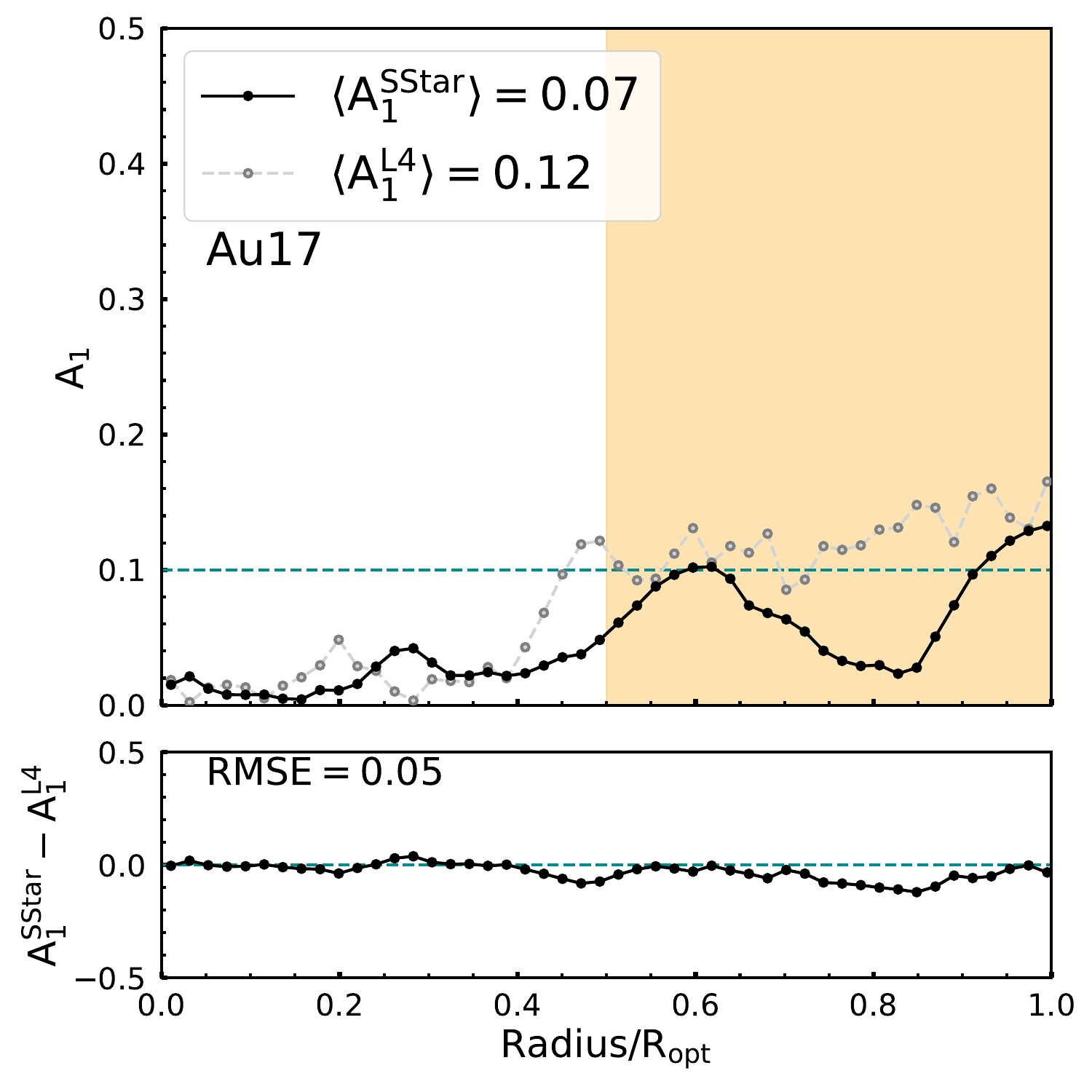}
    \includegraphics[width=0.33\textwidth]{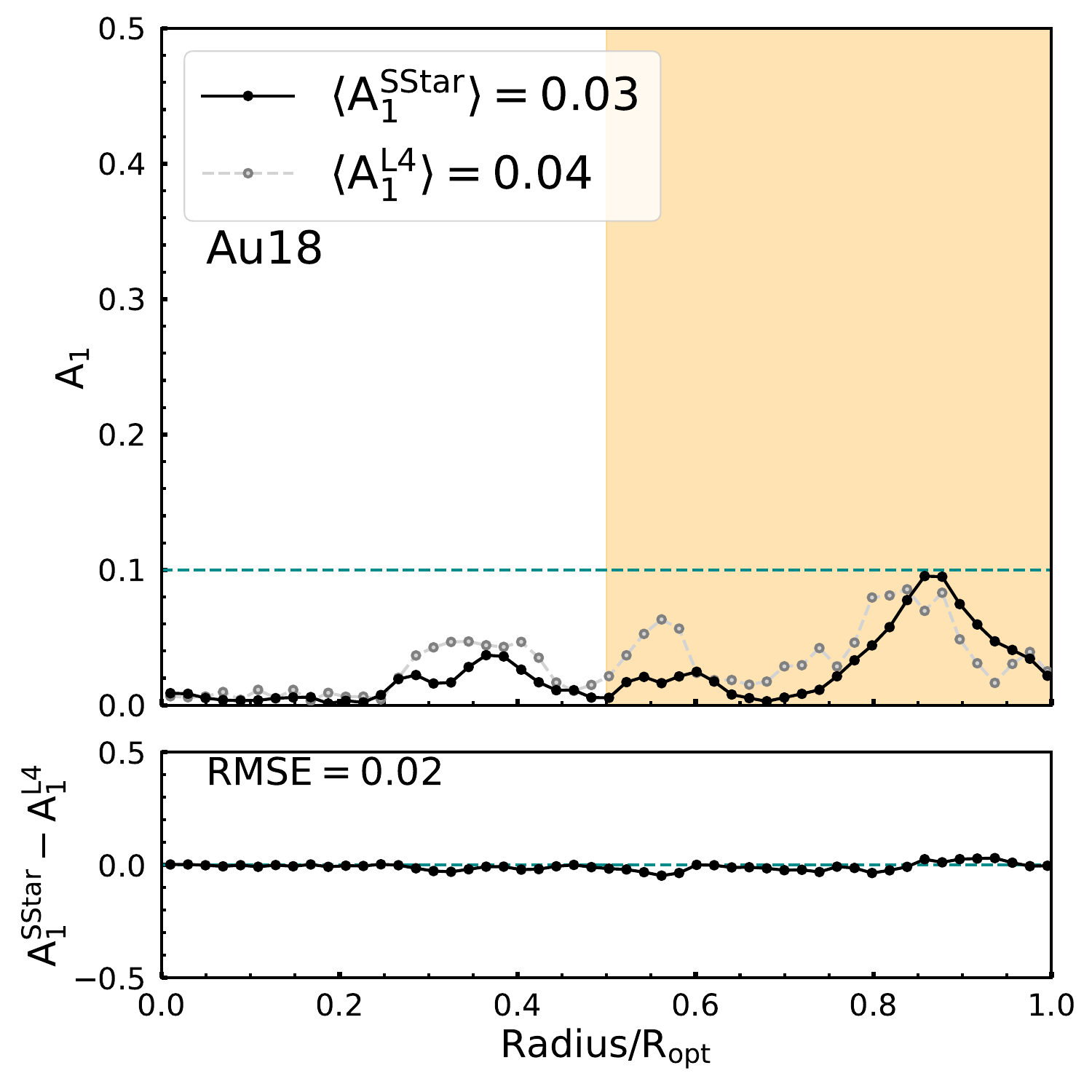}
    \includegraphics[width=0.33\textwidth]{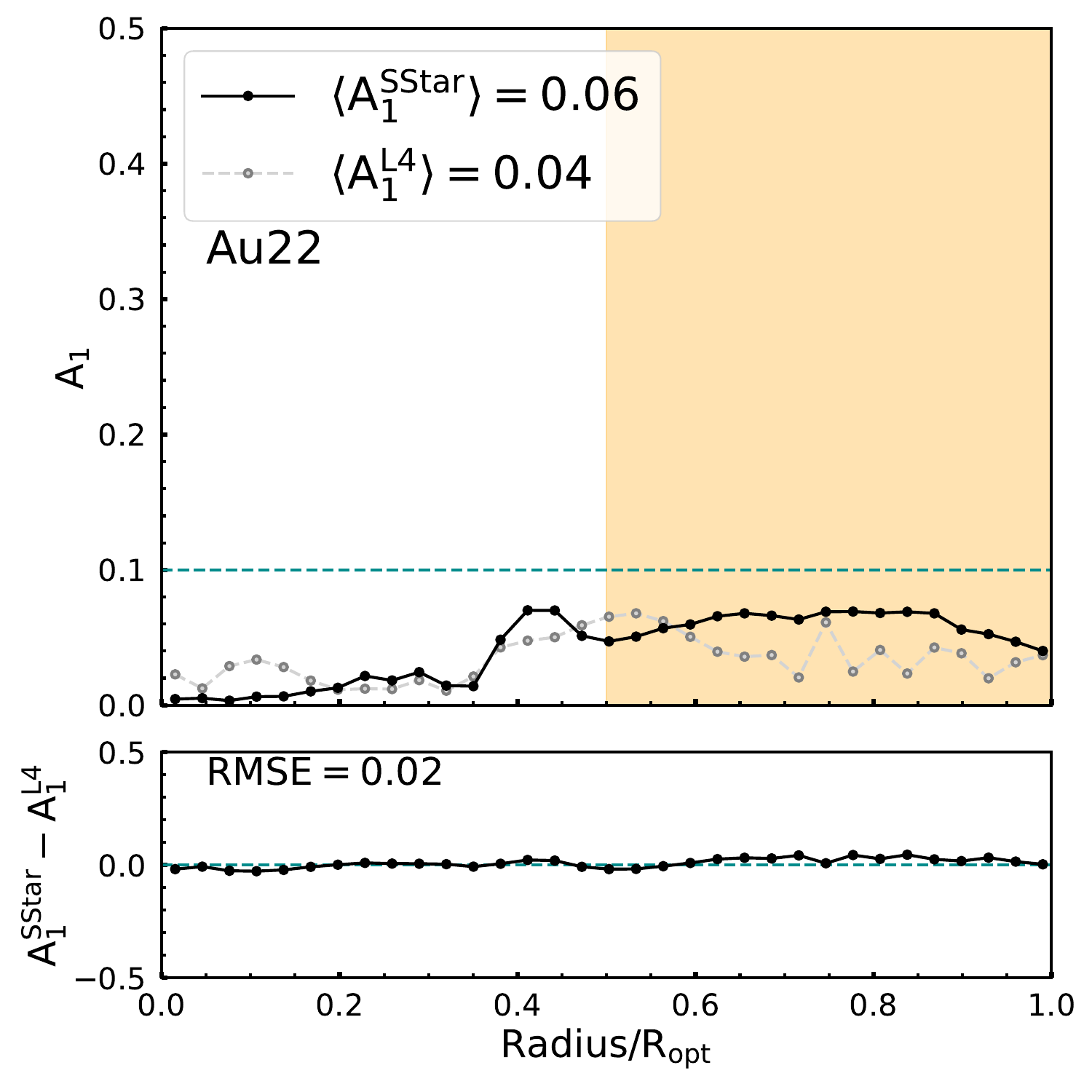}
    \includegraphics[width=0.33\textwidth]{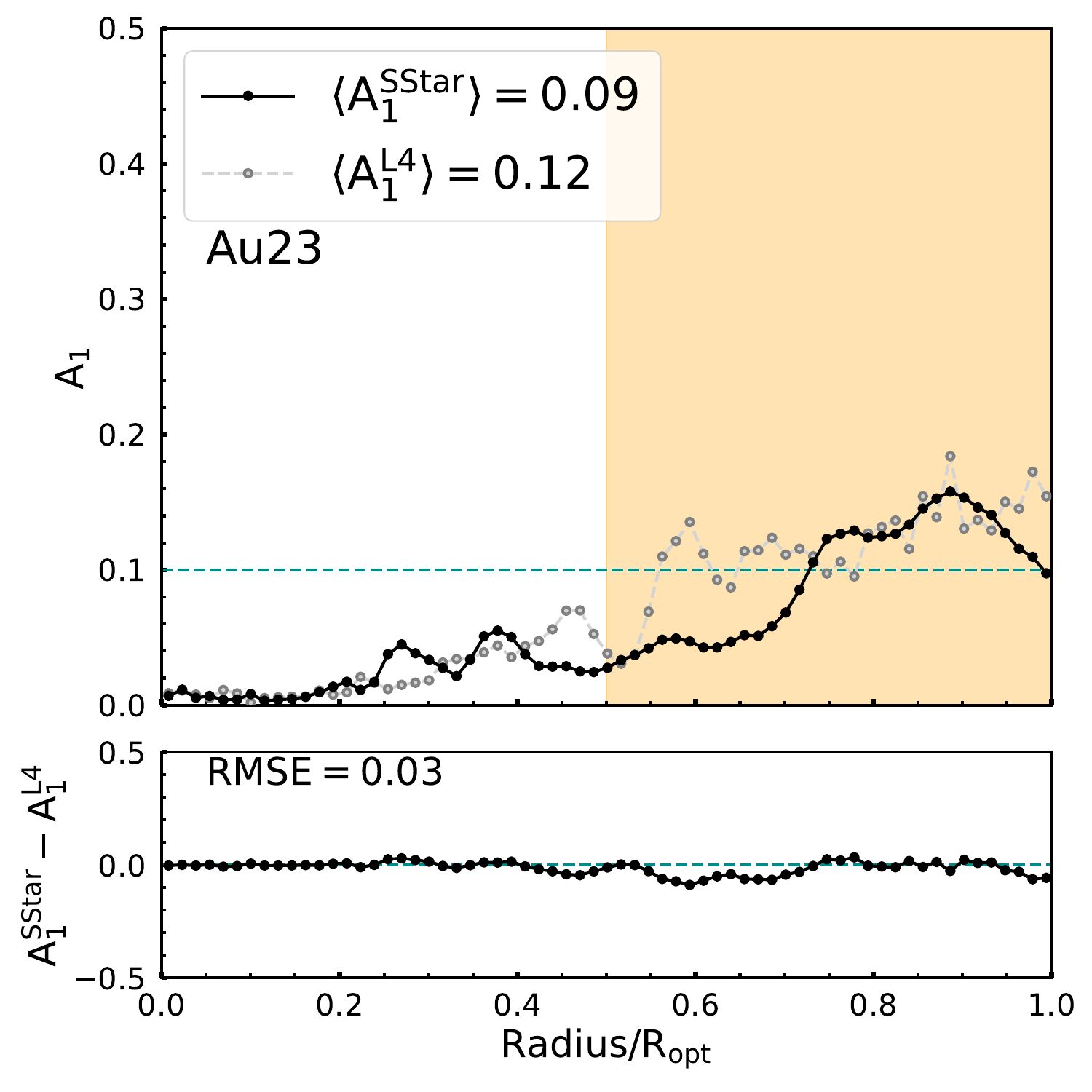}
    \includegraphics[width=0.33\textwidth]{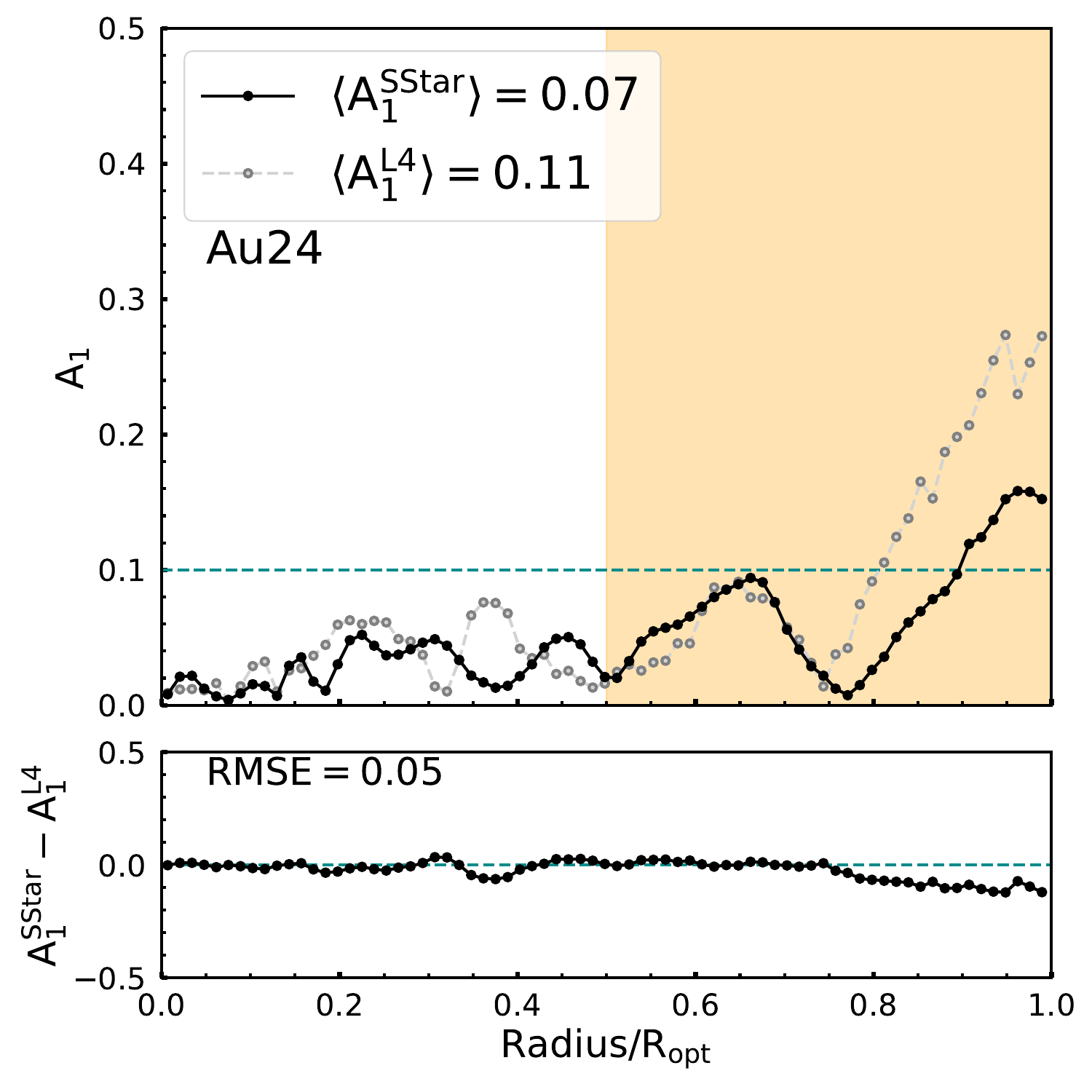}
    \includegraphics[width=0.33\textwidth]{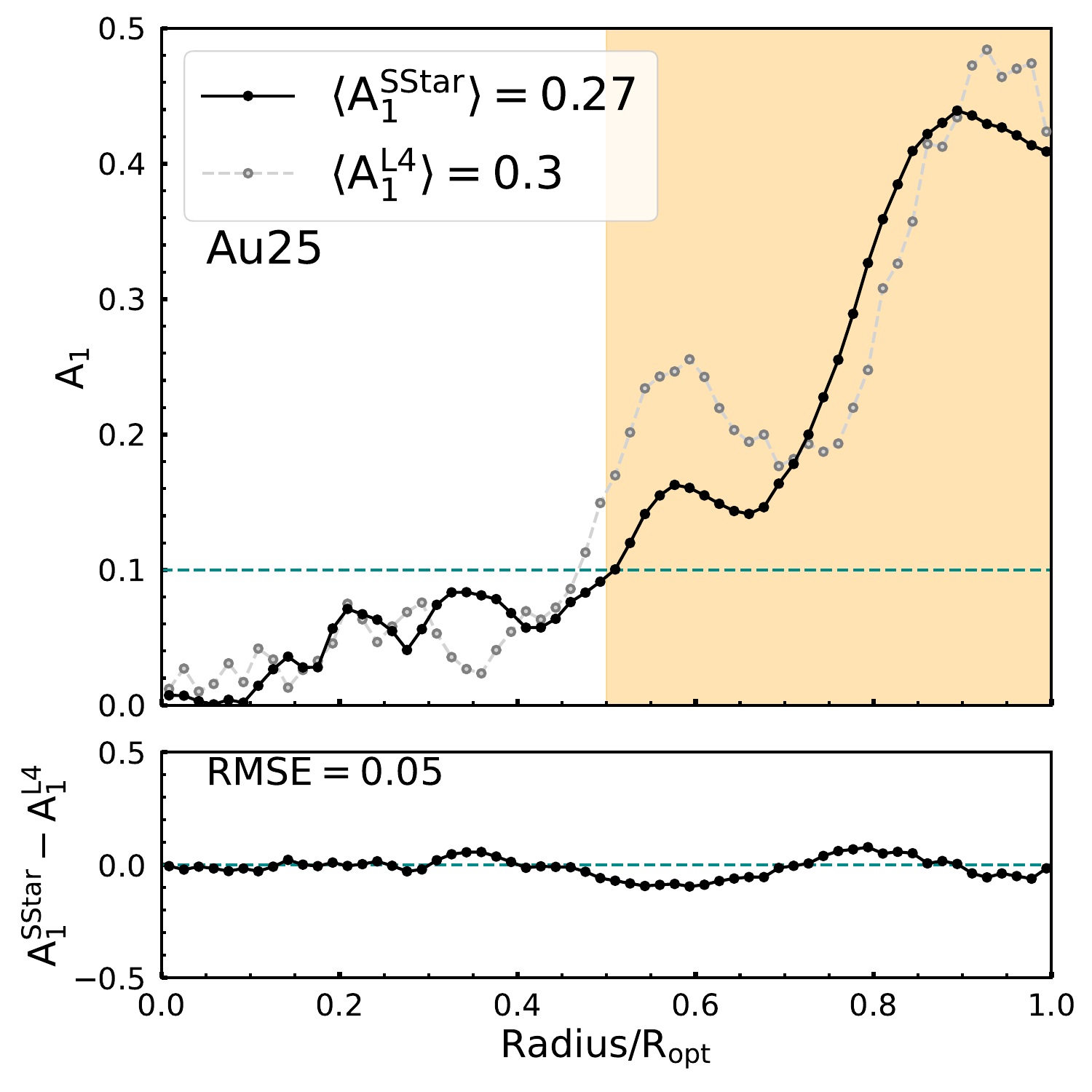}
    \includegraphics[width=0.33\textwidth]{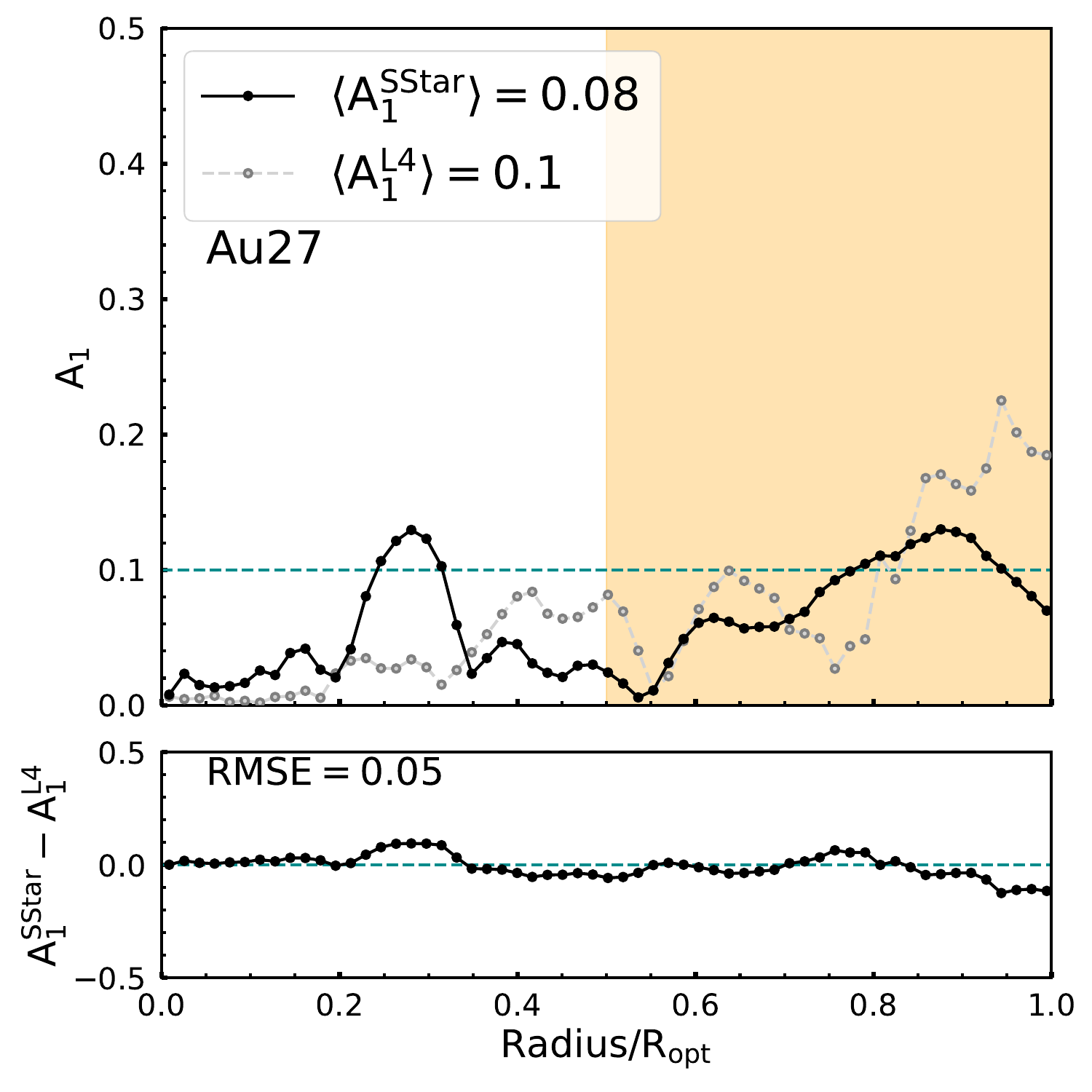}
    \includegraphics[width=0.33\textwidth]{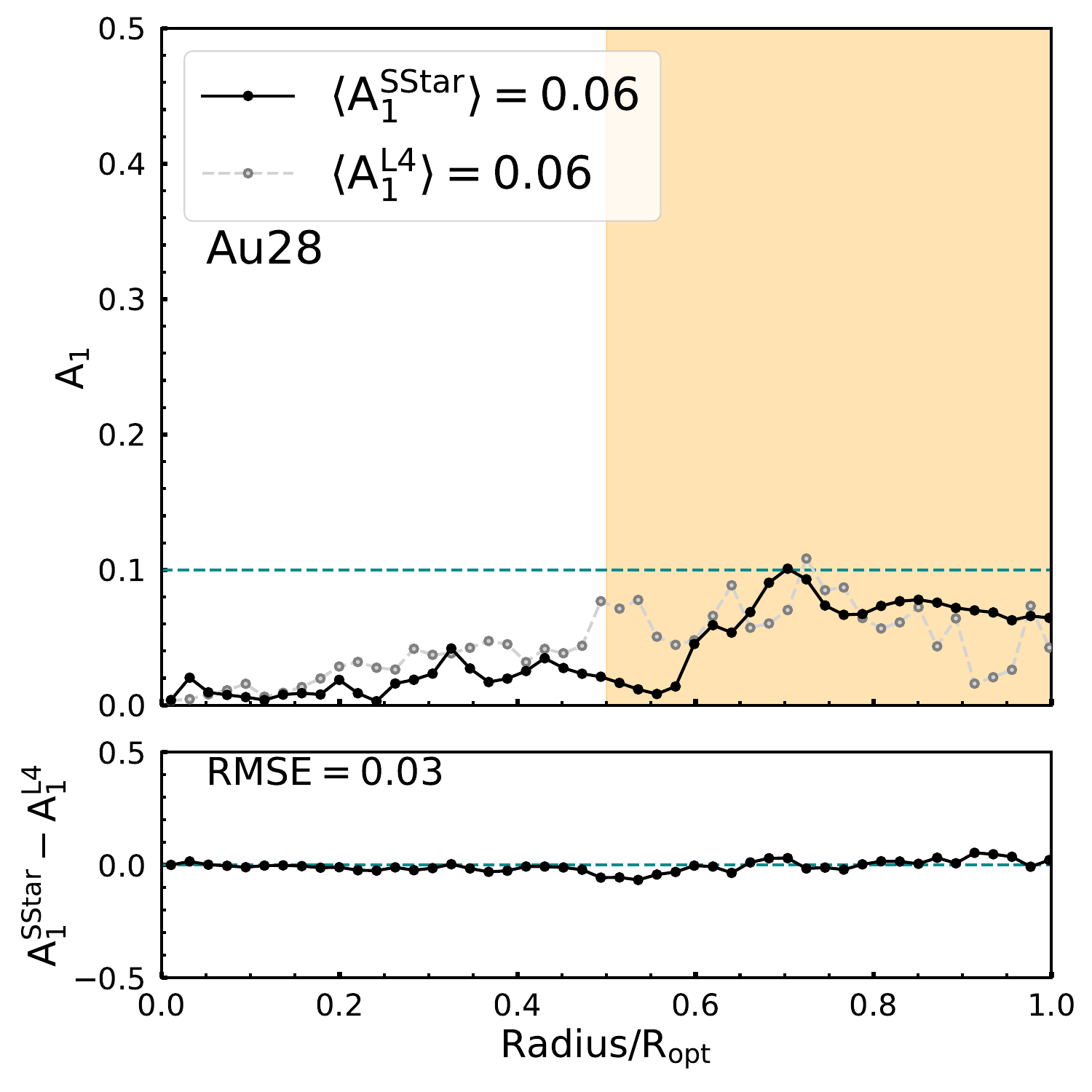}
    \caption{{\it Top panels:} Present-day lopsided profiles of stellar mass density of the nine galaxies in Superstars (black lines) compared with their corresponding low-resolution counterpart in the original Auriga simulations from \citet{Grand2017} (gray lines). The horizontal dashed line indicates the lopsidedness threshold $\langle \mathrm{A_{1}} \rangle = 0.1$, while the shaded orange region represents the radial interval between $0.5$-$1\, R_{\mathrm{opt}}$ used to quantify the global lopsidedness of the stellar disk. The resulting global $\langle \mathrm{A_{1}} \rangle$ is shown in the legend for each galaxy at both the Superstars and L4 resolution. {\it Bottom panels:} For each galaxy, we compute the difference between the lopsided profiles measured in Superstars and L4 at each radius and calculate the root mean square error to quantify the overall variation.}
    \label{fig:convergence_lopsidedness}
\end{figure*}

In this section, we first study whether the improved stellar particle mass resolution of Superstars has a significant impact on the overall stellar mass distribution within the galactic disk at $z=0$.
For each Superstars galaxy (see Sec. \ref{sec:sample_selection}), we consider its low resolution counterpart in L4. For both sets of galaxies, we calculate the optical radius (i.e. $R_{\mathrm{opt}}^{\mathrm{SStar}}$ and $R_{\mathrm{opt}}^{\mathrm{L4}}$) and disk height (i.e. $h_{90}^{\mathrm{SStar}}$ and $h_{90}^{\mathrm{L4}}$), as described in Sec. \ref{sec:stellar_disk}. We then calculate the average values of <$R_{\mathrm{opt}}$> and <$h_{90}$> between the Superstars and L4 resolutions for each galaxy\footnote{We note that the individual values of $R_{\mathrm{opt}}$ and $h_{90}$ are generally similar between the Superstars and L4 simulations}. We use these average values to define the extension of the stellar disk, and the spatial region over which the lopsidedness is measured in both the Superstars and L4 resolutions.  

In the top panels of Fig. \ref{fig:convergence_lopsidedness}, we show the present-day lopsided profiles of each Superstars galaxy at both the Superstars and L4 resolution, while, in the bottom panels, we show the difference between the lopsided profiles measured at the Superstars and L4 resolutions. We also calculate the root-mean-square-error (RMSE) to quantify the overall difference between the lopsided profiles. In the bottom panels of Fig. \ref{fig:convergence_lopsidedness}, we see that the difference $\mathrm{A}_{1}^{\mathrm{SStar}}-\mathrm{A}_{1}^{\mathrm{L4}}$ between the lopsided profiles measured in the Superstars and L4 resolutions show only moderate local variations as a function of radius ($\lesssim0.02$). We find an overall good agreement between the lopsided profiles of the galaxies in Superstars and L4, as suggested by the small $\mathrm{RMSE}\lesssim0.05$. Only one galaxy (i.e. Au6) shows a larger $\mathrm{RMSE}\simeq0.07$, but the shapes of the lopsided profiles in Superstars and L4 still show good agreement. In fact, we find that the strongly lopsided galaxies in Superstars are also strongly lopsided in L4 (e.g. Au25; $\mathrm{A}_{1}>>0.1$), while the strongly symmetric galaxies in Superstars are also strongly symmetric in L4 (e.g. Au18; $\mathrm{A}_{1}<<0.1$), as suggested by the comparable $\langle \mathrm{A_{1}} \rangle$ values calculated for each galaxy. Weakly perturbed galaxies in Superstars with $\langle \mathrm{A_{1}} \rangle$ values close to the threshold $\simeq0.1$ also show similar weakly perturbed lopsided profiles in L4 (i.e. Au17, Au23, Au24, Au27). 
Furthermore, we note that the Superstars galaxies have always generally the same (Au28) or slightly smaller $\langle \mathrm{A_{1}} \rangle$ than the original lower resolution L4 galaxies. This indicates that noise can affect the lopsidedness measurement, as we see that the L4 lopsided profiles are noisier at large radii than the Superstars lopsided profiles.

\begin{figure}[!htbp]
    \centering
    \includegraphics[width=0.45\textwidth]{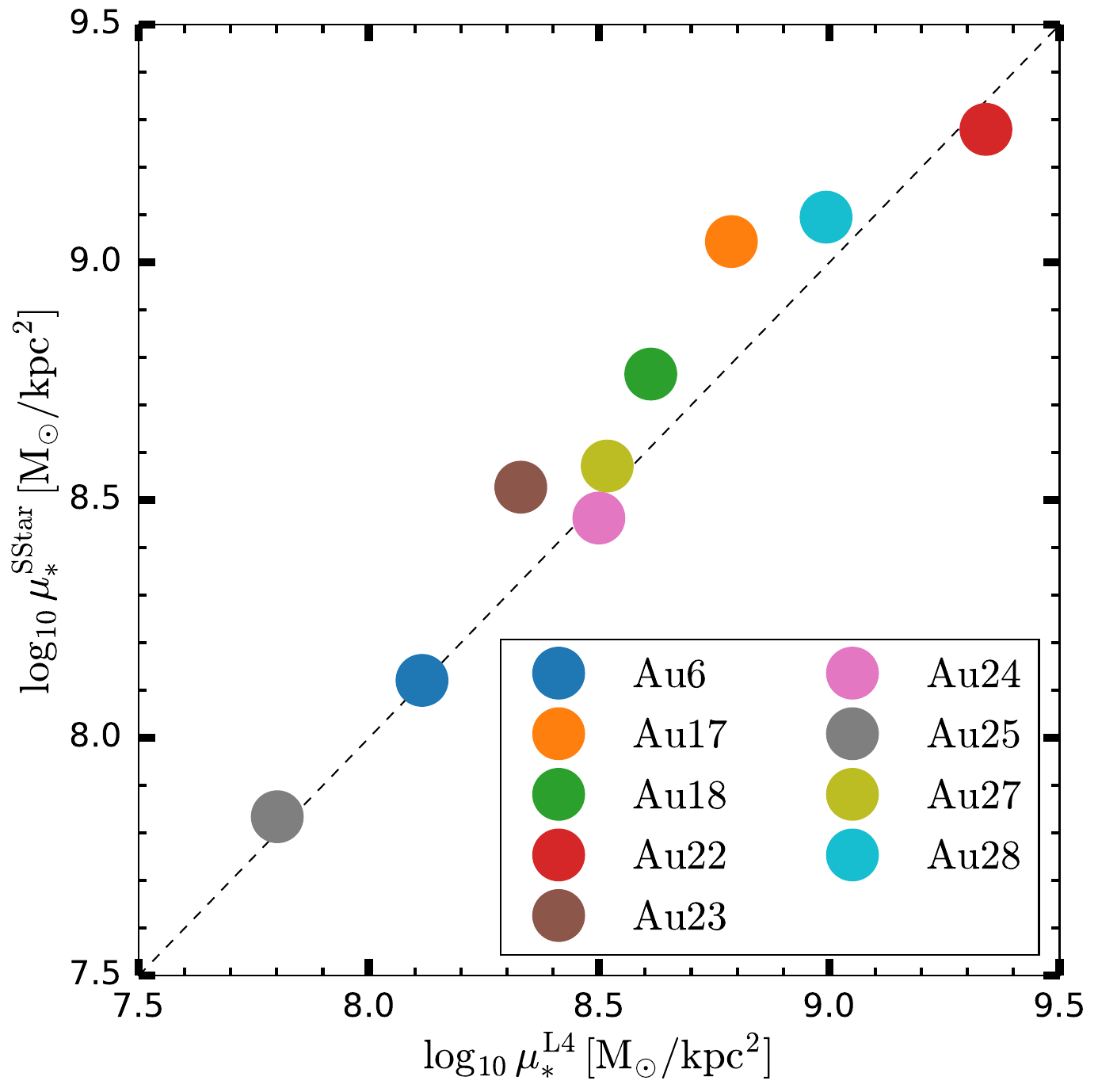}
    \caption{The present-day central stellar mass density of the nine galaxies in Superstars compared with the central stellar mass density of the same galaxies at the L4 resolution. The diagonal dashed black line indicates the one-to-one relation.}
    \label{fig:convergence_stellar_mass_surface_density}
\end{figure}

For each Superstars galaxy, we also calculate the present-day central stellar mass density\footnote{The central stellar mass density, $\mu_{*}=M_{*,\mathrm{h}}/\pi R_{\mathrm{h}}^{2}$, is defined as the stellar mass enclosed within an area delimited by the stellar half-mass radius $R_{\mathrm{h}}$ of the galaxy.} and we compare it with that measured in the L4 counterpart. Similarly to Fig. \ref{fig:convergence_lopsidedness}, we use the average value of the half-mass radius $\langle R_{\mathrm{h}} \rangle$, calculated between the Superstars and L4 resolutions for each galaxy, to define the inner galactic regions where the central stellar mass density is measured in both Superstars and L4 resolutions. The results are shown in Fig. \ref{fig:convergence_stellar_mass_surface_density}. We find that there is a very good agreement between the central stellar mass density in Superstars and L4 at $z=0$ (i.e. all galaxies typically lie along the one-to-one relation). Only two galaxies (i.e. Au17, Au23) show a mild displacement from the one-to-one relation, being characterized by higher central stellar mass density in Superstars than in L4. This might explain their, on average, lower $\langle \mathrm{A_{1}} \rangle$ measured in Superstars than L4 (see Fig. \ref{fig:convergence_lopsidedness}), consistent with previous works that found a strong dependance of the lopsided amplitude on the central stellar mass density of the galaxies \citep{Fontirroig2025}.

Overall, these results show that the implementation of the Superstars method to improve the stellar particle mass resolution with respect to that of the gas cells does not significantly affect the overall stellar mass distribution (i.e. lopsidedness) within the galactic stellar disk, nor the central stellar mass density of the galaxy. 
In Sec. \ref{sec:present_day_lopsidedness_correlation}, \ref{sec:present_day_lopsidedness_bars}, and \ref{sec:kinematics_morphology_lopsidedness_correlation}, we use these results to expand the Superstars galaxy sample by including the L4 disk galaxies (i.e. with $\mathrm{D/T} > 0.45$) that are not modeled with Superstars, thereby improving the statistical significance of the correlations between global galaxy properties.

\section{Temporal evolution}
\label{sec:temporal_evolution}
In this section, we present maps of the stellar and HI density distributions at consecutive output times for three of the Superstars galaxies discussed in Sec. \ref{sec:drivers_lopsidedness}. The corresponding lookback time is indicated in each panel.

\begin{figure*}
    \centering
    \includegraphics[width=0.15\textwidth]{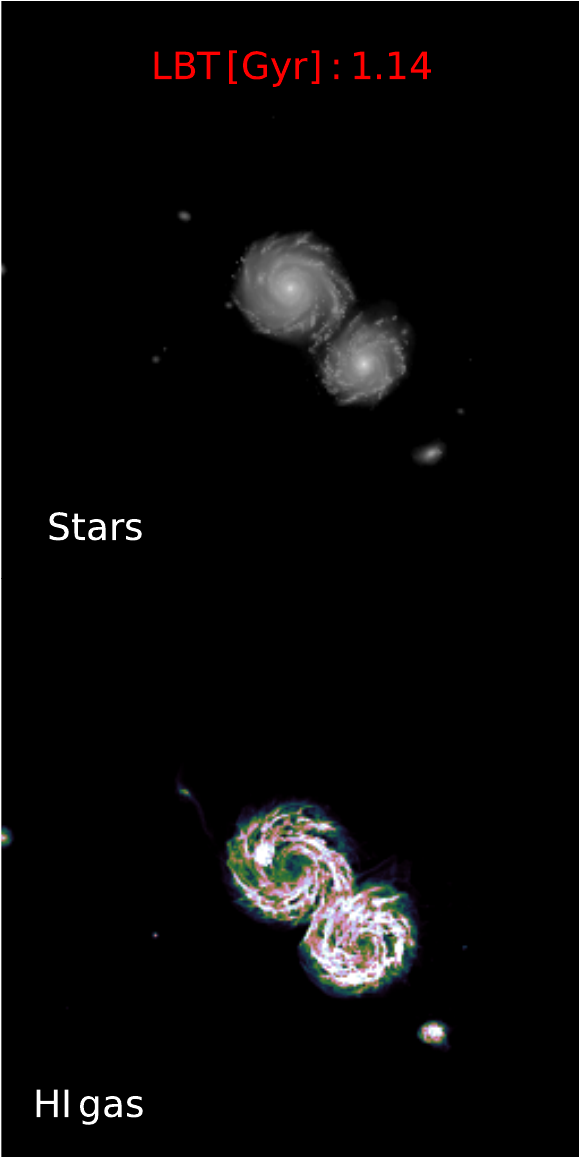}
    \includegraphics[width=0.15\textwidth]{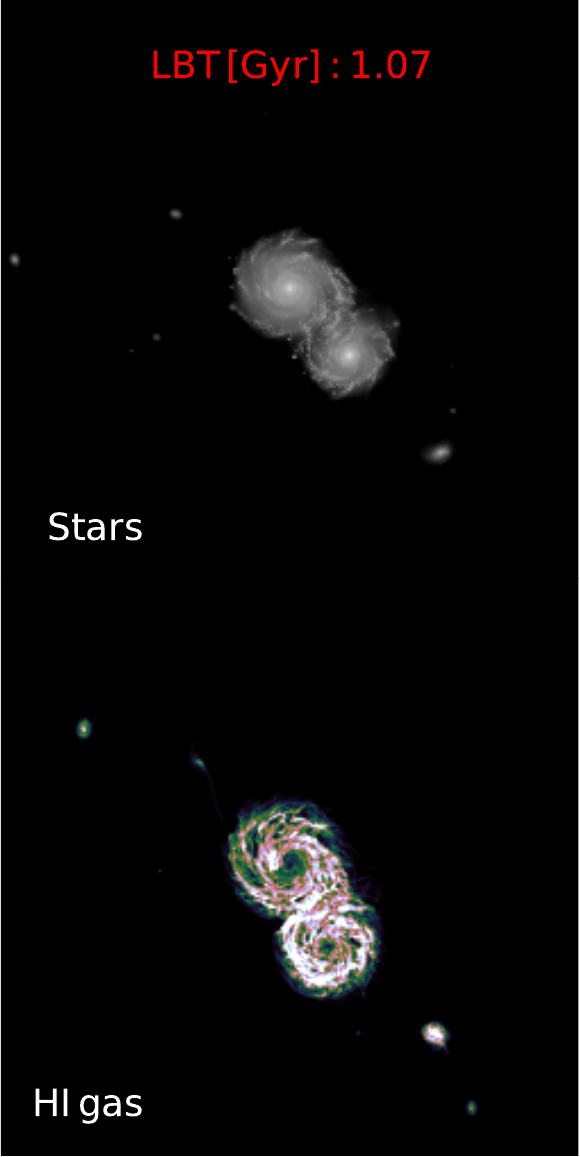}
    \includegraphics[width=0.15\textwidth]{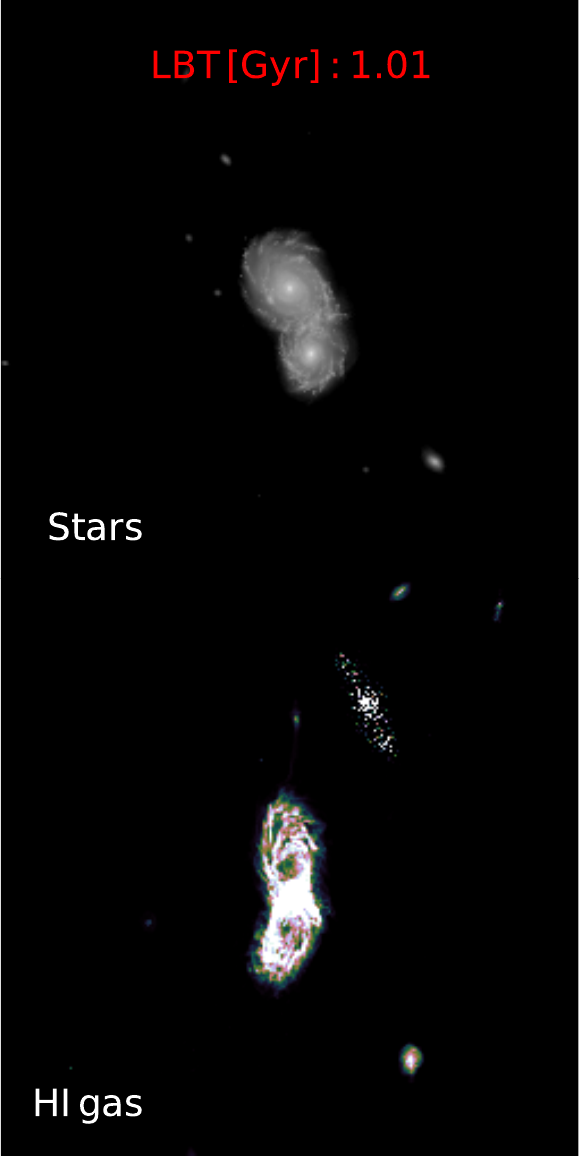}
    \includegraphics[width=0.15\textwidth]{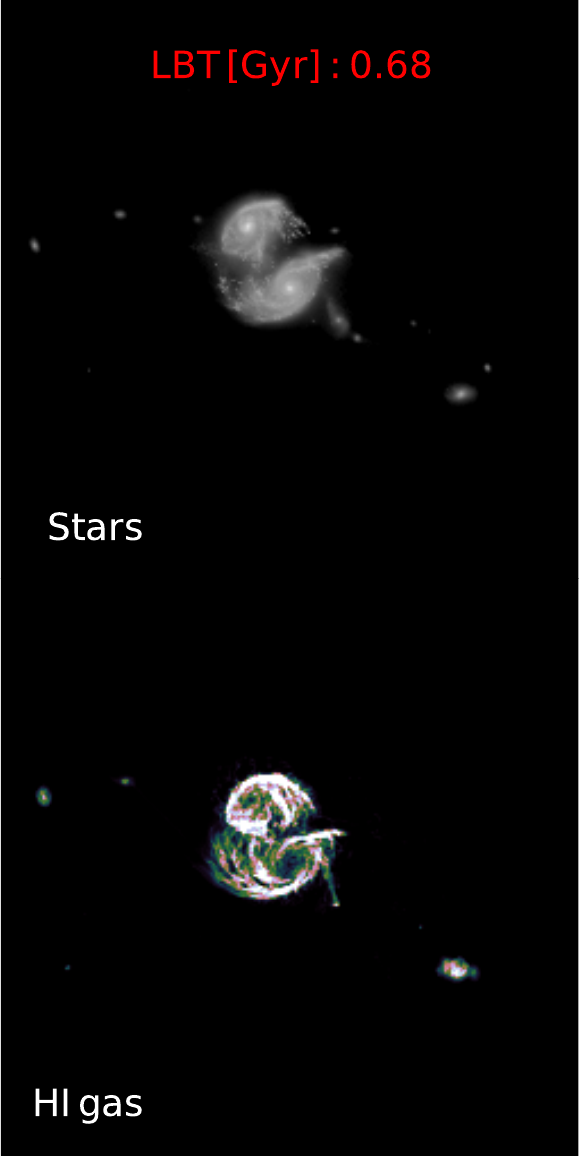}
    \includegraphics[width=0.15\textwidth]{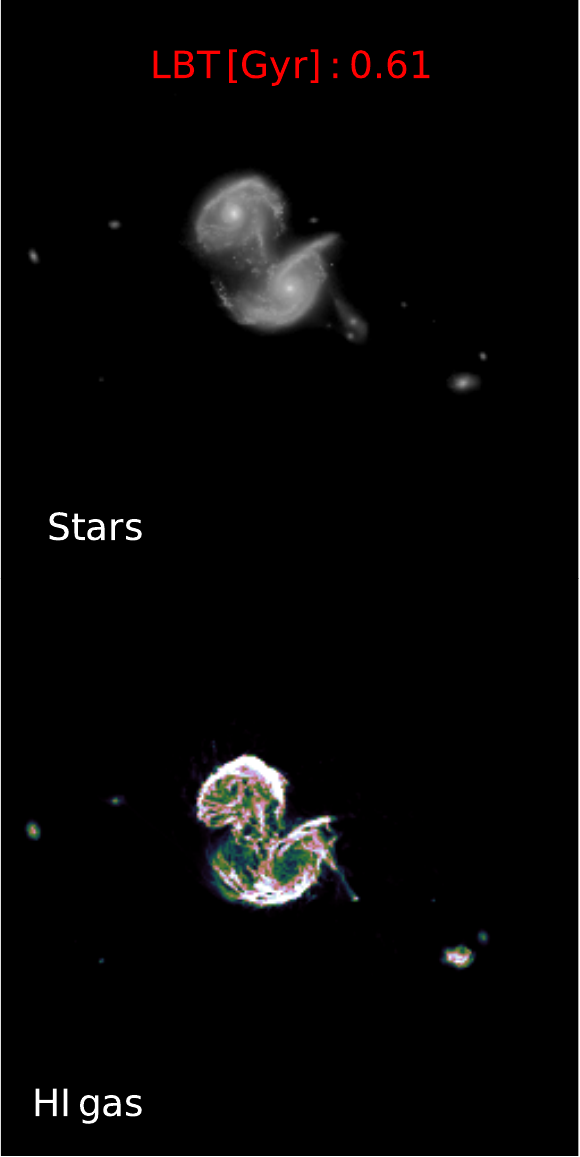}
    \includegraphics[width=0.15\textwidth]{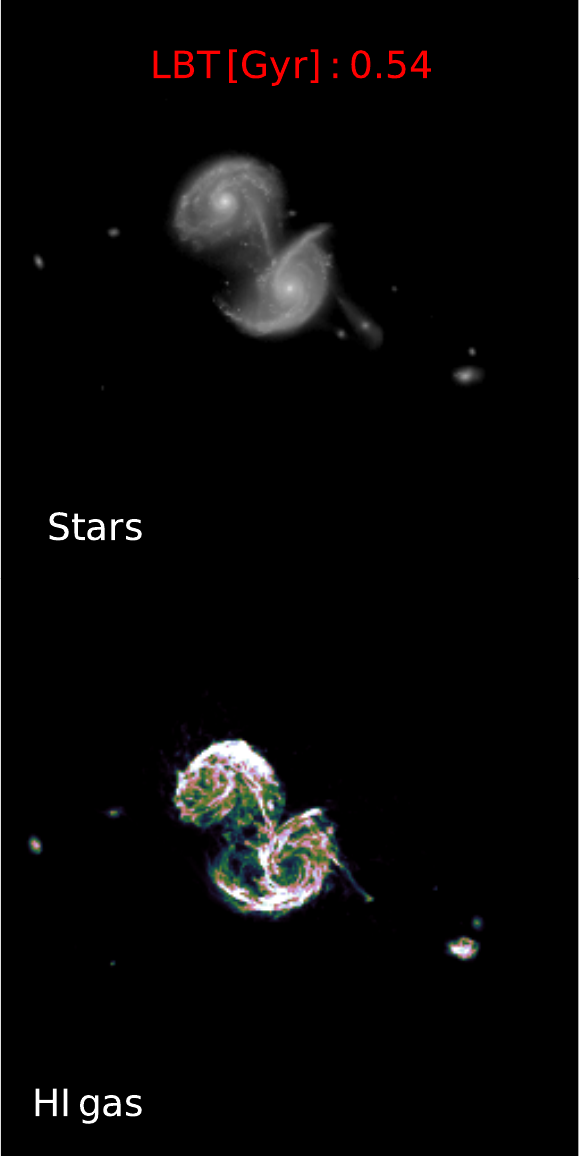}
    \caption{{\bf Au25}: Snapshot sequence of the significant recent interaction experienced with a massive satellite (i.e. total mass-ratio: $1$:$4$-$1$:$3$) over the last $1\, \mathrm{Gyr}$, which produced the coherent perturbation in both the stellar and HI components, as shown in Fig. \ref{fig:drivers_lopsidedness}. The pericentric passage occurs at $\sim1\, \mathrm{Gyr\, ago}$ consistent with the increase in the global lopsidedness $\langle \mathrm{A}_{1} \rangle$ of both tracers. The box length, centered on the main host galaxy, corresponds to twice the virial radius $R_{200}$ at each output time. The panels are shown in order of decreasing lookback time, from left to right.}
    \label{fig:Au25}
\end{figure*}

\begin{figure*}
    \centering
    \includegraphics[width=0.15\textwidth]{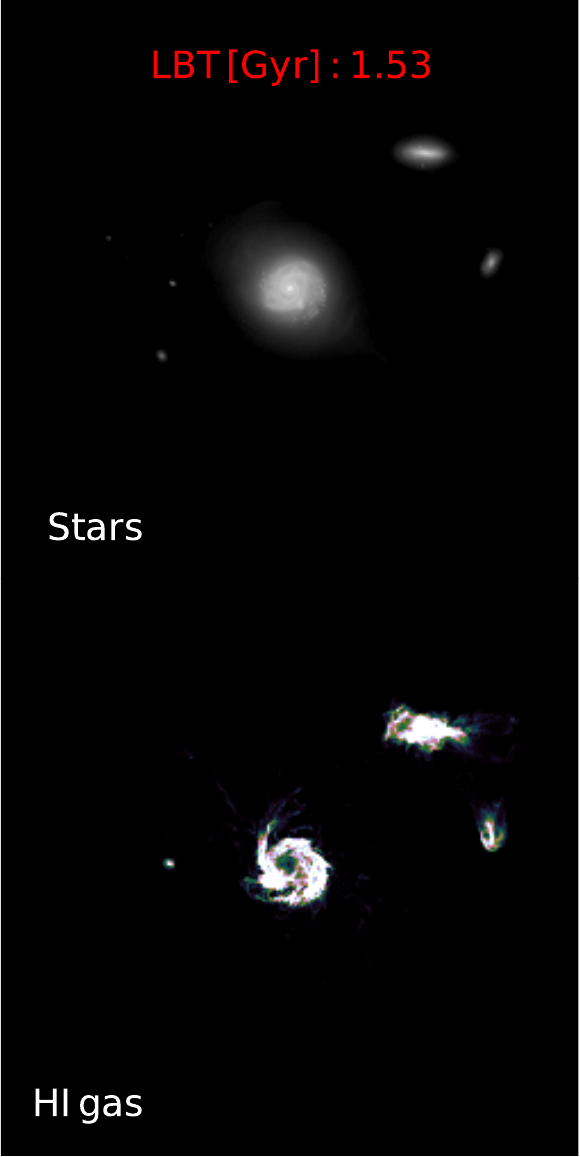}
    \includegraphics[width=0.15\textwidth]{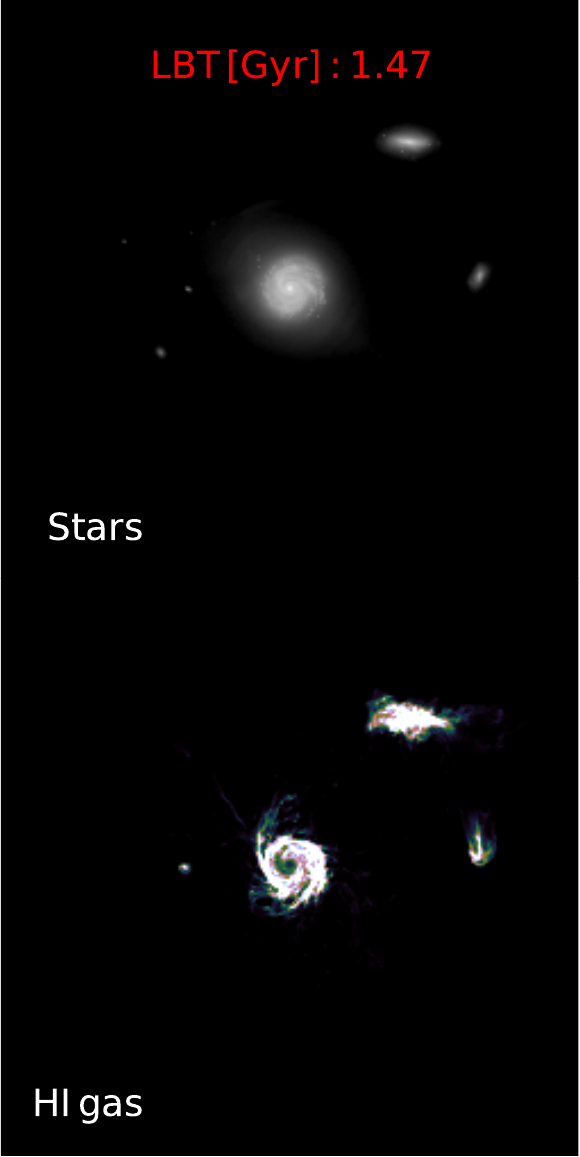}
    \includegraphics[width=0.15\textwidth]{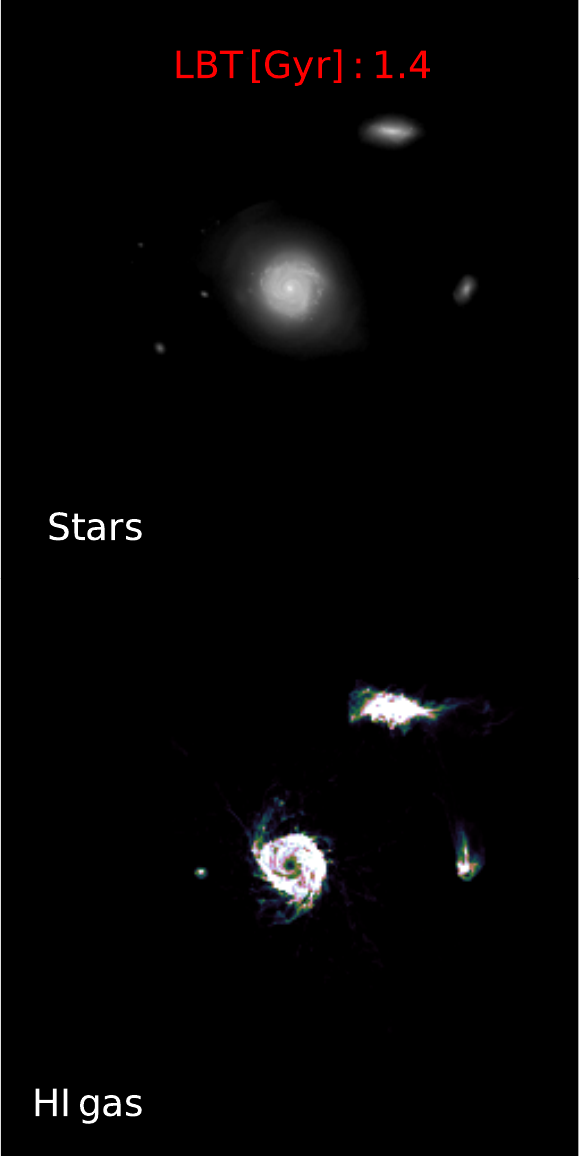}
    \includegraphics[width=0.15\textwidth]{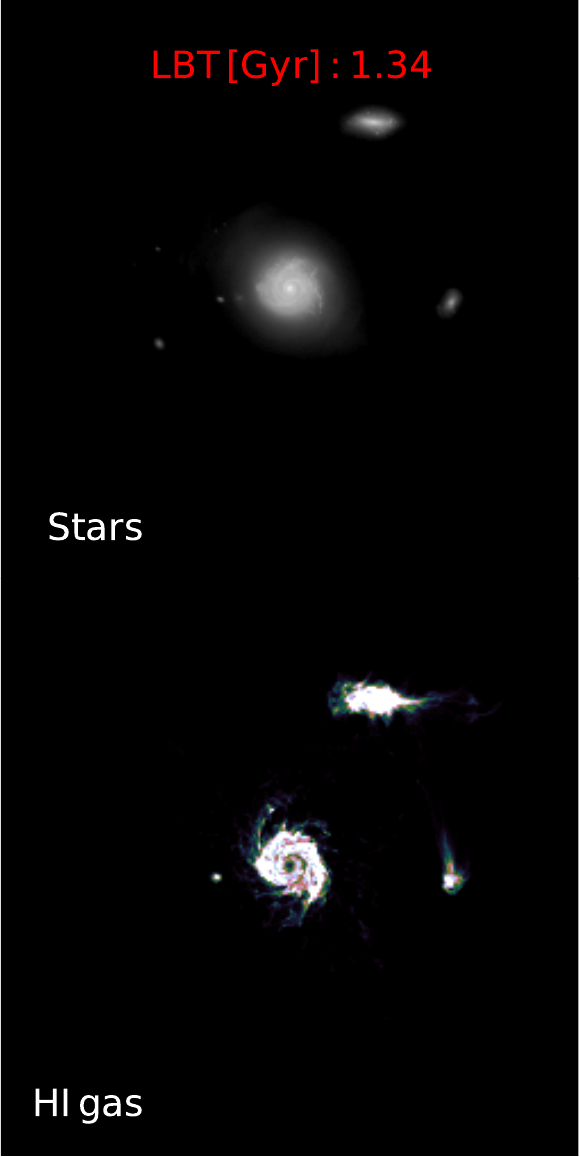}
    \includegraphics[width=0.15\textwidth]{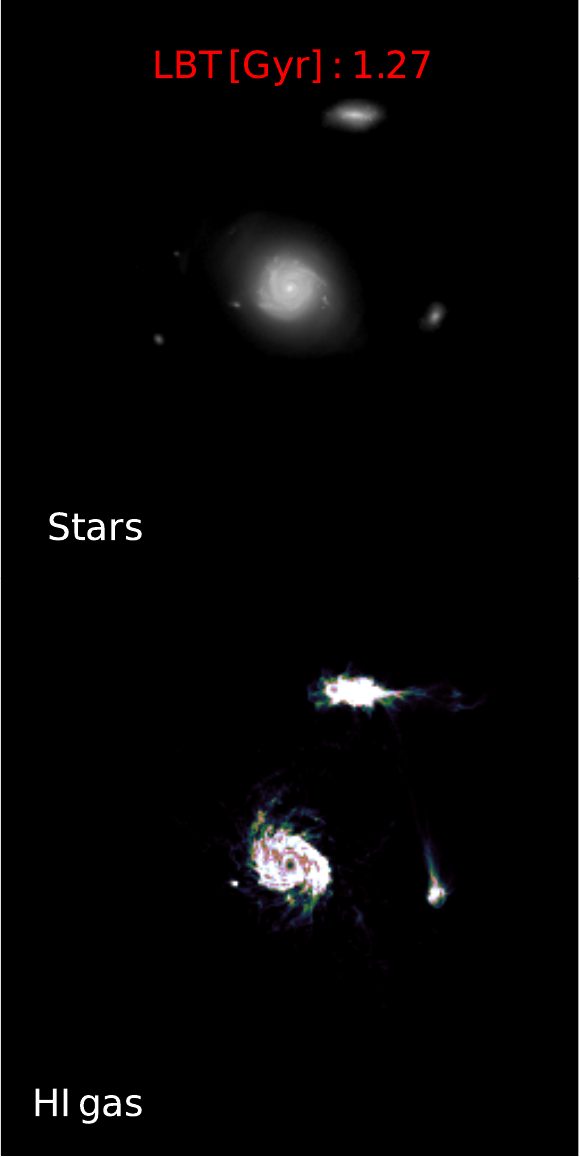}
    \includegraphics[width=0.15\textwidth]{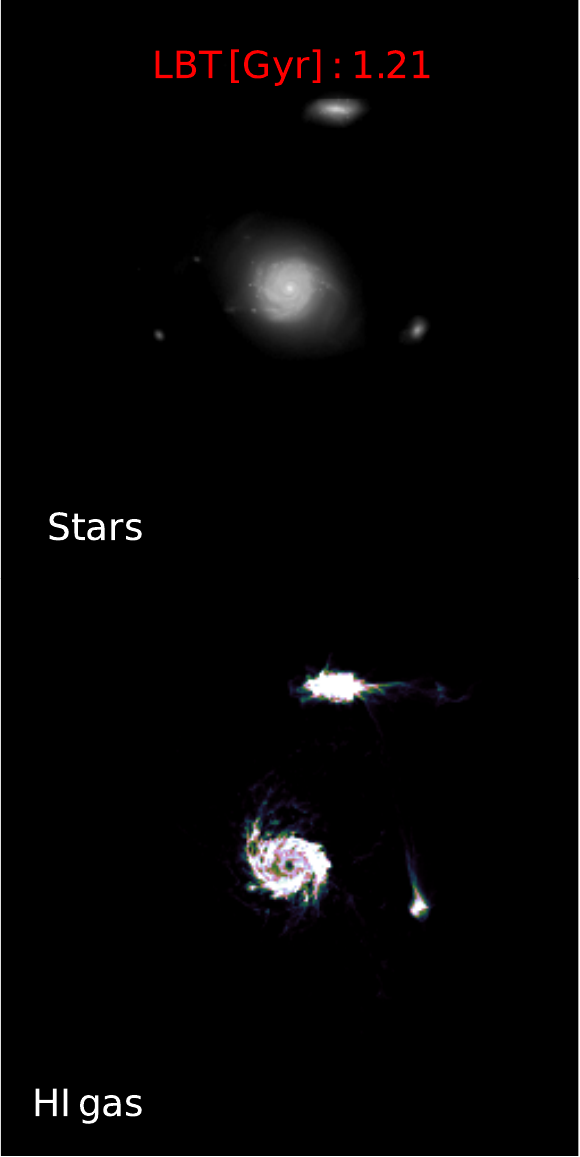}
    \caption{{\bf Au6}: Snapshot sequence of the orbit of a massive fly-by satellite (i.e. total mass-ratio: $\gtrsim1$:$20$). As shown in Fig. \ref{fig:drivers_lopsidedness}, the satellite never penetrates deeper than $0.5\, R_{200}$ and the pericenter is reached at $\sim1.5\, \mathrm{Gyr\, ago}$, consistent with the increase in the global lopsidedness $\langle \mathrm{A}_{1} \rangle$ of both the stellar and HI components. The box length, centered on the main host galaxy, corresponds to twice the virial radius $R_{200}$ at each output time. The panels are shown in order of decreasing lookback time, from left to right.}
    \label{fig:au6}
\end{figure*}

\begin{figure*}
    \centering
    \includegraphics[width=0.15\textwidth]{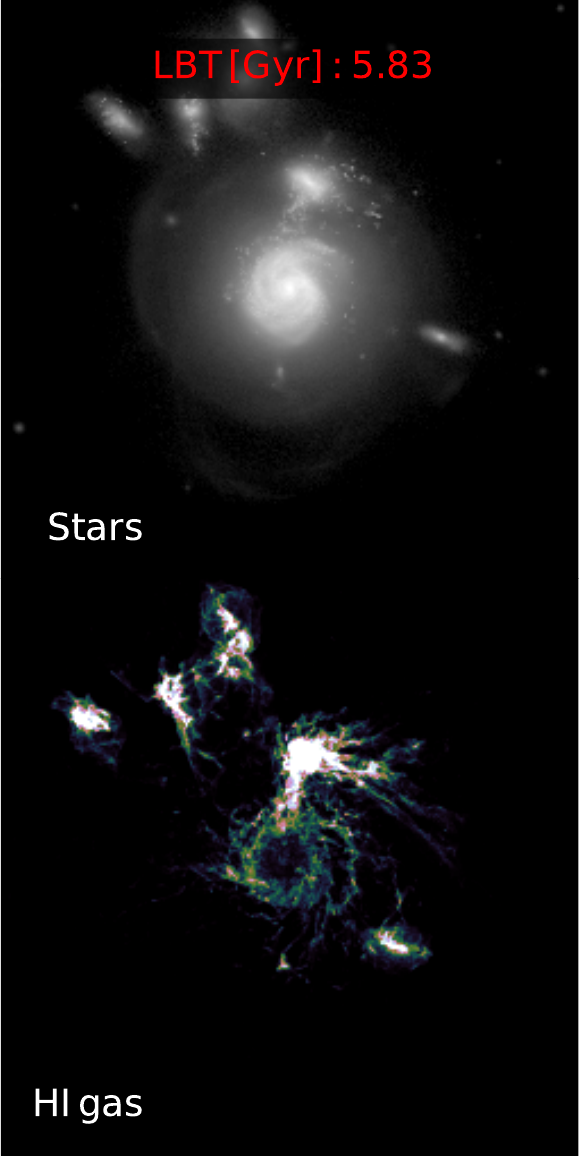}
    \includegraphics[width=0.15\textwidth]{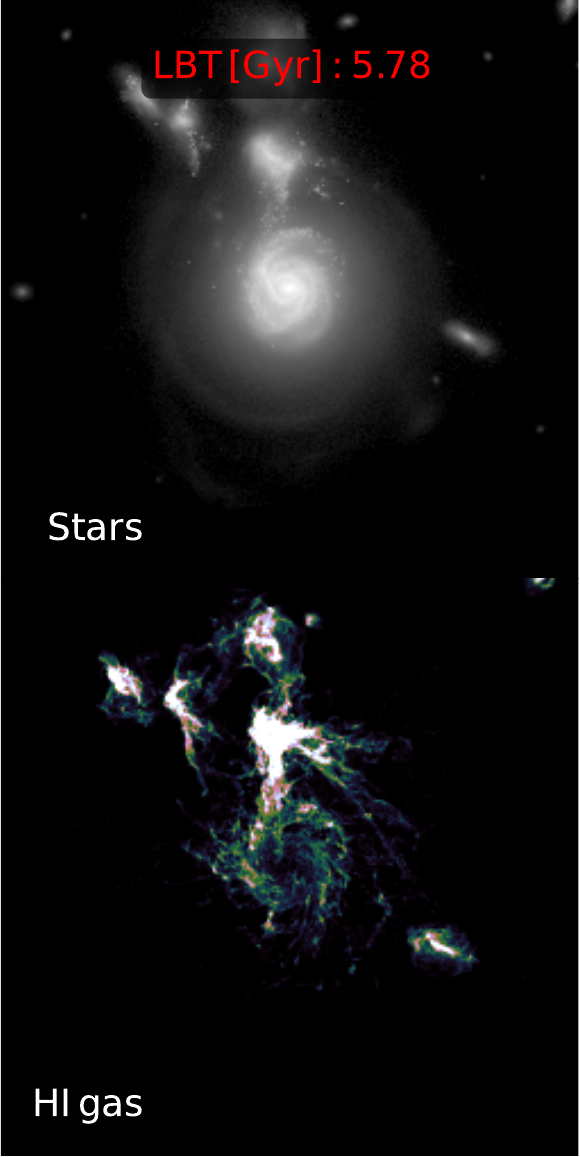}
    \includegraphics[width=0.15\textwidth]{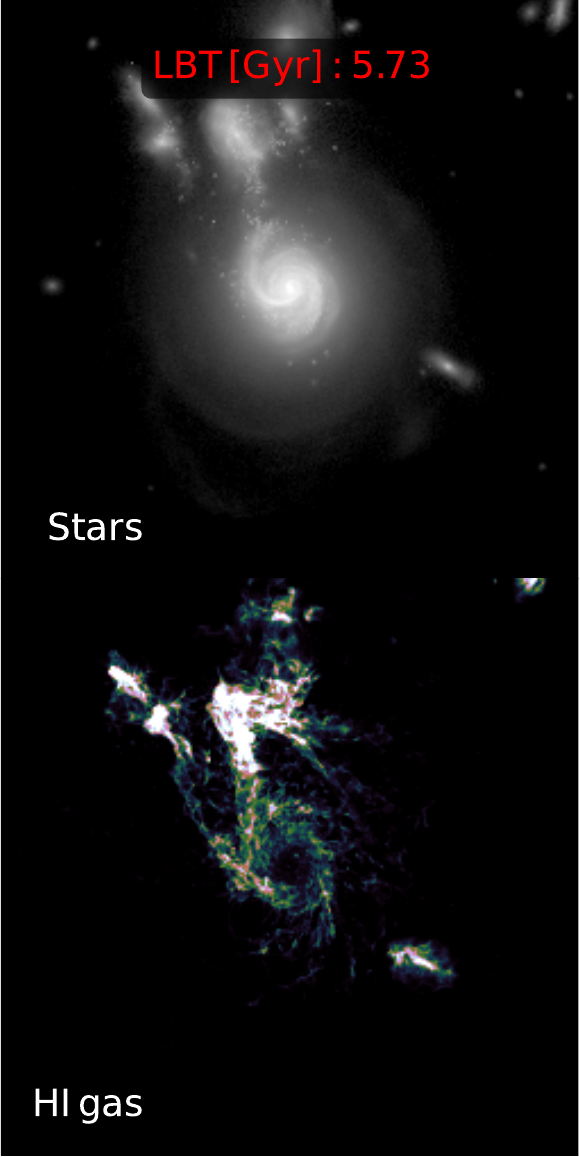}
    \includegraphics[width=0.15\textwidth]{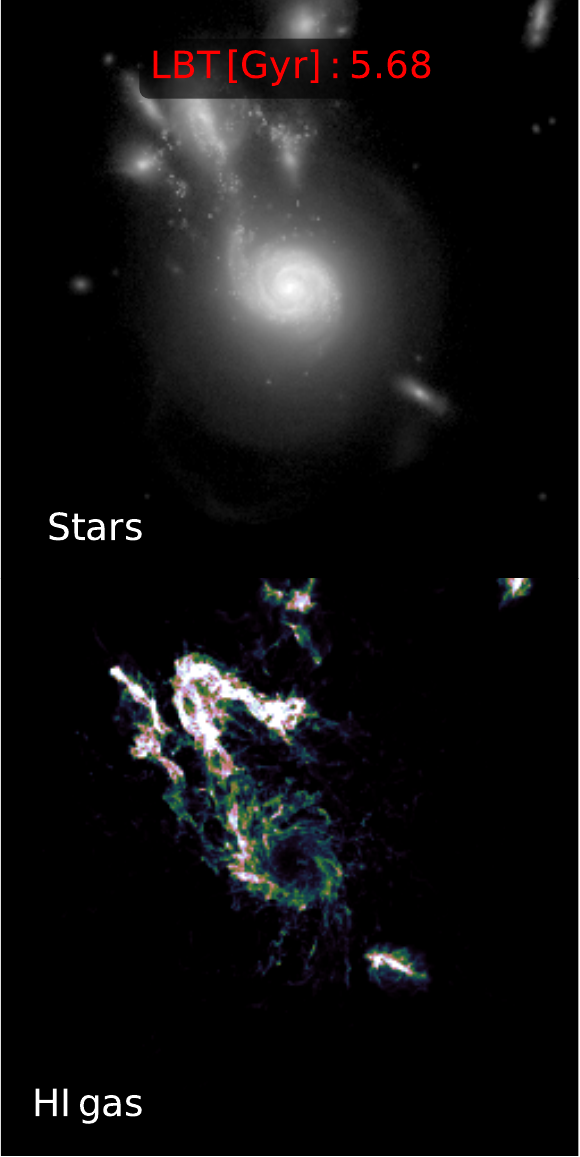}
    \includegraphics[width=0.15\textwidth]{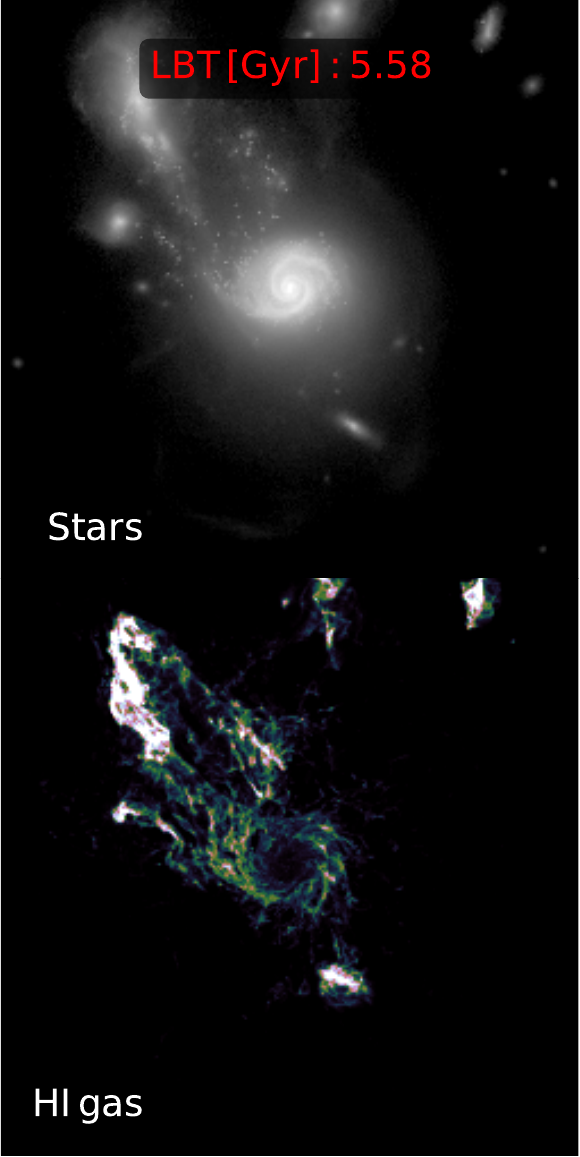}
    \includegraphics[width=0.15\textwidth]{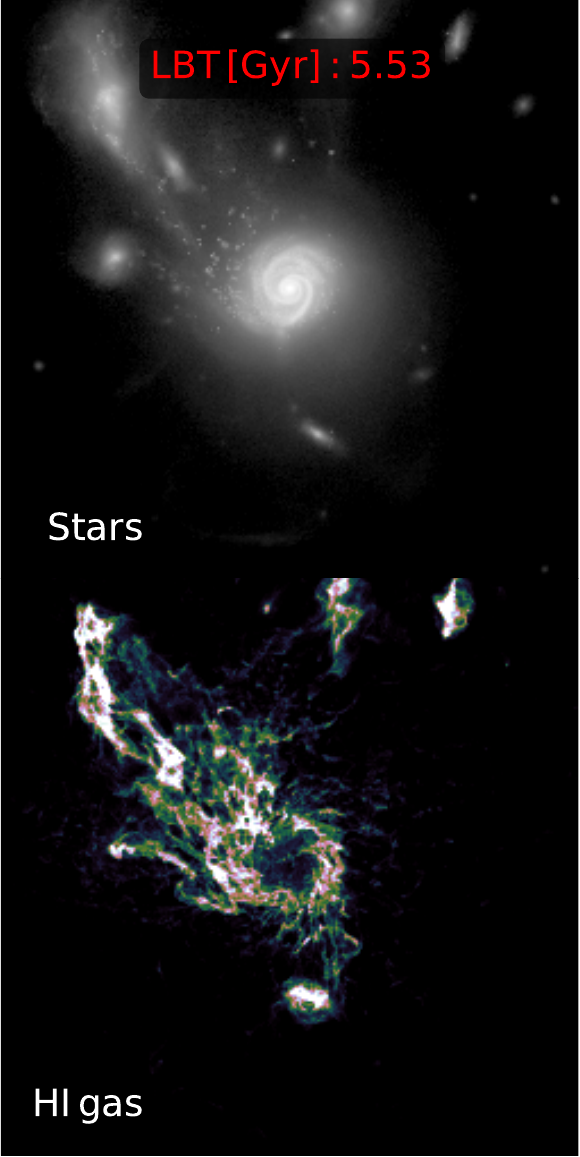}
    
    \includegraphics[width=0.15\textwidth]{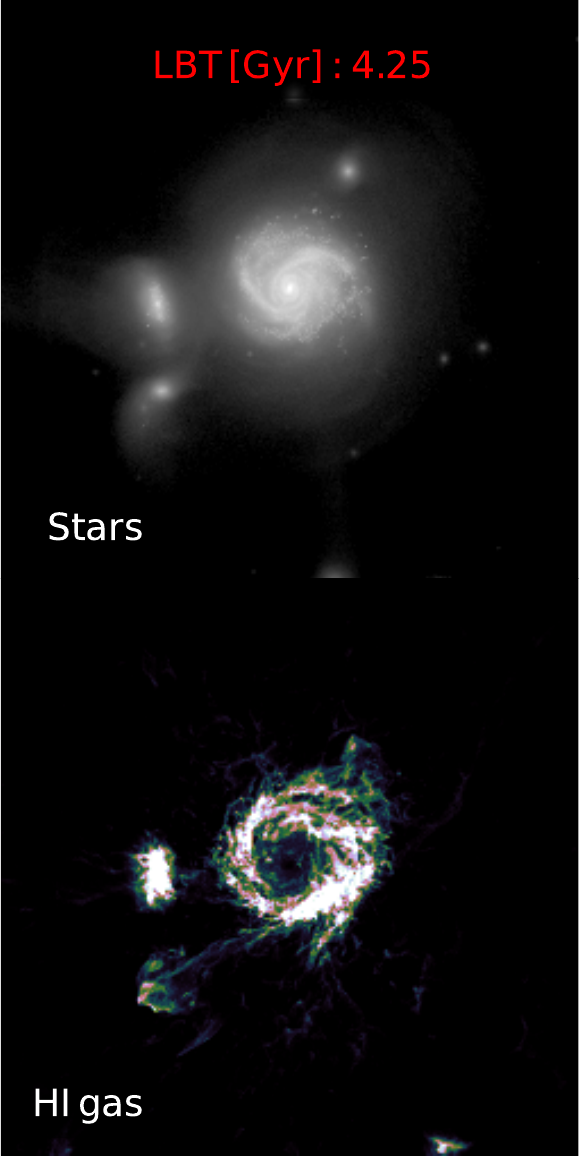}
    \includegraphics[width=0.15\textwidth]{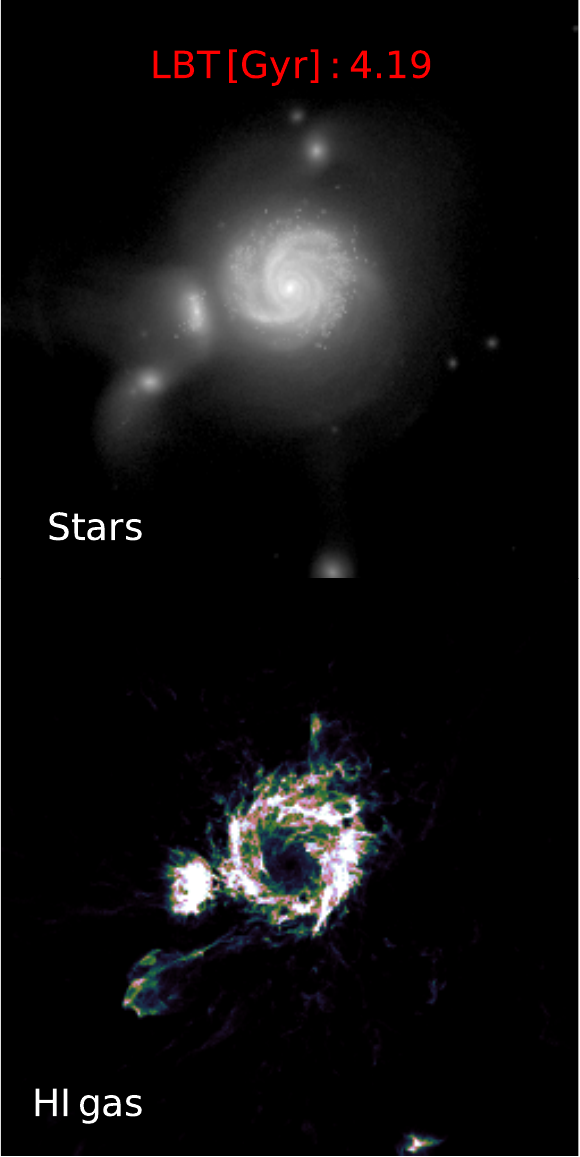}
    \includegraphics[width=0.15\textwidth]{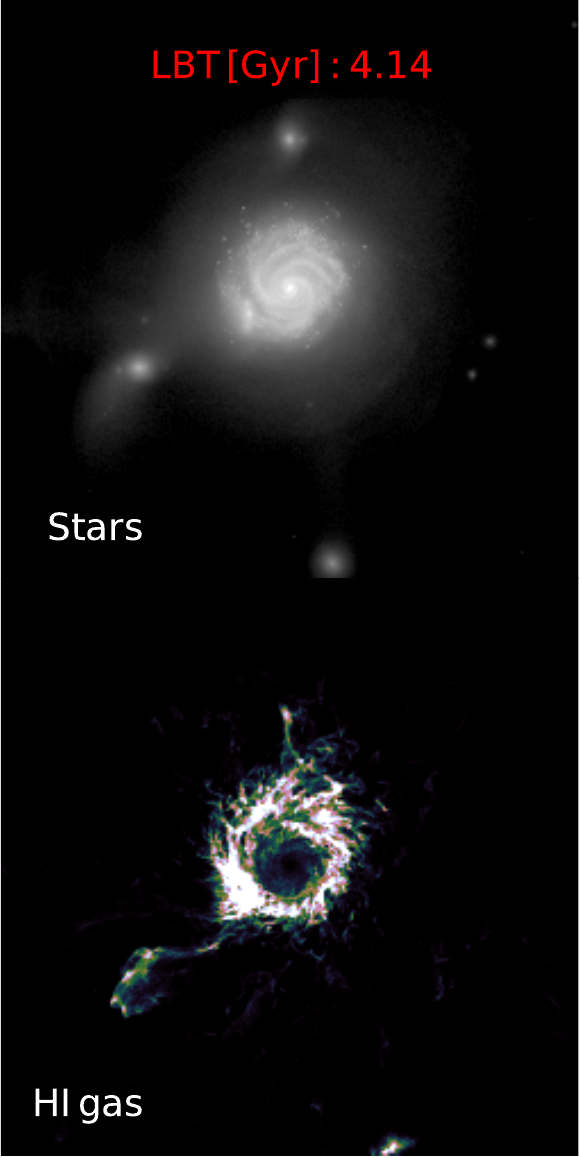}
    \includegraphics[width=0.15\textwidth]{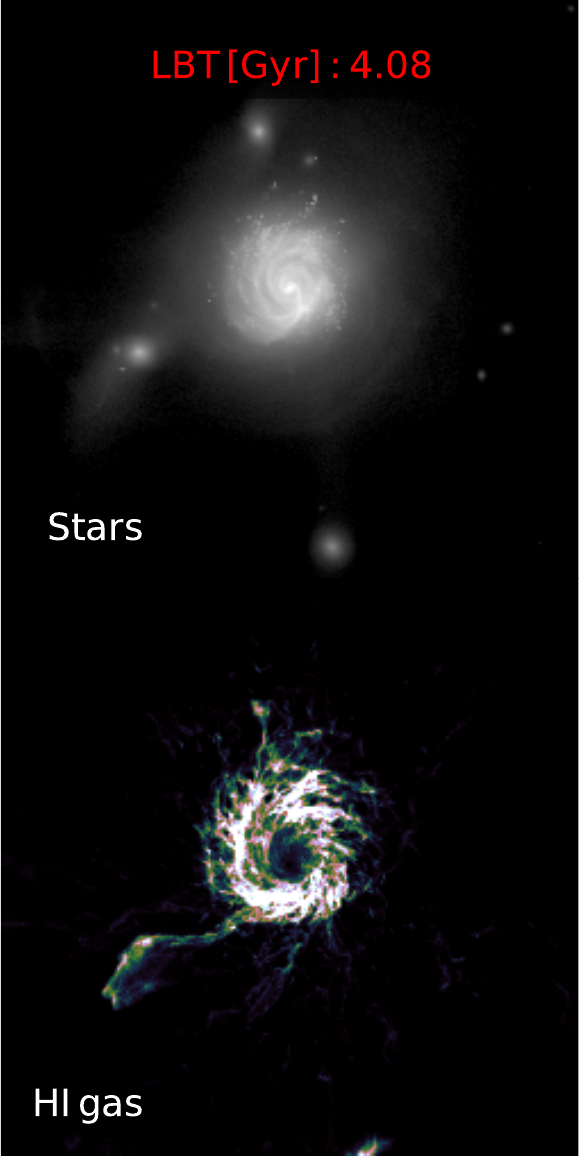}
    \includegraphics[width=0.15\textwidth]{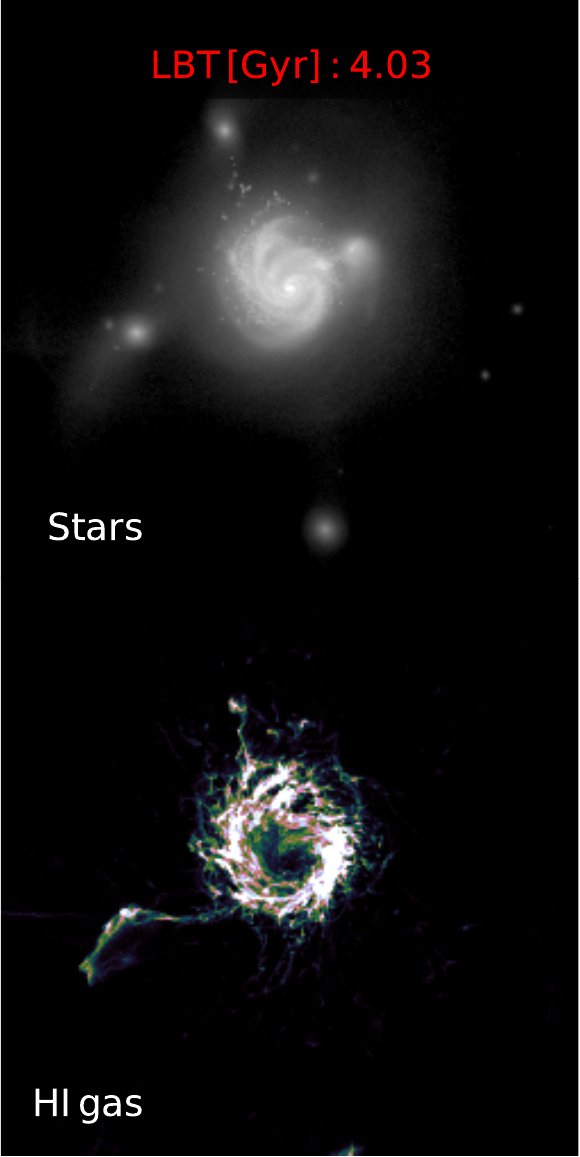}
    \includegraphics[width=0.15\textwidth]{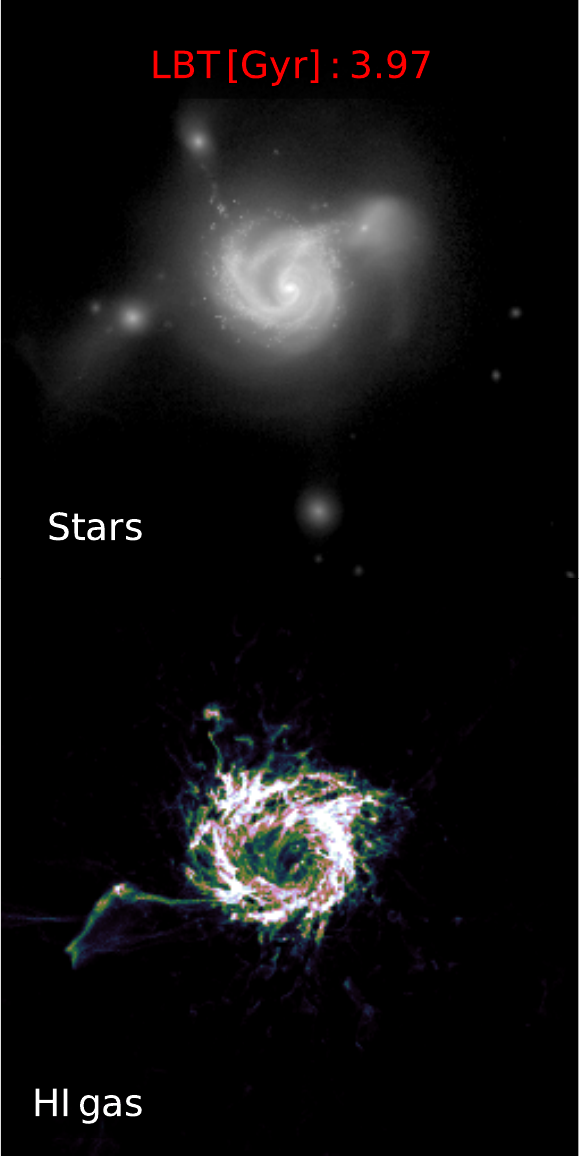}
    \caption{{\bf Au23}: Snapshot sequence showing the early smooth gas accretion from two gas-rich fly-by satellites at $\sim6\, \mathrm{Gyr\, ago}$ ({\it top panels}) and the subsequent significant interaction experienced with a massive satellite (total mass-ratio $\sim1$:$10$) during its pericenter passage at $\sim4\, \mathrm{Gyr\, ago}$ ({\it bottom panels}). As shown in Fig. \ref{fig:drivers_lopsidedness}, the former event only perturbed the HI component, while the latter perturbed both the stellar and HI components. The box length, centered on the main host galaxy, corresponds to one virial radius $R_{200}$ at each output time. The panels are shown in order of decreasing lookback time, from left to right.}
    \label{fig:au23}
\end{figure*}

We first highlight Au25 (Fig. \ref{fig:Au25}, Sec. \ref{sec:sample1}), which is an example of a tidally-induced lopsided perturbation in both its stellar and HI components, triggered by an interaction with a massive satellite within the last $1\, \mathrm{Gyr}$. Similarly, Au6 (Fig. \ref{fig:au6}, Sec. \ref{sec:sample1}) shows a recent peak in lopsidedness across both its stellar and HI components at $\sim1.5\, \mathrm{Gyr}$. This was likely caused by the disk responding to a dark matter halo perturbed by the fly-by of a massive satellite. Finally, we show Au23 (Fig. \ref{fig:au23}, Sec. \ref{sec:sample2}), which experienced gas accretion from a fly-by satellite at early times ($\sim6\, \mathrm{Gyr\, ago}$). This event induced lopsidedness in the HI gas while leaving the stellar component unperturbed (top panels of Fig. \ref{fig:au23}). Subsequently, a further interaction with another massive satellite at $\sim4\, \mathrm{Gyr\, ago}$ perturbed both the stellar and HI distributions (bottom panels of Fig. \ref{fig:au23}).

\end{document}